\documentclass[fleqn,usenatbib,useAMS]{mnras}

\usepackage{graphicx}	% Including figure files
\usepackage{amsmath}	% Advanced maths commands
\usepackage{amssymb}	% Extra maths symbols
\usepackage{multicol}        % Multi-column entries in tables
\usepackage{bm}		% Bold maths symbols, including upright Greek
\usepackage{pdflscape}	% Landscape pages
\usepackage{hyperref}
\usepackage{dcolumn}

\usepackage[T1]{fontenc}
\usepackage{ae,aecompl}

\usepackage{newtxtext,newtxmath}
\usepackage{eso-pic}% http://ctan.org/pkg/eso-pic

\AddToShipoutPictureBG*{%
  \AtPageUpperLeft{%
    \hspace{0.75\paperwidth}%
    \raisebox{-3.5\baselineskip}{%
      \makebox[0pt][l]{\textnormal{DES-2022-0691}}
}}}%

\AddToShipoutPictureBG*{%
  \AtPageUpperLeft{%
    \hspace{0.75\paperwidth}%
    \raisebox{-4.5\baselineskip}{%
      \makebox[0pt][l]{\textnormal{FERMILAB-PUB-22-261-PPD}}
}}}%

\title[Automated Uniform Lens Modeling]{STRIDES: Automated uniform models for 30 quadruply imaged quasars}

\author[T. Schmidt et al.]{
\parbox{\textwidth}{
\href{https://orcid.org/0000-0002-2772-8160}{T.~Schmidt,$^{1}$\thanks{Contact e-mail: \href{mailto:thomas@astro.ucla.edu}{thomas@astro.ucla.edu}}}
\href{https://orcid.org/0000-0002-8460-0390}{T.~Treu,$^{1}\dagger$}
\href{https://orcid.org/0000-0003-3195-5507}{S.~Birrer,$^{2,3,4}$}
\href{https://orcid.org/0000-0002-5558-888X}{A.~J.~Shajib,$^{5,1}\ddagger$}
\href{https://orcid.org/0000-0003-2456-9317}{C.~Lemon,$^{6}$} \href{https://orcid.org/0000-0001-7051-497X}{M.~Millon,$^{6}$}
\href{https://orcid.org/0000-0001-6116-2095}{D.~Sluse,$^{7}$} \href{https://orcid.org/0000-0001-9775-0331}{A.~Agnello,$^{8}$}
\href{https://orcid.org/0000-0003-0930-5815}{T.~Anguita,$^{9,10}$}
\href{}{M.~W. Auger-Williams,$^{11,12}$}
\href{https://orcid.org/0000-0001-8447-8869}{R.~G.~McMahon,$^{11,12}$}
\href{https://orcid.org/0000-0003-4446-7465}{V.~Motta,$^{13}$}
\href{https://orcid.org/0000-0002-5665-4172}{P.~Schechter,$^{14}$} \href{https://orcid.org/0000-0002-3909-6359}{C.~Spiniello,$^{15,16}$} \href{https://orcid.org/0000-0002-9267-2677}{I.~Kayo,$^{17}$} 
\href{https://orcid.org/0000-0003-0758-6510}{F.~Courbin,$^{6}$}
\href{https://orcid.org/0000-0002-5085-2143}{S.~Ertl,$^{18,19}$} 
\href{https://orcid.org/0000-0002-4030-5461}{C.~D.~Fassnacht,$^{20}$} \href{https://orcid.org/0000-0003-4079-3263}{J.~A.~Frieman,$^{21,5}$}
\href{https://orcid.org/0000-0001-7714-7076}{A.~More,$^{22,23}$}
\href{https://orcid.org/0000-0003-2497-6334}{S.~Schuldt,$^{18,19}$}
\href{https://orcid.org/0000-0001-5568-6052}{S.~H.~Suyu,$^{18,19,24}$}
M.~Aguena,$^{25}$
F.~Andrade-Oliveira,$^{26}$
J.~Annis,$^{21}$
D.~Bacon,$^{27}$
E.~Bertin,$^{28,29}$
D.~Brooks,$^{30}$
D.~L.~Burke,$^{2,3}$
A.~Carnero~Rosell,$^{31,25,32}$
M.~Carrasco~Kind,$^{33,34}$
J.~Carretero,$^{35}$
C.~Conselice,$^{36,37}$
M.~Costanzi,$^{38,39,40}$
L.~N.~da Costa,$^{25,41}$
M.~E.~S.~Pereira,$^{42}$
J.~De~Vicente,$^{43}$
S.~Desai,$^{44}$
P.~Doel,$^{30}$
S.~Everett,$^{45}$
I.~Ferrero,$^{46}$
D.~Friedel,$^{33}$
J.~Garc\'ia-Bellido,$^{47}$
E.~Gaztanaga,$^{48,49}$
D.~Gruen,$^{50}$
R.~A.~Gruendl,$^{33,34}$
J.~Gschwend,$^{25,41}$
G.~Gutierrez,$^{21}$
S.~R.~Hinton,$^{51}$
D.~L.~Hollowood,$^{45}$
K.~Honscheid,$^{52,53}$
D.~J.~James,$^{54}$
K.~Kuehn,$^{55,56}$
O.~Lahav,$^{30}$
F.~Menanteau,$^{33,34}$
R.~Miquel,$^{57,35}$
A.~Palmese,$^{58}$
F.~Paz-Chinch\'{o}n,$^{33,11}$
A.~Pieres,$^{25,41}$
A.~A.~Plazas~Malag\'on,$^{59}$
J.~Prat,$^{60,5}$
M.~Rodriguez-Monroy,$^{61}$
A.~K.~Romer,$^{62}$
E.~Sanchez,$^{43}$
V.~Scarpine,$^{21}$
I.~Sevilla-Noarbe,$^{43}$
M.~Smith,$^{63}$
E.~Suchyta,$^{64}$
G.~Tarle,$^{26}$
C.~To,$^{52}$
and T.~N.~Varga$^{65,66,67}$
\begin{center} (DES Collaboration) \end{center}
}
\\\\
Affiliations can be found at the end of the manuscript.
}

\date{\today}

\pubyear{2022}

\begin{document}
\label{firstpage}
\pagerange{\pageref{firstpage}--\pageref{lastpage}}
\maketitle

\begin{abstract}
Gravitational time delays provide a powerful one step measurement of $H_0$, independent of all other probes.
One key ingredient in time delay cosmography are high accuracy lens models.
Those are currently expensive to obtain, both, in terms of computing and investigator time (10$^{5-6}$ CPU hours and $\sim$ 0.5-1 year, respectively). Major improvements in modeling speed are therefore necessary to exploit the large number of lenses that are forecast to be discovered over the current decade. In order to bypass this roadblock, building on the work by \citet{Shajib19}, we develop an automated modeling pipeline and apply it to a sample of 30 quadruply imaged quasars and one lensed compact galaxy, observed by the Hubble Space Telescope in multiple bands. Our automated pipeline can derive models for 30/31 lenses with few hours of human time and $<100$  CPU hours of computing time for a typical system.
For each lens, we provide measurements of key parameters and predictions of magnification as well as time delays for the multiple images.
We characterize the cosmography-readiness of our models using the stability of differences in Fermat potential (proportional to time delay) w.r.t. modeling choices. We find that for 10/30 lenses our models are cosmography or nearly cosmography grade ($<3$\% and 3-5\% variations). For 6/30 lenses the models are close to cosmography grade (5-10\%).  These results are based on informative priors and will need to be confirmed by further analysis. However, they are also likely to improve by extending the pipeline modeling sequence and options. In conclusion, we show that uniform cosmography grade modeling of large strong lens samples is within reach.

\end{abstract}


%%%%%%%%%%%%%%%%%%%%%%%%%%%%%%%%%%%%%%%%%%%%%%%%%%

%%%%%%%%%%%%%%%%% BODY OF PAPER %%%%%%%%%%%%%%%%%%

% The MNRAS class isn't designed to include a table of contents, but for this document one is useful.
% I therefore have to do some kludging to make it work without masses of blank space.
% \begingroup
% \let\clearpage\relax
% \tableofcontents
% \endgroup

\section{Introduction}

Our most successful cosmological model to date, the $\Lambda$ Cold Dark Matter ($\Lambda$CDM) model, has been able to accurately explain a plethora of cosmological observations in the early and late universe, including observations of the cosmic microwave background (CMB) radiation, the Big Bang nucleosynthesis, the formation of large scale structures and galaxy clustering, and the acceleration in the expansion of our universe \citep[e.g.,][]{PlanckCollaboration18, Eisenstein05, Riess98, Perlmutter99}. However, over the last few years, the tension in the measurements of the Hubble constant, which quantifies the Universe's current expansion rate, has been increasing between probes of the early Universe, i.e. measurements using the information contained within the CMB, and the probes of the late Universe, such as methods using the local distance ladder. Early-Universe measurements of the CMB give a Hubble constant of $67.4 \pm{0.5}$ km s$^{-1}$ Mpc$^{-1}$ \citep{PlanckCollaboration18} while observations of the late Universe measure $H_0$ at a higher value of $73.0\pm{1.4}$ km s$^{-1}$ Mpc$^{-1}$ \citep{Riess21}, resulting in a currently 5-6 $\sigma$ tension between the two measurements \citep{Verde19, Wong20}. Solving this tension, if confirmed, would require new Physics, for example changing the sound horizon at recombination via the introduction of a new relativistic particle or a form of early dark energy \citep{Knox20, DiValentino21}. Given the importance of the tension, it is  imperative to develop multiple independent methods with sufficiently high precision to confirm the $H_0$ tension or possibly rule it out. 

Strong gravitational lensing, where the lensed source is a  multiply-imaged quasar, provides a powerful cosmological probe that can be used to determine the Hubble constant, independent of measurements relying on the local distance ladder \citep{Refsdal64}. Light rays from a variable point source, the quasar, traverse the gravitational potential of a foreground galaxy, the lens or deflector, with paths of different lengths and through different points in the gravitational field of the deflector. Therefore, we observe different images of the same quasar in the plane of the lens, the image plane. High cadence, long-term observations of the lensed source allow us to use the intrinsic quasar variability to measure the time delay of the variations between the four observed images in the plane of the lens \citep{Tewes13}. Since the angular diameter distances from the observer to the main deflector, $D_{\rm{d}}$, from the observer to the source, $D_{\rm{s}}$, and from the deflector to the source, $D_{\rm{ds}}$, are much greater than the physical extent of the lensing galaxy, we can simplify the geometry of the problem by considering a two dimensional deflector, which leads to the following expression for the measurable time-delay distance between two images A and B:

\begin{equation}
    \Delta t_{\rm{AB}} = \frac{D_{\Delta t}}{c}  \Delta \Phi _{\rm{AB}} ,
\end{equation}

\noindent
where the time-delay distance, $D_{\Delta t}$, is related to the angular diameter distances and the main deflector's redshift, $z_{\rm{d}}$, by:

\begin{equation}
    D_{\Delta t} = (1 + z_{\rm{d}}) \frac{D_{\rm{d}} D_{\rm{s}}}{D_{\rm{ds}}}.
\end{equation}

\noindent
$\Delta \Phi_{\rm{AB}}$ represents the difference in the Fermat potential of the lens at the position of the images A and B, and $c$ is the speed of light. The Fermat potential for an image position, {\bm{$\theta$}}, and source position, {\bm{$\beta$}}, is given by:

\begin{equation}
    \Phi(\bm{\theta; \beta}) = \frac{1}{2}(\bm{\theta - \beta})^2 - \psi(\bm{\theta}) ,
\end{equation}

\noindent
where the deflection potential, $\psi(\bm{\theta})$, is related to the projected surface mass density (or convergence), $\kappa$, by:

\begin{equation}
    \nabla^2 \psi = 2 \kappa.
\end{equation}

\noindent
Therefore, if we can recover the Fermat potential for a given lens configuration by reconstructing a model that matches high resolution imaging of the system, we are able use the measured time delays between lensed quasar image positions to determined the time-delay distance, which is inversely proportional to the Hubble constant.

Achieving $\sim1$ \% precision in the Hubble constant requires a sample size of at least $\sim$40 systems  \citep{Treu16, Shajib18, Birrer20b}. Fortunately, ongoing and future wide-field deep-sky surveys are expected to rapidly increase the number of known quadruply imaged quasars \citep[e.g.,][]{Oguri10, Collett15}. Indeed, in recent years, the discovery rate has accelerated owing to the large dataset and the development of automatic detection algorithms \citep[e.g.,][]{Agnello15,Williams17, Williams18, Lemon18}. Thus, the prospect of precise and accurate  Hubble constant measurements from strong gravitational lensing is bright, provided sufficient resources can be devoted to follow-up and model the systems.  High-precision models of strong lens systems are currently very time consuming, with an approximate 6 to 12 months of investigator time required per lens, depending on the complexity of the deflectors involved. Therefore, major improvements in modeling speed are required to scientifically exploit the anticipated influx of newly discovered strong lenses.

This paper takes an important step towards relieving the bottleneck created by time limitations in the modeling speed. Using an improved version of the uniform lens modeling framework set forth by \citet{Shajib19}, we built an automated pipeline to model strong gravitational lenses expanded around an elliptical power-law mass profile for a system's central main deflector. To facilitate the reconstruction of a wide array of lenses with varying intricacies, the pipeline makes modeling choices selected from a uniform set of components for mass and light profiles to iteratively increase each lens model's complexity until a good fit is found that accurately matches the observational data for the object. With this automated approach, we are able to process sets of strong lenses that are much larger than in previous studies and reduce the requirement of an investigator's involvement to ancillary tasks, such as data reduction and addressing failure modes. These advantages make the automated pipeline a powerful springboard for the scientific analysis of the expected increase in newly discovered lensed systems.

We apply our automated lens modeling pipeline to a sample of 31 strong gravitational lenses imaged by the Hubble Space Telescope ({\it{HST}}) during cycles 25 and 26 between the years 2017 and 2020 in filters F160W, F475X, and F814W. To assess the stability of the difference in the Fermat potential at the position of the lensed quasar in the image plane, we introduce a new metric that allows us to visualize and test the impact of the pipeline's modeling choices on the Fermat potential at the population level. To demonstrate its usefulness, we further use this new metric to address the impact of the source complexity level in a model by introducing small perturbations in the source light structure and evaluate if, and by how much, the introduced perturbations change the stability of the Fermat potential difference between image positions.

The paper is organized in the following manner: Section~\ref{HST_Sample} gives a description of our sample, highlights the data reduction, and discusses {\it{HST}} cycle 26 lenses. Detail on our lens modeling procedures, along with the parameterization of mass and light profiles, are listed in Section~\ref{Uniform_Lens_Modeling}, uniform lens modeling. The results of our analysis are presented in Section~\ref{Results}, results. We address the impact of modeling choices and underlying systematic uncertainties in source complexity in Section~\ref{Discussion} and conclude with a summary in Section~\ref{Summary}.
Magnitudes are reported in the AB system and whenever necessary we use a cosmological concordance model with parameters $H_0 = 70$ km s$^{-1}$ Mpc$^{-1}$, $\Omega_{\rm{m,0}} = 0.3$, and $\Omega_{\rm{\Lambda, 0}} = 0.7$.
\\

\section{{\emph{HST}} Sample} \label{HST_Sample}

Our sample consists of 31 lenses from {\it HST} cycle 25 and cycle 26, with the cycle 25 lenses consiting of the same sample as modelled by \citet{Shajib19}. The targets that were observed during {\it HST} cycle 26 consist of 16 quads and two five-image systems with two main deflectors for a total of 18 lenses. While information about cycle 25 targets are listed in \citet{Shajib19}, a brief description of the main characteristics and respective discovery of the cycle 26 sample can be found below in Section~\ref{lens_descriptions}.\\

\subsection{Data and Data Reduction} \label{Data_Reduction}

The observations of the lenses in our sample were taken by the Hubble Space Telescope under cycle 25 and cycle 26 programs HST-GO-15320 and HST-GO-15652 (PI: Treu), respectively, using the  Wide Field Camera 3 (WFC3). With the exception of one lens, W2M J1042+1641, exposures for each lens were taken in three filters, F160W for infrared (IR) data and F475X, as well as F814W, for ultraviolet-visual (UVIS) data. For W2M J1042+1641 the two programs did not obtain data in the IR channel and restricted the observation to the UVIS bands, as IR images are available from a previous {\it HST} visit as explained in the description of the lens below. In order to improve the sampling of the data, 
we adopted a 4-point dither pattern in the IR channel, while for the UVIS channel observations we adopted a 2-point dither pattern. To properly sample the full dynamic range of the data, including areas around the bright quasar images, we took a long and short exposure at each dither point. The total exposure times per filter band are comparable to the exposure times of the 13 lenses listed in Table 1 of \citet{Shajib19}, since observations took place with the same instrument under an identical strategy. 
For our data reduction, as well as alignment and combination of the various exposures in each filter, we use the Python package \textsc{AstroDrizzle} \citep{Avila15}. The pixel size in the final reduced and combined images is $0.08\arcsec/ \rm{pix}$ for IR exposures and $0.04\arcsec/ \rm{pix}$ for exposures in the UVIS bands.\\

\subsection{Notes on individual quads} \label{lens_descriptions}

This section gives a brief description of each quadruply imaged quasar in our sample, regardless of whether it was successfully modeled by the automated pipeline or whether a model needs further work.\\

\subsubsection{J0029-3814}

J0029-3814 was discovered among extragalactic objects with astrometric anomalies between the optical and infrared in VEXAS \citep{Spiniello-Agnello19}, further prioritized as a "naked cusp" candidate from model-based deblending of its image cutouts \citep[following][]{Morgan04}, and its spectroscopic confirmation at the 3.5m ESO-NTT (PI T.~Anguita) determined a preliminary source redshift z=2.821, while the deflector redshift needs deeper follow-up with larger facilities (Schechter et al. in prep).

\subsubsection{PS J0030-1525}

This lens was discovered by \citet{Lemon18} by cross-matching multiple catalogued detections in \textit{Gaia} Data Release 1 (DR1) against photometric quasar candidates from the Wide-field Infrared Spectroscopic Explorer. The imaging in Pan-STARRS shows just two blue point sources offset from a galaxy, and follow-spectroscopy confirmed these to both be quasars at z=3.36. An archival VST-ATLAS $r$-band image revealed a likely counterimage, and \citet{Lemon18} suggested that this system is likely a fold quad, with image A composed of a merging pair. They modeled the system as an SIE + shear, predicting flux ratios of 7:7:3:1 (ABCD), yet only measuring 7:0.5:4:1, suggesting a strong demagnification of image B. They report a particularly large best-fit total model magnification of 71.

\subsubsection{DES J0053-2012}

DES J0053-2012 was discovered and confirmed by \citet{lemon2020}, after being selected in \textit{Gaia} DR1 as a double detection associated to a red {\it WISE} detection. The source redshift is $\approx$3.8, however this is uncertain due to absorption and possible blueshift of the broad quasar emission lines. \citet{lemon2020} find that a SIE + shear model is insufficient to reproduce the image positions, but including an SIE for the galaxy 4 arcseconds to the South-East provides a good fit to the system.

\subsubsection{WG0214-2105}

WG0214-2105 was discovered by \citet{Agnello18d} as a \textit{Gaia} multiplet corresponding to an extragalactic candidate from its {\it WISE} magnitudes. It has a high UV deficit and "blue" {\it WISE} colours, which are more similar to those of known white dwarfs and may explain why it was discovered only once the ESA-\textit{Gaia} mission pipeline resolved it into multiple source detections.
Its source redshift is  3.229$\pm$0.004, and the deflector's photometric redshit is  0.22$\pm$0.09, as it was too faint to obtain a secure spectroscopic redshift on the 10m SALT follow-up \citep[PI L.~Marchetti;][]{Spiniello19}.

\subsubsection{DES J0530-3730}

This system was discovered using the method described by \citet{Ostrovski17}, and by Lemon et al.\ (in prep) as a triple detection in \textit{Gaia} DR2 around a photometric quasar candidate. The coordinates are RA=05:30:36.984, DEC=-37:30:11.16 (J2000). It was  confirmed as a quasar at z=$2.838$ from spectra obtained at the NTT in December 2016 during the run described by \citet{Anguita18}.

\subsubsection{J0659+1629}

This system was discovered by Lemon et al. in prep as a triple detection in \textit{Gaia} DR2 around a photometric quasar candidate. They confirm the source redshift to be 3.09, and an SIE + shear model requires only a modest shear of 0.06, however the predicted flux of image D is 60\% fainter than observed, suggesting variability over the time delay as a possible cause for this discrepancy. The system was also independently selected by \citet{delchambre2019} using the astrometry of the three \textit{Gaia} DR2 detections, and \citet{stern2021} also spectroscopically confirm that the source redshift is 3.083, and the lens redshift is 0.766. \citet{stern2021} model the system as an SIS + shear, however their flux ratios are poorly reproduced and the ellipticity is unrealistic. \citet{stern2021} suggest this is indicative of a missing nearby galaxy.

\subsubsection{J0818-2613}

This system was discovered by Lemon et al. in prep as four detections in \textit{Gaia} DR2 around a photometric quasar candidate. They confirm the source redshift to be a BAL quasar at z=2.155. Their SIE + shear model recovers the image positions, but is highly unphysical with perpendicular shear and mass ellipticity, suggesting that the system is likely lensed by a complex mass distribution composed of several galaxies. The system was also independently confirmed by \citet{stern2021} who measure a source redshift of 2.164. They reach the same conclusion as Lemon et al. in prep. regarding the likely presence of a galaxy group or cluster.

\subsubsection{W2M J1042+1641}

Information about the discovery, main characteristics, and measured redshifts for this system can be found in the paper by \citet{Glikman18}.
For this target we obtained UVIS data only. The IR data used to model this lens was observed with {\it HST} Proposal 14706 (PI: E. Glikman), which is publicly available from the {\it HST} archive.

\subsubsection{J1131-4419}

This system was found using \textit{Gaia} catalogue positions as potential quad configurations using extremely randomized trees by \citet{krone-martins2018} and \citet{delchambre2019} as  GRAL113100-441959. It was spectroscopically confirmed by \citet{wertz2019}, who measure a source redshift of 1.09, and present models in the absence of the lensing galaxy position.

\subsubsection{2M1134-2103}

This bright quad was discovered serendipitously by \citet{lucey2018} while visually inspecting the target catalogue of the Taipan Galaxy Survey. \citet{rusu2019} obtained spectra for this system, confirming the source to be at 2.77. Both papers confirm that a large shear is required to model the system. \citet{rusu2019} detect a companion object in the Pan-STARRS r and i PSF-subtracted images $\approx$ 4 arcseconds South-East of the system, which they suggest could be partly responsible for the shear. The strong shear could also be due to a galaxy group $\approx$1 arcminute North-West of the system.

\subsubsection{J1537-3010}
This system was discovered by \citet{Lemon19}, who obtained a source redshift of 1.72. They are able to fit the system well with an SIE + shear model. The system was also independently selected using \textit{Gaia} astrometry by \citet{delchambre2019} and spectroscopically confirmed by \citet{stern2021}, who corroborate a source redshift of 1.721.

\subsubsection{J1721+8842}

The system was originally discovered by \citet{Lemon19} who confirmed the source to be at z$\approx$2.37, with strong absorption features. \textit{Gaia} DR2 catalogues 5 detections, and an in-depth study of this system by \citet{lemon2022} show that there are two quasar sources at similar redshifts, with one being lensed into four images, and one into two images. They provide several mass models for the system, which we will compare to in Section \ref{lensmodels}. The source is also unique in that the bright images A and C are confirmed to have a proximate damped Lyman alpha absorber.

\subsubsection{J1817+2729}
This system was discovered by \citet{Lemon18} as a \textit{Gaia} quartet associated with a photometric quasar candidate in {\it WISE}. Only three of these detections were due to the images of the system, with the fourth due to a nearby star. They measure a source redshift of 3.07. The system was independently selected by \citep{delchambre2019} and confirmed spectroscopically by \citet{stern2021}, who measured a source redshift of 3.074. \citet{rusu2018} present a detailed model of this system based on Subaru-FOCAS $i$-band imaging, showing that the lens is an edge-on disk galaxy.

\subsubsection{WG2100-4452}

WG2100-4452 was discovered\footnote{This discovery was first reported in 05/2018, arxiv:1805.11103} by \citet{Agnello19} as an extragalactic candidate with astrometric anomalies between the optical and infrared in VEXAS \citep{Spiniello-Agnello19}. Its source redshift is  0.920$\pm$0.002 and its deflector redshift is 0.203$\pm$0.002 \citep{Spiniello19}.

\subsubsection{J2145+6345}

J2145+6345 was discovered by \citet{Lemon19} as a quartet in \textit{Gaia} associated with a {\it WISE} photometric quasar candidate. The images are particularly bright (\textit{Gaia} magnitudes of 16.86, 17.26, 18.34, 18.56) and has X-ray (ROSAT) and radio (VLASS) detections. \citet{Lemon19} did not report the lensing galaxy position, as it was not detected in the Pan-STARRS PSF subtracted images; either since it was too faint or the PSF model was not sufficient to correctly subtract the four nearby bright PSFs.

\subsubsection{J2205-3727}

This quad was discovered by Lemon et al. in prep. by searching photometric quasar candidates from {\it WISE} for multiple \textit{Gaia} detections following \citet{Lemon19}. They confirm the source to be at redshift 1.848.

\subsection{Notes on individual Five-image systems}

This section gives a brief description of the quads in our sample that hold a fifth image due to a lens configuration that includes two primary deflectors.\\

\subsubsection{J0343-2828}

This system was discovered by Lemon et al. in prep. by searching for single \textit{Gaia} detections offset from galaxies, as possible lensed quasars, following \citet{Lemon17}. The system was selected for {\it HST} follow-up imaging due to the image colours and point-source nature, however follow-up spectroscopy reveals no quasar emission lines, but absorption features of a galaxy at z=1.655. The lens redshift is 0.385.

\subsubsection{2M1310-1714}

This system was discovered serendipitously by \citet{lucey2018} while visually inspecting the target catalogue of the Taipan Galaxy Survey. They report the presence of two lensing galaxies at z=0.293, and the source to be at z=1.975. Their mass model of two SIEs fixed to the galaxy positions with position angles both matching that of the extended halo light, and a shear fixed at 45 deg to this, predicts a fifth image five magnitudes fainter than the outer images. They also note the presence of a possible Einstein ring in VISTA Hemisphere Survey $K_{\textrm{s}}$ band imaging.

\begin{figure*}
 \centering
\includegraphics[width=\textwidth]{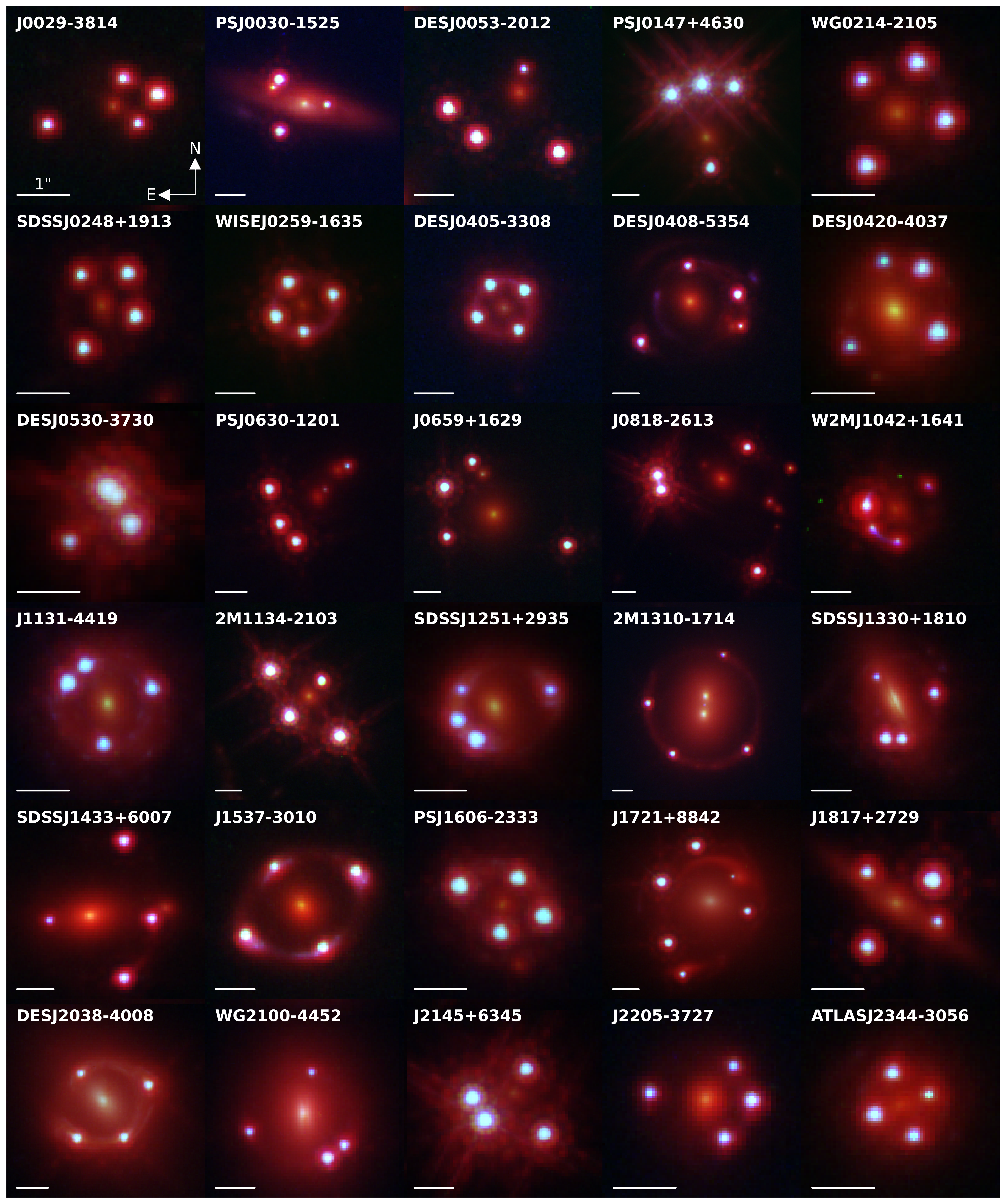}
 \caption{Sample of quadruply lensed quasar used in our analysis. The figure shows a composite red-green-blue (RGB) image for each lens, generated from {\it HST} observation in bands F160W (red channel), F475X (blue channel), and F814W (green channel). For visualization purposes, the intensities for each band vary between systems and are adjusted to emphasize each lens' individual configuration.}
 \label{fig:sample}
\end{figure*}

\begin{figure}
 \includegraphics[width=0.35\textwidth]{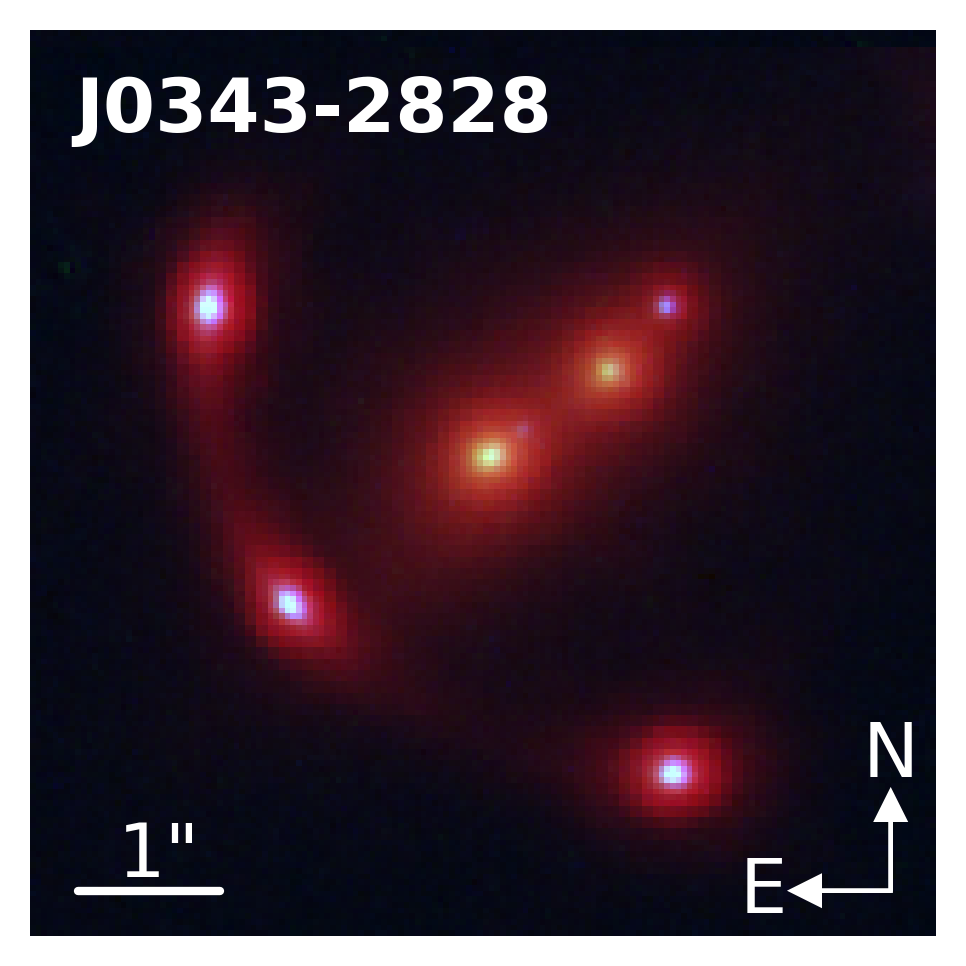}
 \caption{Quintuply lensed galaxy J0343-2828 used in our analysis. The figure shows a composite red-green-blue (RGB) image for each lens, generated from {\it HST} observation in bands F160W (red channel), F475X (blue channel), and F814W (green channel). For visualization purposes, the intensities for each band is adjusted to emphasize the system's configuration.}
 \label{fig:sample_J0343}
\end{figure}

\begin{figure*}
 \includegraphics[width=\textwidth]{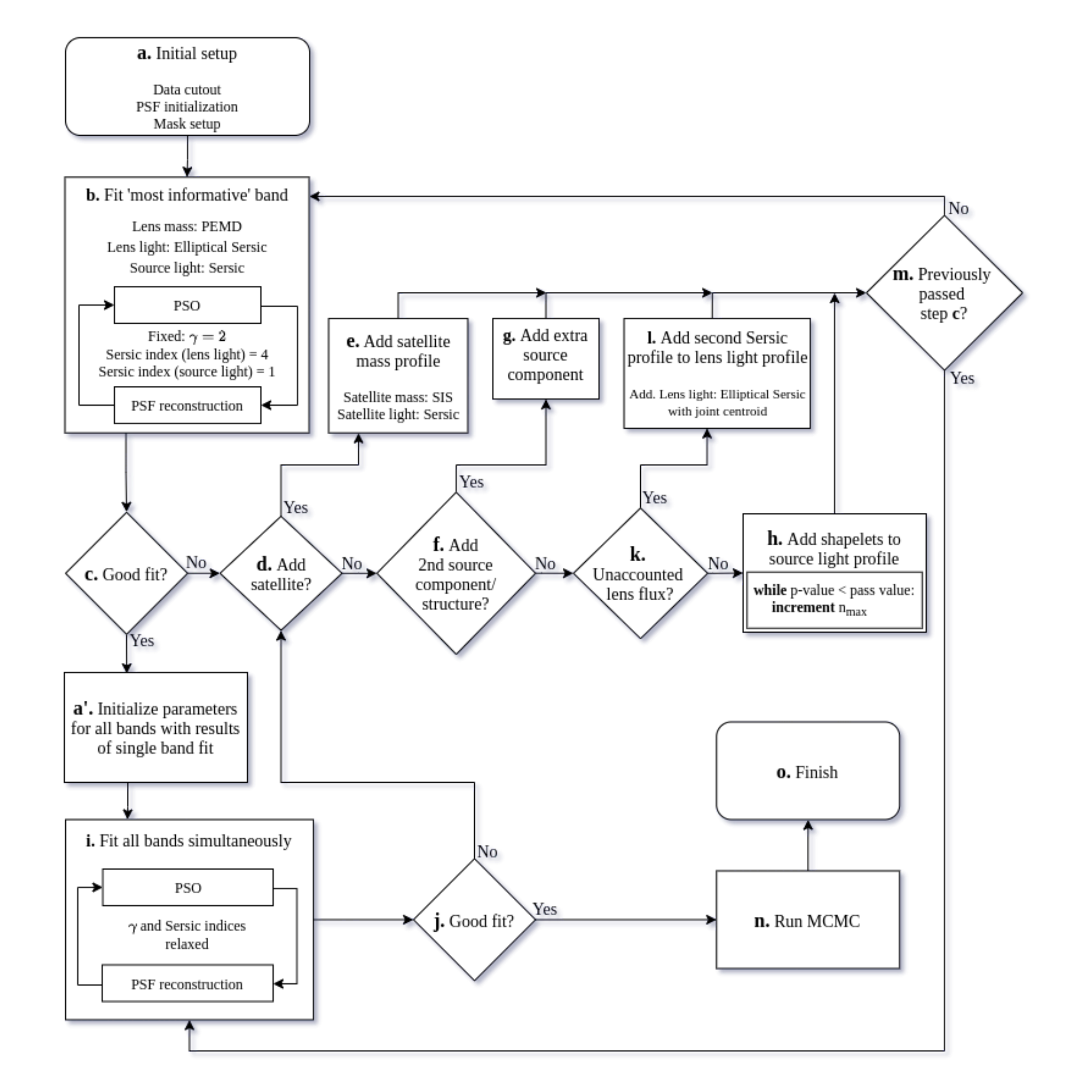}
 \caption{Flowchart illustrating individual modeling choices that are made by the pipeline along the process of lens model reconstruction. After being set up in node a., the pipeline traverses this decision tree, iteratively adding model complexity until the adopted minimum acceptance threshold  for the p-value or associated reduced $\chi^2$-value is achieved. Steps c., j., and h., initially use the mask that includes the lens light. For these three steps (c., j., h.), the lens light flux is only excluded if there are remaining residuals in the lens light after a second light profile was added in step l. In the last step, node n., the pipeline probes the posterior distribution of each free model parameter until convergence is reached.}
 \label{fig:flow_chart}
\end{figure*}

\section{Uniform Lens Modeling} \label{Uniform_Lens_Modeling}

We develop and apply an automated pipeline that is based on the uniform lens modeling process that was originally set forth by \citet{Shajib19} and further improved as detailed in Section~\ref{Modeling_Procedure}. Except for the initial setup of a lens, outlined in step a. of Section~\ref{Modeling_Procedure}, all model component decisions, e.g. to increase necessary model complexity, are made during runtime by the automated pipeline.

Our pipeline is based on the gravitational lens modeling software \textsc{Lenstronomy} \citep{Birrer18} \footnote{\href{https://github.com/sibirrer/lenstronomy}{https://github.com/sibirrer/lenstronomy}}, which is a publicly available open source distribution written in Python. \textsc{Lenstronomy} is the foundation in many strong lens analyses and is also used in time-delay cosmography \citep{Birrer16, Birrer19, Shajib20}. Additionally, \textsc{Lenstronomy} is an \textsc{Astropy} \citep{AstropyCollaboration13, AstropyCollaboration18} affiliated package. Explicit details on the modeling choices and analysis procedures to probe the parameter space for our models are presented in Section~\ref{Modeling_Procedure}, modeling procedure.
We refer to e.g. \cite{Shajib:2021, Etherington:2022} for automated pipelines analysing galaxy-galaxy lenses without lensed quasars.\\

\subsection{Mass profile parameterization} \label{Mass_Profile_Param}

The mass profile of the main deflector is modeled with a power-law elliptical mass distribution (PEMD), which corresponds to a radial mass density profile of $\rho \propto r^{-\gamma}$, where $\gamma$ is the power-law slope. The convergence, or dimensionless projected surface mass density, for the profile at position $\theta$ is parameterized as
\begin{equation}
\kappa(\theta_1, \theta_2) = \frac{3-\gamma}{2} \left(\frac{\theta_{E}}{\sqrt{q \theta_1^2 + \theta_2^2/q}} \right)^{\gamma-1}
\end{equation}
%
%\noindent
where $\theta_1$ and $\theta_2$ are aligned along the semi-major and semi-minor axis through the rotational position angle $\phi = \arctan(\theta_2, q\theta_1)$, and where $q$ is the corresponding axis ratio.

If our data show a second main deflector, resulting in a fifth image, or a satellite to the main deflector, we model the secondary object using a Singular Isothermal Sphere (SIS), which is a PEMD with a fixed power-law slope, $\gamma$, of $2.0$ and an axis ratio, $q$ of $1.0$.
Any additional linear distortions to the lensed structure, resulting from line-of-sight perturbers, are modelled through an external shear profile with strength 
\begin{equation}
    \gamma_{\rm{ext}} = \sqrt{\gamma_{\rm{ext},1}^2 + \gamma_{\rm{ext},2}^2} ,
\end{equation}
%
%\noindent
and position angle
\begin{equation}
    \phi_{\rm{ext}} = \frac{1}{2}\arctan \left(\gamma_{\rm{ext},2}, \gamma_{\rm{ext},1} \right).
\end{equation}\\

\subsection{Light profile parameterization} \label{Light_Profile_Param}

The light profile of the main deflector is modeled with an elliptical S\'ersic function \citep{Sersic68}, which is parameterized as:
\begin{equation}
    I(\theta) = I(\theta_{\rm{e}}) \: \exp \left\{ -C(n) \left[ \left( \frac{(q_{\rm{L}}\theta_1)^2 + \theta_2^2}{q_{\rm{L}}\theta_{\rm{e}}^2} \right)^{\frac{1}{2n}} -1 \right] \right\},
\label{elliptical_sersic}
\end{equation}
%
%\noindent
where $C(n)$ is a normalization constant so that at the effective radius, $\theta_{\rm{e}}$, the profile includes half of the deflector's light. $n$ represents the S\'ersic index, $\theta_1$ and $\theta_2$ are the angular coordinates aligned along the semi-major and semi-minor axis through the rotational position angle $\phi_{\rm{L}} = \arctan(\theta_2, q\theta_1)$ of the light profile, and $q_{\rm{L}}$ represents the corresponding axis ratio. Each main deflector in our sample is initially modeled with one elliptical S\'ersic, however, as further detailed in node l. of the modeling procedure below, the pipeline adds an additional S\'ersic with a fixed S\'ersic index in the case of unaccounted lens flux.

If the main deflector is accompanied by a satellite, or if the lens has a secondary main deflector, the light of the additional perturber is modelled as a circular S\'ersic function, which corresponds to an elliptical S\'ersic function (\ref{elliptical_sersic}) with a fixed axis ratio at $q_{\rm{L}}=1.0$.
We restrict our analysis to circular secondary light distributions in order to limit the number of free parameters in our models. In nearly all cases the circular S\'ersic function models the light of the additional perturber with sufficient precision.

The images of the lensed quasar are modeled by a point spread function (PSF) in the image plane. To model the light of the lensed source, or host galaxy of the lensed quasar, we choose a circular S\'ersic function in the source plane as described in the light profile parameterization of additional perturbers above. If additional lensed source light is identified that is not part of the primary source hosting the quasar, we adopt a second circular S\'ersic to model the extra source light separately from light profile of the host galaxy. If the S\'ersic functions are insufficient to describe the complexity of the source, we add a set of two-dimensional Cartesian shapelets \citep{ Refregier03, Birrer15}. The shapelet number, or number of basis functions which form an orthonormal basis, is given by
\begin{equation}
    N_{\rm shapelet} = \frac{(n_{\rm{max}}+1)(n_{\rm{max}}+2)}{2},
\label{n_max}
\end{equation}

\noindent
where $n_{\rm{max}}$ represents the highest shapelet order, or maximum source complexity, and is linked to the maximum spatial scale, $l_{\rm{max}}$, and the characteristic scale, $\beta$, by $l_{\rm{max}} = \beta \sqrt{n_{\rm{max}} + 1}$. Increasing the parameter $n_{\rm{max}}$ corresponds to the reconstruction of additional smaller features in the lensed source.\\

\subsection{Priors}\label{Priors}

\begin{figure}
 \includegraphics[width=0.45\textwidth]{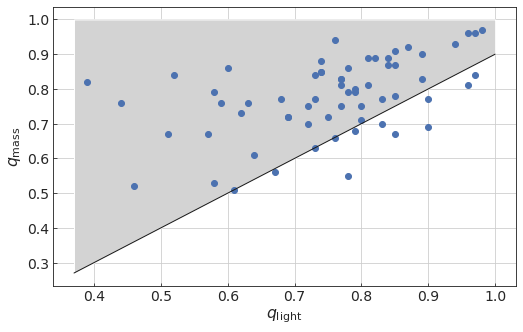}
 \caption{Linear prior on axis ratio for main deflector's mass profile (shaded area), motivated by the analysis of 63 lenses from the SLACS sample, and chosen with 95\% of the 63 SLACS lenses meeting the constraint, given a 0.05 tolerance in the axis ratio. For each lens, we compare the axis ratio of the mass profile, $q_{\rm{mass}}$, to the respective light profile's axis ratio, $q_{\rm{light}}$.}
 \label{fig:axis_ratio_prior}
\end{figure}

A number of well known degeneracies affect lens modeling \citep[see, e.g.,][]{Falco85, Schneider14}. 
To avoid non-physical results, we impose priors on the axis ratio, $q$, and the position angle, $\phi$, of the primary deflector's mass profile, motivated by the analysis of 63 lenses from the SLACS sample \citep{Bolton06, Bolton08, Auger10}. For each SLACS lens we compare the axis ratio of the deflector's mass profile to the corresponding axis ratio of the light profile, with the results of this comparison shown in Figure~\ref{fig:axis_ratio_prior}.  Given a 5\% error and a requirement that 95\% of the sample to fall within the constraint, we then determine a linear prior whereby the lower limit of the mass profile's axis ratio is given by $q \geq q_{\rm{L}} - 0.1$. If during the fitting process a model instance produces an axis ratio below this limit, the pipelines discards the likelihood of the model. This prior avoids nonphysical solutions, such as extreme ellipticity in a deflector's mass profile, and guides the model to increase the strength in the external shear instead.

\begin{figure}
 \includegraphics[width=0.45\textwidth]{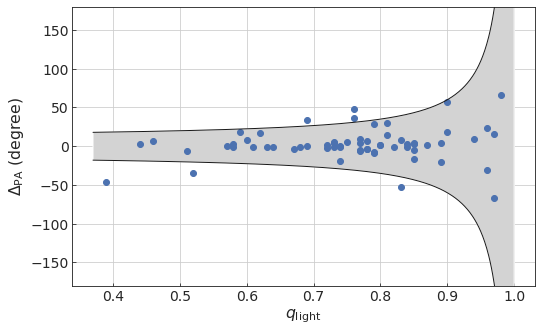}
 \caption{Prior on the position angle for main deflector's mass profile (shaded area), based on axis ratio and position angle of the deflector's respective light profile. The y-axis shows the difference in the position angle between the mass and light profile, $\Delta_{\rm{PA}}$, as a function of the respective light profile's axis ratio, $q_{\rm{light}}$. The prior is set with 95\% of the 63 strong lenses in the SLACS sample meeting the criterion, given a 10 degree tolerance. Values outside the gray shaded are are exluded a priori in our analysis.}
 \label{fig:pa_prior}
\end{figure}

To find a suitable restriction on the convergence's position angle, we plot the absolute difference between the position angles of the mass and light profiles, $\Delta_{\rm{PA}}$, as a function of the light profile's axis ratio, $q_{\rm{L}}$, for the 63 lenses in the SLACS sample. Due to symmetry, any position angle difference greater or less than 90 degrees is shifted by 180 degrees with the results shown in Figure~\ref{fig:pa_prior}. Following a requirement for 95\% of the sample to fall within the constraint, given an error margin of 10 degrees, we arrive at a prior for the upper limit of the position angle difference given by $\Delta_{\rm{PA}} \leq 10 - 5/(q_{\rm{L}} - 1)$. Models with  angle difference exceeding this limit are excluded a priori. Although our prior is well justified and prevents unphysical solution, it is of course not a unique choice. It is thus important that this as well as other informative  priors adopted in our analysis are to be kept in mind when interpreting our results.

To place a constraint on the centroid of the main deflector's mass profile, we use a Gaussian prior for each axis that depends on the centroid coordinates of the deflector's light profile and a standard deviation of $0.04 \arcsec$, which corresponds to 1 pixel in UVIS. If a lens model includes a secondary deflector, or satellite, we join the centroid of the sattelilte's mass profile with the centroid of the corresponding light profile.

For some of our targets the lensed host galaxies of the multiply imaged quasars do not a have sufficiently high signal to noise ratio and therefore provide insufficient radial information to constrain the slope of the mass density profile. For that reason, we adopt an informative prior to constrain the power-law slope of the main deflector's mass density profile. Due to a degeneracy between the slope and the characteristic scale, $\beta$, in the shapelets used to describe higher source complexity, the prior prevents nonphysical results when the slope is not well constrained by the data \citep{Birrer16}.
In their analysis of early-type galaxy strong gravitational lenses from the SLACS sample, \citet{Auger10b} find a distribution of the power-law slope with a mean of $2.078 \pm{0.027}$, which agrees with the findings of \citet{Koopmans09}. For all lenses in our sample we use these results in a Gaussian distributed prior and additionally reject the likelihood of any model that produces a slope with $12 \sigma$ above or below the aforementioned mean.

We note that the slope of the radial mass density profile is a key parameter for determining the time delay distance and hence $H_0$ \citep[e.g.,][]{Wucknitz02}. Therefore, if one wishes to use the results of this work as a starting point for cosmographic work, the prior needs to be accounted for in order to avoid underestimating the errors or biasing the results.

\subsection{Modeling procedure} \label{Modeling_Procedure}

To fit the observed data from all {\it HST} filters, all lenses in our sample are modeled using \textsc{Lenstronomy}'s particle swarm optimization (PSO). We probe the posterior distribution of each model via Markov chain Monte Carlo (MCMC) sampling, built on the \textsc{emcee} package \citep{Foreman-Mackey13}, which is an affine-invariant ensemble sampler \citep{Goodman10}. Since the effectiveness of an optimization routine depends the initial starting point, we implemented a three step process to effectively find the global maximum likelihood for our models, with each step in our fitting routine building on the results of the previous optimization. Should an optimization routine produce an unsatisfactory fit to the data, we increase the model complexity to account for additional features. During each step, we evaluate the difference in the Fermat potential at the image positions in order to track the lens model's evolution. Using \textsc{Lenstronomy}'s PSO, we first find the best fit for a single band (F814W), which we deem the most informative band as most or all features that are visible in other bands also appear in the F814W filter, and as it has a higher resolution than WFC3-IR. Once an acceptable model has been established, we fit all three bands simultaneously using the results from the previous fitting routine for each model parameter, again using \textsc{Lenstronomy}'s PSO. After an acceptable lens model has been established, we probe the model's posterior probability distribution with \textsc{Lenstronomy}'s above described MCMC routine.

Figure~\ref{fig:flow_chart} gives a general overview to our uniform modeling procedure while a detailed description of each node in the flow chart can be found in the following subsections:\\

\textbf{a. Initial setup:}\\

First, we pre-process the images in each filter band. After our data reduction process, as described in Section~\ref{Data_Reduction}, we select a cutout for each {\it HST} filter, large enough to encompass the lens, the lensed quasar images, and any satellite or perturbers that are to be included in the model. We then subtract the mean of the background flux, which is determined by running \textsc{SourceExtractor} \citep{Bertin96} on the full {\it HST} image. Afterwards, we make preliminary guesses for the position of the lensed quasar images and for the main deflector's centroid. If the model were to include another perturber or additional source components, initial guesses for the location of these features are determined as well. To differentiate additional source components, lensed by the main perturber, we look for structure with conjugate components that are situated near the lensed primary source. We then apply a circular mask to the cutout with a radius appropriate to exclude unwanted nearby features. If we identify additional attributes within the circular mask that are not deemed to be part of the lens model, we exclude them by applying further masking. A second circular mask is separately applied to the cutout to separate the lens flux, which allows the pipeline to determine the goodness of a uniform lens light profile fit.  Additionally we estimate a radius up to which neighboring QSO images will be blocked during the iterative PSF fitting process, as described below in step b. Lastly, we select a set of five or more small bright stars in each reduced {\it HST} image to obtain an initial estimate of the point spread function for each band \citep{Birrer19, Shajib22}.\\

\textbf{b. Fit the 'most informative' band:}\\

For a typical system in our sample, the pollution in the arc and in the lensed images, caused by the lens light contribution, decreases in the bluer bands. At the same time, in the bluer filters the arc light intensity from the lensed source decreases compared the redder bands. We therefore designate the F814W filter as the 'most informative' band, as the signal to noise ratio for the lensed source is typically highest in filter F814W, compared to the other two bands used in our observations. 

Since even our simplest starting models include the deflector's mass and light profile in addition to the four point source locations and a light profile for the lensed source, the fitting routines have to traverse a large parameter space to find the maximum likelihood to fit the lens model to our data. We therefore follow the procedure as set forth by \citet{Shajib19} and fit the most informative band and increase the model's complexity before fitting the data in all filters simultaneously. For this step, we hold the power-law slope of the main deflector's mass profile constant at a value of 2.0, effectively fitting the profile for an isothermal mass distribution. To further limit the number of free parameters in the initial fitting process, and moreover effectively decreasing the computation time, we also hold the S\'ersic index of the source and lens light profile fixed at $1.0$ and $4.0$, representative of an exponential and de Vaucouleurs light profile, respectively. Due to the strong degeneracy between the light profile's effective radius and S\'ersic index, holding these settings constant furthermore prevents the half light radii from reaching on nonphysical values. 
Because we start each model with the same set of initial parameters, we first sample the parameter space with a broad search region. Within the same fitting sequence, after the completion of each PSO run, we optimize the PSF to best fit the model's quasar images after accounting for extended source light. We perform this iterative PSF reconstruction with 90 degree symmetry and update the PSF's error map with each new iteration \citep[see][]{Chen16, Birrer19b, Shajib20}. In order to avoid corrections that have already been included in the error map of a nearby quasar image, we block any neighboring images around their centroid up to a radius that is determined in the initial setup for the lens (step a.). The alternating PSO/PSF fitting is then repeated with a narrower search region, corresponding to 1/10 of the previous iteration and centered around the results of the maximum likelihood for the previous PSO. This process is continued until the search region has been reduced to probe the parameter space within 1/1000 of the first PSO sampling range. Further details on the iterative approach to reconstructing the PSF and finding the maximum likelihood of models by probing the parameter space with PSOs can be found in the paper by \citet{Birrer18}.\\ %(Lenstronomy).\\

\textbf{c. Good fit?}\\

To determine how well our data fit the current model, we compute the p-value for the masked circular region in the most informative band, using the reduced $\chi^2$ value resulting from the best fit and the degree of freedom represented by the pixels in the applied mask. We follow the acceptance criterion as set forth by \citet{Shajib19} and deem the fit to be acceptable if the computed p-value is greater than $10^{-8}$, which, given the diversity of lenses in our sample, should be beyond sufficient to indicate missing features in our models without modeling noise in the data. As an alternative acceptance criterion, we use the reduced $\chi^2$ value and test if it is smaller than $1.1$ for the masking region. 

If node c. is being visited after a second S\'ersic function was added to the description of the main deflector's light profile in step l. and there are remaining residuals in the lens center that would require a higher lens light complexity, then we subtract the lens center mask from the fitting region and re-evaluate the above discussed acceptance criteria to determine the goodness of the fit. This exclusion of the lens light from the fitting mask is necessary, since additional descriptions to the lens light flux would be needed and the pipeline, in its current stage, is limited to a double S\'ersic as most complex light profile. \\

\textbf{d. \& e. Add satellite to mass profile:}\\

If the acceptance criteria in the goodness tests of step c. or step j. are not met, indicating the current model is missing components or complexity, and a satellite has been identified in the initial setup (step a.) but is not yet included in the model, we add an SIS profile, as outlined in the mass profile parameterization, to the description of the main deflector's mass profile. The light profile of this additional pertuber is modeled by a spherical S\'ersic as described in Section~\ref{Light_Profile_Param} for the light profile parameterization. The joint centroid for both, the satellite's mass and light profile, is initialized with the guess that is made during the model setup (see step a.) and the pipeline returns to the iterative fitting process of step b. or step i., depending on the evaluation of node m.\\

\textbf{f. \& g. Add additional source component:}\\

If steps c. or j. for the current model indicate missing complexity and an additional source was identified in step a., we add a separate source light profile using a circular S\'ersic function as outlined on the section on the light profile parameterization. The centroid for this additional source light profile is initialized with the guess determined in the model setup (node a.) before the iterative fitting process is restarted in steps b. or i. For the centroids location we use \textsc{Lenstronomy}'s \textsc{bijective} mode, whereby the location of the additional source is indentified and constrained in the lens plane and then ray-traced back to its position in the source plane. \\

\textbf{k. Check for unaccounted lens flux:}\\

To check our models for flux, not captured by the current lens light profile, we again compute the reduced $\chi^2$ and associated p-value for the latest fit, only using the mask that singles out the lens flux as described in the initial setup procedures. We compare this p-value and chi square result, which only pertains to the lens light profile, with the fitting results computed in step c. In the case of a lower p-value, or larger reduced $\chi^2$ result, for the lens light mask, which would indicate missing lens light flux, the pipeline proceeds to step l. and adds an additional S\'ersic profile to the description of the lens light, given that node l. has not been previously visited. In all other cases the pipeline proceeds to the next node in the decision tree.\\

\textbf{l. Add second S\'ersic function to lens light profile:}\\

Should node k. call for the addition of a lens light to account for missing flux in the main deflector's light profile, we add a second elliptical S\'ersic profile to the existing description of the lens light model, with a joint centroid. We follow \citet{Shajib19} by setting the S\'ersic indices, as described in the light profile parameterization, to constant values of n = $1.0$ and n = $4.0$, representative of an exponential and de Vaucouleurs light profile, respectively. As discussed by \citet{Shajib19}, we hold the S\'ersic indices fixed for numerical stability in our models only; therefore the two light profiles are not to be understood as individual galactic components of the main deflector. If, however, the addition of a second lens light profile results in a fit, after steps b. or i., with a larger overall reduced chi square or smaller associated p-value, the addition of the second S\'ersic profile to the lens light description is reversed and the previous fitting result is used for the remainder of the modeling process.\\

\textbf{h. Add shapelets to source light profile:}\\

Additional complexity in the source light and not accounted by the source's S\'ersic profile is modelled through a basis set of shapelets, which shares the same centroid as the primary source's light profile. To find the proper shapelet order we iteratively increase the maximum order and guess the characteristic scale, $\beta$, using the primary source's S\'ersic radius. Running a \textsc{SciPy} minimization routine, the pipeline proceeds to find the $\beta$ value to the current maximum shapelet order that results in the best p-value, and lowest associated chi square number, effectively performing a linear minimization of the shapelet coefficient, and then tests if the acceptance criteria as set forth in step c are reached. If the p-value for the best $\beta$ scale lies below the threshold, the shapelet order is incremented and the minimization steps are repeated until the shapelet order was raised by $6$ for a newly added basis set, or raised by $5$ for a previously fitted basis set, in which case the pipeline returns to the PSO/PSF fitting step (b. or i.) that lead to this node. If the $\chi^2$ result, or associated p-value, meets the acceptance threshold, the pipeline proceeds to  the simultaneous fitting of all bands with the shapelet order starting values determined from the minimization routine. This iterative approach to raising the source complexity is performed for each band in which the p-value of the corresponding filter's cutout mask lies below our acceptance criterion. \\

\textbf{m. Completed fit for most informative band?}\\

Since it is possible for nodes e., g., h., and l. to be reached after fitting the single, most informative, band or after fitting all bands simultaneously, we check if a previous iteration has already achieved a good fit for a single filter, in which case we continue with the simultaneous fitting of all bands in step i.\\

\textbf{i. Fit all bands simultaneously:}\\

On the first visit of this node we align the data from all filters to the data of the most informative band. For this step we use \textsc{Lenstronomy}'s iterative alignment routine, as described by \citet{Birrer18}, to match the coordinate frames of different filters using the astrometric positions of the lensed quasar images. We estimate this alignment to be accurate within 1 milliarcsecond. After the alignment we initialize each free parameter with the results of the best fit for the most informative band and continue to simultaneously fit all filters using \textsc{Lenstronomy}'s PSO routine iteratively. For this step, we relax the power-law slope of the main deflector's mass profile as well as the S\'ersic indices of the light profiles, as these parameters were held constant during the fitting described by step b. Due to the strong correlation between the effective radius and the S\'ersic index in the light profile parameterization and to further avoid nonphysical fitting results, the upper boundaries of the S\'ersic indices are set to a limit of $6.0$ and $4.0$ for the lens light and source light profile, respectively. 

We begin the sampling of the parameter space with 1/10 of the initial search region used for fitting the most informative band. As in step b., we continue to optimize the PSF within the same fitting sequence to obtain the best fit for our model's quasar images. Again, this iterative PSF reconstruction is performed for each filter with a 90 degree symmetry in the PSF and the PSF's error map is updated for each band. In each filter we block neighboring images around their centroid position to avoid the double counting of corrections from nearby quasar images. As previously outlined in step c., the alternating PSO/PSF fitting is repeated for all bands simultaneously with 1/10 of the former search region and around the results of the maximum likelihood for the previous PSO iteration. This is continued until the search region has been reduced to probe the parameter space down to 1/100 of the first PSO sampling range in this step. For the simultaneous fitting approach of all bands we follow \citet{Shajib19} and hold the following lens light, additional perturber light, and source light profile parameters common across all filters: S\'ersic radius, S\'ersic index, centroid, ellipticity, and position angle. This choice greatly simplifies the computational cost of the fit, and it is commonly adopted in the literature when large dataset need to be fit (e.g. SDSS) - see \citet{Stoughton02} and \citet{Lackner12}. \citet{Shajib19} find that this common parameter approach across various filters results in fits that are within the estimated uncertainties compared to fits obtained from the fitting using unlinked parameters. Therefore, in our automated uniform approach, we deem this approximation  to be  acceptable for the purpose of this work. All other model parameters not specifically mentioned to be held common (e.g. maximum shapelet order) are allowed to vary across filters.\\

\textbf{j. Good fit?}\\

To test the fit of our model for the bands that have been fit simultaneously, we repeat the procedures described in node c., namely computing the p-value for the masking region in each filter and test of it is above $10^{-8}$ or if the associated reduced $\chi^2$ meets the acceptance criterion of being lower than $1.1$. This acceptance procedure is performed for each filter separately, with the pipeline proceeding to add higher complexity to the model if one of the bands fails these tests. As a third alternative to the two acceptance criteria (outlined immediately above), we also compute the overall reduced $\chi^2$ value for the fit combining all bands and accept the current model if the overall result lies below $1.1$. As described in the single band fitness test (step c.), if we detect residuals in the lens flux after a second S\'ersic profile has been added to the lens light description, we exclude the masking region that encompasses the lens center for the purpose of calculating the $\chi^2$ and associated p-values.\\

\textbf{n. Run MCMC:}\\

Once the alternating PSO/PSF fitting routine finds a good model, meeting our acceptance criteria, we probe the posterior distribution for each free model parameter using \textsc{Lenstronomy}'s MCMC routine. We first initialize each free parameter with the best fit found by the final PSO run and then run a burn-in cycle for $1500$ iterations to assure the chain reaches an equilibrium distribution. The total number of likelihood evaluations corresponding to the burn-in cycle is given by the product of the number of free parameters in the model, the number of walkers per parameter, and the number of iterations. After the burn-in, we stop the MCMC run every $100$ iterations to compute the mean as well as the spread in the distribution for each free model parameter, using the corresponding distribution's $16$- and $84$-th percentiles. The pipeline continues by comparing the current mean of each parameter with the mean computed during the previous $100$ iterations. If the change in the mean value is less than $1/100$ of the full spread for the respective parameter, we consider the value to be converged. Only if this convergence criterion has been reached simultaneously for all free parameters in our model, the pipeline considers the reconstruction completed.\\

\textbf{o. Finish}\\

Given the large diversity of lenses in our sample, we visually inspect each model after the successful completion of the pipeline's reconstruction process, to assess how well the pipeline performed. We also check if model parameters have diverged towards their corresponding upper or lower bounds. Additionally, we track the evolution of the difference in a model's Fermat potential at the position of the quasar images to ensure stability in our models. Further details relating to this stability metric can be found in Section~\ref{stability_of_fermat_pot}.\\

\section{Results} \label{Results}

\begin{table*}
  \caption{Model parameters for lens mass distributions, which are median values. The associated uncertainties are statistical in nature and were computed using 84th and 16th percentiles.}
  \label{tab:lens_mass_results}
  \begin{tabular}{lccccccc}
    \hline
    Name of &  $\theta_\mathrm{E}$  & $\gamma$ & $q$   & $\phi$ & $\gamma_\mathrm{ext}$ & $\phi_\mathrm{ext}$ & Area of \\
    Lens System & & & & (N of E) & & (N of E) & Inner Caustic\\
      &  (arcsec)    &          &     & (degree)        &                & (degree) & (arcsec$^2$)\\
    \hline
    \\[-0.75em]
J0029-3814 & $0.769^{+0.007}_{-0.010}$ & $1.99^{+0.02}_{-0.02}$ & $0.54^{+0.06}_{-0.03}$ & $73.4^{+0.2}_{-0.3}$ & $0.252^{+0.017}_{-0.014}$ & $-15.2^{+0.1}_{-0.1}$ & $0.610^{+0.041}_{-0.059}$\\[+0.75em]
PS J0030-1525 & $0.996^{+0.003}_{-0.003}$ & $1.97^{+0.02}_{-0.03}$ & $0.72^{+0.02}_{-0.02}$ & $9.6^{+0.6}_{-0.7}$ & $0.071^{+0.004}_{-0.004}$ & $-11.0^{+2.1}_{-2.0}$ & $0.057^{+0.005}_{-0.005}$\\[+0.75em]
DES J0053-2012 & $1.380^{+0.005}_{-0.006}$ & $2.03^{+0.02}_{-0.02}$ & $0.69^{+0.03}_{-0.02}$ & $-58.1^{+0.6}_{-0.4}$ & $0.215^{+0.009}_{-0.007}$ & $21.7^{+0.2}_{-0.2}$ & $0.250^{+0.028}_{-0.030}$\\[+0.75em]
PS J0147+4630 & $1.886^{+0.005}_{-0.004}$ & $2.08^{+0.02}_{-0.02}$ & $0.80^{+0.01}_{-0.01}$ & $-85.8^{+0.4}_{-0.4}$ & $0.147^{+0.005}_{-0.006}$ & $-12.2^{+0.2}_{-0.2}$ & $0.463^{+0.005}_{-0.006}$\\[+0.75em]
WG0214-2105 & $0.849^{+0.001}_{-0.001}$ & $2.08^{+0.02}_{-0.03}$ & $0.86^{+0.01}_{-0.01}$ & $-17.8^{+2.4}_{-2.5}$ & $0.101^{+0.003}_{-0.003}$ & $-50.3^{+0.4}_{-0.3}$ & $0.029^{+0.002}_{-0.002}$\\[+0.75em]
SDSS J0248+1913 & $0.767^{+0.001}_{-0.001}$ & $2.01^{+0.06}_{-0.06}$ & $0.54^{+0.01}_{-0.01}$ & $80.2^{+0.8}_{-0.7}$ & $0.222^{+0.003}_{-0.004}$ & $-86.8^{+0.6}_{-0.7}$ & $0.026^{+0.004}_{-0.004}$\\[+0.75em]
WISE J0259-1635 & $0.742^{+0.001}_{-0.001}$ & $2.20^{+0.03}_{-0.02}$ & $0.79^{+0.01}_{-0.01}$ & $78.8^{+0.5}_{-0.5}$ & $0.058^{+0.003}_{-0.003}$ & $-28.8^{+0.6}_{-0.6}$ & $0.038^{+0.002}_{-0.002}$\\[+0.75em]
J0343-2828 & $0.900^{+0.002}_{-0.002}$ & $1.99^{+0.01}_{-0.01}$ & $0.50^{+0.01}_{-0.01}$ & $-44.3^{+0.1}_{-0.2}$ & $0.150^{+0.002}_{-0.002}$ & $46.8^{+0.1}_{-0.1}$ & $0.013^{+0.002}_{-0.002}$\\[+0.75em]
DES J0405-3308 & $0.705^{+0.001}_{-0.001}$ & $2.15^{+0.03}_{-0.03}$ & $0.70^{+0.01}_{-0.01}$ & $49.4^{+0.4}_{-0.5}$ & $0.039^{+0.002}_{-0.001}$ & $27.2^{+1.6}_{-2.0}$ & $0.015^{+0.001}_{-0.001}$\\[+0.75em]
DES J0420-4037 & $0.839^{+0.001}_{-0.001}$ & $2.02^{+0.03}_{-0.03}$ & $0.78^{+0.01}_{-0.01}$ & $61.6^{+0.4}_{-0.5}$ & $0.038^{+0.001}_{-0.001}$ & $88.1^{+1.7}_{-1.7}$ & $0.015^{+0.001}_{-0.001}$\\[+0.75em]
DES J0530-3730 & $0.557^{+0.010}_{-0.008}$ & $2.07^{+0.03}_{-0.03}$ & $0.68^{+0.11}_{-0.11}$ & $73.0^{+6.7}_{-20.0}$ & $0.107^{+0.039}_{-0.044}$ & $76.2^{+5.7}_{-23.7}$ & $0.001^{+0.002}_{-0.001}$\\[+0.75em]
PS J0630-1201 & $1.574^{+0.004}_{-0.010}$ & $2.11^{+0.02}_{-0.02}$ & $0.56^{+0.01}_{-0.01}$ & $-77.3^{+1.0}_{-1.1}$ & $0.209^{+0.002}_{-0.002}$ & $85.1^{+0.7}_{-0.7}$ & $0.129^{+0.002}_{-0.002}$\\[+0.75em]
J0659+1629 & $2.124^{+0.016}_{-0.017}$ & $1.89^{+0.03}_{-0.03}$ & $0.85^{+0.01}_{-0.01}$ & $-59.7^{+1.7}_{-1.9}$ & $0.069^{+0.005}_{-0.005}$ & $25.8^{+1.0}_{-1.0}$ & $0.010^{+0.001}_{-0.001}$\\[+0.75em]
J0818-2613 & $2.896^{+0.001}_{-0.001}$ & $2.07^{+0.01}_{-0.01}$ & $0.60^{+0.01}_{-0.01}$ & $76.0^{+0.6}_{-1.1}$ & $0.317^{+0.003}_{-0.002}$ & $59.7^{+0.1}_{-0.1}$ & $0.000^{+0.001}_{-0.000}$\\[+0.75em]
W2M J1042+1641 & $0.892^{+0.001}_{-0.001}$ & $2.17^{+0.02}_{-0.03}$ & $0.68^{+0.01}_{-0.01}$ & $65.8^{+0.6}_{-0.7}$ & $0.055^{+0.003}_{-0.003}$ & $85.9^{+1.7}_{-1.7}$ & $0.023^{+0.002}_{-0.002}$\\[+0.75em]
J1131-4419 & $0.876^{+0.001}_{-0.001}$ & $2.02^{+0.02}_{-0.02}$ & $0.58^{+0.01}_{-0.01}$ & $81.4^{+0.2}_{-0.2}$ & $0.057^{+0.004}_{-0.003}$ & $74.0^{+0.7}_{-0.8}$ & $0.052^{+0.003}_{-0.003}$\\[+0.75em]
2M1134-2103 & $1.264^{+0.003}_{-0.004}$ & $2.15^{+0.02}_{-0.02}$ & $0.66^{+0.02}_{-0.01}$ & $-55.6^{+0.7}_{-0.8}$ & $0.338^{+0.008}_{-0.007}$ & $45.5^{+0.1}_{-0.1}$ & $0.444^{+0.015}_{-0.013}$\\[+0.75em]
SDSS J1251+2935 & $0.841^{+0.001}_{-0.001}$ & $2.09^{+0.01}_{-0.01}$ & $0.81^{+0.01}_{-0.01}$ & $63.0^{+0.5}_{-0.5}$ & $0.090^{+0.002}_{-0.002}$ & $-11.5^{+0.3}_{-0.4}$ & $0.077^{+0.001}_{-0.001}$\\[+0.75em]
2M1310-1714 & $1.465^{+0.002}_{-0.002}$ & $2.01^{+0.01}_{-0.01}$ & $0.65^{+0.01}_{-0.01}$ & $-72.5^{+0.4}_{-0.1}$ & $0.024^{+0.001}_{-0.001}$ & $80.5^{+1.6}_{-0.6}$ & $0.983^{+0.040}_{-0.027}$\\[+0.75em]
SDSS J1330+1810 & $0.996^{+0.007}_{-0.007}$ & $2.06^{+0.03}_{-0.03}$ & $0.37^{+0.02}_{-0.02}$ & $65.7^{+0.2}_{-0.2}$ & $0.124^{+0.007}_{-0.006}$ & $78.2^{+1.1}_{-1.1}$ & $0.184^{+0.022}_{-0.021}$\\[+0.75em]
SDSS J1433+6007 & $1.581^{+0.002}_{-0.003}$ & $1.92^{+0.03}_{-0.03}$ & $0.96^{+0.01}_{-0.01}$ & $-28.1^{+4.5}_{-2.6}$ & $0.127^{+0.004}_{-0.004}$ & $-82.4^{+0.4}_{-0.4}$ & $0.002^{+0.002}_{-0.001}$\\[+0.75em]
J1537-3010 & $1.408^{+0.001}_{-0.001}$ & $2.02^{+0.02}_{-0.02}$ & $0.85^{+0.01}_{-0.01}$ & $55.3^{+0.2}_{-0.3}$ & $0.124^{+0.003}_{-0.004}$ & $-28.3^{+0.1}_{-0.1}$ & $0.167^{+0.005}_{-0.005}$\\[+0.75em]
PS J1606-2333 & $0.700^{+0.003}_{-0.003}$ & $1.93^{+0.01}_{-0.01}$ & $0.54^{+0.01}_{-0.01}$ & $-76.6^{+0.2}_{-0.3}$ & $0.088^{+0.005}_{-0.004}$ & $39.1^{+1.3}_{-1.1}$ & $0.197^{+0.007}_{-0.007}$\\[+0.75em]
J1721+8842 & $1.947^{+0.001}_{-0.001}$ & $1.97^{+0.01}_{-0.01}$ & $0.80^{+0.01}_{-0.01}$ & $19.7^{+0.2}_{-0.1}$ & $0.075^{+0.001}_{-0.001}$ & $-78.8^{+0.1}_{-0.1}$ & $0.199^{+0.002}_{-0.002}$\\[+0.75em]
J1817+2729 & $0.893^{+0.001}_{-0.001}$ & $2.03^{+0.02}_{-0.02}$ & $0.84^{+0.01}_{-0.01}$ & $14.5^{+1.8}_{-0.9}$ & $0.044^{+0.001}_{-0.001}$ & $-12.0^{+1.3}_{-1.1}$ & $0.008^{+0.001}_{-0.001}$\\[+0.75em]
DES J2038-4008 & $1.376^{+0.001}_{-0.001}$ & $2.33^{+0.01}_{-0.01}$ & $0.64^{+0.01}_{-0.01}$ & $52.3^{+0.1}_{-0.1}$ & $0.086^{+0.002}_{-0.002}$ & $-32.5^{+0.2}_{-0.1}$ & $0.297^{+0.005}_{-0.005}$\\[+0.75em]
WG2100-4452 & $1.322^{+0.003}_{-0.002}$ & $2.19^{+0.03}_{-0.04}$ & $0.51^{+0.01}_{-0.01}$ & $87.2^{+0.1}_{-0.1}$ & $0.012^{+0.003}_{-0.002}$ & $30.6^{+25.6}_{-10.5}$ & $0.071^{+0.003}_{-0.004}$\\[+0.75em]
J2145+6345 & $1.013^{+0.004}_{-0.003}$ & $2.03^{+0.03}_{-0.03}$ & $0.71^{+0.04}_{-0.03}$ & $-64.3^{+1.0}_{-1.3}$ & $0.104^{+0.010}_{-0.011}$ & $36.8^{+0.6}_{-0.6}$ & $0.173^{+0.014}_{-0.014}$\\[+0.75em]
J2205-3727 & $0.772^{+0.001}_{-0.001}$ & $2.04^{+0.02}_{-0.02}$ & $0.66^{+0.01}_{-0.01}$ & $-82.9^{+0.5}_{-0.4}$ & $0.017^{+0.005}_{-0.005}$ & $-5.1^{+4.2}_{-5.7}$ & $0.064^{+0.004}_{-0.004}$\\[+0.75em]
ATLAS J2344-3056 & $0.501^{+0.001}_{-0.001}$ & $2.02^{+0.02}_{-0.03}$ & $0.74^{+0.01}_{-0.01}$ & $-24.4^{+0.3}_{-0.4}$ & $0.028^{+0.003}_{-0.003}$ & $89.2^{+2.7}_{-2.2}$ & $0.017^{+0.001}_{-0.001}$\\[+0.75em]
    \hline
  \end{tabular}
\end{table*}

\begin{table*}
  \caption{Model parameters for lens light distributions, which are median values. The associated uncertainties are statistical in nature and were computed using 84th and 16th percentiles.}
  \label{tab:lens_light_results}
  \scriptsize
%   \small
%   \begin{threeparttable}
  \begin{tabular}{lccccccc}
    \hline
    Name of &  $n_{\textrm{S\'{e}rsic}}$ & $\theta_{\rm{e}}$ & $q_{\rm{L}}$ & $\phi_{\rm{L}}$ (N of E) & $I_{\rm{e}}$ (F814W) & $I_{\rm{e}}$ (F475X) & $I_{\rm{e}}$ (F160W) \\
    Lens System  &    & (arcsec)  &     & (degree) & (mag/arcsec$^2$) & (mag/arcsec$^2$) & (mag/arcsec$^2$) \\
    \hline
    \\[-0.75em]
J0029-3814 & $5.99^{+0.01}_{-0.02}$ & $1.07^{+0.04}_{-0.04}$ & $0.59^{+0.01}_{-0.01}$ & $51.4^{+0.4}_{-0.4}$ & $24.12^{+0.02}_{-0.02}$ & $25.15^{+0.02}_{-0.02}$ & $22.75^{+0.05}_{-0.05}$\\[+0.75em]
PS J0030-1525 & $2.32^{+0.02}_{-0.02}$ & $0.71^{+0.01}_{-0.01}$ & $0.38^{+0.01}_{-0.01}$ & $12.3^{+0.2}_{-0.2}$ & $25.06^{+0.01}_{-0.01}$ & $26.10^{+0.01}_{-0.01}$ & $19.16^{+0.01}_{-0.01}$\\[+0.75em]
DES J0053-2012 & $5.98^{+0.02}_{-0.03}$ & $0.66^{+0.01}_{-0.01}$ & $0.76^{+0.01}_{-0.01}$ & $-88.7^{+0.6}_{-0.5}$ & $23.81^{+0.01}_{-0.01}$ & $24.83^{+0.01}_{-0.01}$ & $20.86^{+0.03}_{-0.04}$\\[+0.75em]
PS J0147+4630 & $4.0$ & $1.57^{+0.01}_{-0.01}$ & $0.90^{+0.01}_{-0.01}$ & $34.9^{+1.2}_{-1.2}$ & $23.86^{+0.01}_{-0.01}$ & $24.90^{+0.01}_{-0.01}$ & $19.89^{+0.01}_{-0.01}$\\[+0.50em]
 & $1.0$ & $0.82^{+0.00}_{-0.00}$ & $0.78^{+0.00}_{-0.01}$ & $10.3^{+1.3}_{-1.1}$ & $24.71^{+0.01}_{-0.01}$ & $25.75^{+0.01}_{-0.01}$ & $-$ \\[+0.75em]
WG0214-2105 & $6.00^{+0.01}_{-0.01}$ & $1.86^{+0.02}_{-0.02}$ & $0.86^{+0.01}_{-0.01}$ & $26.6^{+1.1}_{-1.1}$ & $23.67^{+0.01}_{-0.01}$ & $24.70^{+0.01}_{-0.01}$ & $21.96^{+0.01}_{-0.01}$\\[+0.75em]
SDSS J0248+1913 \textsuperscript{$\dagger$} & $2.94^{+0.21}_{-0.17}$ & $0.27^{+0.01}_{-0.01}$ & $0.44^{+0.01}_{-0.01}$ & $79.7^{+0.6}_{-0.7}$ & $24.79^{+0.04}_{-0.04}$ & $25.83^{+0.04}_{-0.04}$ & $19.87^{+0.08}_{-0.07}$\\[+0.75em]
WISE J0259-1635 & $4.0$ & $0.26^{+0.03}_{-0.02}$ & $0.27^{+0.01}_{-0.01}$ & $73.5^{+0.8}_{-1.0}$ & $25.16^{+0.05}_{-0.06}$ & $26.20^{+0.05}_{-0.06}$ & $20.36^{+0.16}_{-0.18}$\\[+0.50em]
 & $1.0$ & $1.07^{+0.01}_{-0.01}$ & $0.98^{+0.01}_{-0.01}$ & $-63.6^{+18.2}_{-14.0}$ & $24.46^{+0.01}_{-0.01}$ & $25.50^{+0.01}_{-0.01}$ & $20.79^{+0.03}_{-0.03}$\\[+0.75em]
J0343-2828 & $4.0$ & $0.34^{+0.01}_{-0.01}$ & $0.60^{+0.01}_{-0.01}$ & $-39.7^{+0.3}_{-0.4}$ & $24.30^{+0.01}_{-0.01}$ & $25.34^{+0.01}_{-0.01}$ & $19.06^{+0.02}_{-0.02}$\\[+0.50em]
 & $1.0$ & $5.00^{+0.00}_{-0.01}$ & $0.37^{+0.00}_{-0.00}$ & $-41.6^{+0.3}_{-0.3}$ & $25.52^{+0.01}_{-0.01}$ & $26.56^{+0.01}_{-0.01}$ & $23.61^{+0.01}_{-0.01}$\\[+0.75em]
DES J0405-3308 & $5.94^{+0.04}_{-0.08}$ & $1.11^{+0.03}_{-0.03}$ & $0.74^{+0.01}_{-0.01}$ & $56.4^{+1.4}_{-1.5}$ & $23.87^{+0.02}_{-0.02}$ & $24.91^{+0.02}_{-0.02}$ & $21.52^{+0.04}_{-0.05}$\\[+0.75em]
DES J0420-4037 & $4.0$ & $0.46^{+0.01}_{-0.01}$ & $0.73^{+0.01}_{-0.01}$ & $61.0^{+0.5}_{-0.5}$ & $24.09^{+0.01}_{-0.01}$ & $25.13^{+0.01}_{-0.01}$ & $18.12^{+0.05}_{-0.05}$\\[+0.50em]
 & $1.0$ & $0.23^{+0.00}_{-0.00}$ & $0.84^{+0.01}_{-0.01}$ & $61.9^{+1.5}_{-1.6}$ & $24.63^{+0.01}_{-0.01}$ & $25.67^{+0.01}_{-0.01}$ & $-$ \\[+0.75em]
DES J0530-3730 & $5.53^{+0.34}_{-0.72}$ & $0.11^{+0.02}_{-0.01}$ & $0.52^{+0.15}_{-0.16}$ & $60.1^{+16.5}_{-19.1}$ & $24.30^{+0.39}_{-0.27}$ & $25.34^{+0.39}_{-0.27}$ & $-$ \\[+0.75em]
PS J0630-1201 & $4.0$ & $0.36^{+0.01}_{-0.01}$ & $0.58^{+0.01}_{-0.01}$ & $-56.0^{+0.5}_{-0.4}$ & $24.33^{+0.01}_{-0.01}$ & $25.37^{+0.01}_{-0.01}$ & $19.61^{+0.06}_{-0.03}$\\[+0.50em]
 & $1.0$ & $0.10^{+0.00}_{-0.00}$ & $0.27^{+0.01}_{-0.00}$ & $61.1^{+0.5}_{-0.9}$ & $25.85^{+0.02}_{-0.04}$ & $26.89^{+0.02}_{-0.04}$ & $-$ \\[+0.75em]
J0659+1629 & $5.75^{+0.05}_{-0.05}$ & $1.48^{+0.03}_{-0.03}$ & $0.95^{+0.01}_{-0.01}$ & $-72.2^{+1.6}_{-1.6}$ & $23.61^{+0.01}_{-0.01}$ & $24.65^{+0.01}_{-0.01}$ & $20.58^{+0.03}_{-0.03}$\\[+0.75em]
J0818-2613 & $4.0$ & $2.43^{+0.01}_{-0.03}$ & $0.70^{+0.01}_{-0.01}$ & $49.2^{+0.4}_{-0.6}$ & $24.13^{+0.01}_{-0.01}$ & $25.17^{+0.01}_{-0.01}$ & $20.83^{+0.01}_{-0.02}$\\[+0.50em]
 & $1.0$ & $0.98^{+0.01}_{-0.01}$ & $0.67^{+0.01}_{-0.01}$ & $49.9^{+0.6}_{-0.5}$ & $24.87^{+0.01}_{-0.01}$ & $25.91^{+0.01}_{-0.01}$ & $-$ \\[+0.75em]
W2M J1042+1641 & $5.61^{+0.15}_{-0.15}$ & $2.22^{+0.10}_{-0.10}$ & $0.75^{+0.01}_{-0.01}$ & $67.6^{+0.9}_{-0.8}$ & $23.89^{+0.02}_{-0.02}$ & $24.93^{+0.02}_{-0.02}$ & $22.31^{+0.08}_{-0.08}$\\[+0.75em]
J1131-4419 & $4.0$ & $0.29^{+0.01}_{-0.01}$ & $0.61^{+0.01}_{-0.01}$ & $83.4^{+0.7}_{-0.7}$ & $24.28^{+0.01}_{-0.01}$ & $25.32^{+0.01}_{-0.01}$ & $18.31^{+0.03}_{-0.03}$\\[+0.50em]
 & $1.0$ & $0.96^{+0.03}_{-0.03}$ & $0.36^{+0.01}_{-0.01}$ & $49.4^{+0.4}_{-0.5}$ & $25.54^{+0.03}_{-0.03}$ & $26.58^{+0.03}_{-0.03}$ & $21.36^{+0.07}_{-0.08}$\\[+0.75em]
2M1134-2103 & $6.00^{+0.01}_{-0.01}$ & $1.13^{+0.02}_{-0.04}$ & $0.74^{+0.01}_{-0.01}$ & $-58.1^{+0.9}_{-0.8}$ & $23.86^{+0.01}_{-0.01}$ & $24.90^{+0.01}_{-0.01}$ & $20.94^{+0.03}_{-0.04}$\\[+0.75em]
SDSS J1251+2935 \textsuperscript{$\dagger$} & $4.0$ & $1.18^{+0.02}_{-0.02}$ & $0.68^{+0.01}_{-0.01}$ & $63.0^{+0.6}_{-0.6}$ & $24.16^{+0.01}_{-0.01}$ & $25.20^{+0.01}_{-0.01}$ & $20.51^{+0.05}_{-0.06}$\\[+0.50em]
 & $1.0$ & $0.83^{+0.01}_{-0.01}$ & $0.66^{+0.01}_{-0.01}$ & $55.4^{+1.2}_{-1.2}$ & $24.89^{+0.02}_{-0.02}$ & $25.93^{+0.02}_{-0.02}$ & $19.66^{+0.04}_{-0.04}$\\[+0.75em]
2M1310-1714 & $4.0$ & $0.84^{+0.01}_{-0.01}$ & $0.63^{+0.01}_{-0.01}$ & $-87.6^{+0.1}_{-0.1}$ & $24.25^{+0.01}_{-0.01}$ & $25.29^{+0.01}_{-0.01}$ & $18.80^{+0.01}_{-0.01}$\\[+0.50em]
 & $1.0$ & $5.00^{+0.00}_{-0.00}$ & $0.63^{+0.00}_{-0.00}$ & $-59.7^{+0.3}_{-0.3}$ & $24.95^{+0.01}_{-0.01}$ & $25.99^{+0.01}_{-0.01}$ & $22.73^{+0.01}_{-0.01}$\\[+0.75em]
SDSS J1330+1810 & $4.0$ & $1.47^{+0.02}_{-0.02}$ & $0.36^{+0.01}_{-0.01}$ & $65.4^{+0.1}_{-0.1}$ & $24.85^{+0.01}_{-0.01}$ & $25.89^{+0.01}_{-0.01}$ & $20.53^{+0.03}_{-0.03}$\\[+0.50em]
 & $1.0$ & $0.30^{+0.00}_{-0.00}$ & $0.21^{+0.00}_{-0.00}$ & $64.7^{+0.2}_{-0.2}$ & $26.11^{+0.01}_{-0.01}$ & $27.15^{+0.01}_{-0.01}$ & $19.15^{+0.04}_{-0.04}$\\[+0.75em]
SDSS J1433+6007 & $4.0$ & $0.58^{+0.01}_{-0.01}$ & $0.59^{+0.01}_{-0.01}$ & $-10.0^{+0.2}_{-0.2}$ & $24.33^{+0.01}_{-0.01}$ & $25.37^{+0.01}_{-0.01}$ & $19.25^{+0.01}_{-0.01}$\\[+0.50em]
 & $1.0$ & $3.63^{+0.05}_{-0.04}$ & $0.51^{+0.01}_{-0.01}$ & $-2.2^{+0.3}_{-0.3}$ & $25.18^{+0.01}_{-0.01}$ & $26.22^{+0.01}_{-0.01}$ & $23.42^{+0.03}_{-0.03}$\\[+0.75em]
J1537-3010 \textsuperscript{$\dagger$} & $7.16^{+0.08}_{-0.08}$ & $2.55^{+0.05}_{-0.05}$ & $0.76^{+0.01}_{-0.01}$ & $57.5^{+0.3}_{-0.3}$ & $23.74^{+0.01}_{-0.01}$ & $24.78^{+0.01}_{-0.01}$ & $21.90^{+0.04}_{-0.03}$\\[+0.75em]
PS J1606-2333 & $5.97^{+0.02}_{-0.04}$ & $1.48^{+0.08}_{-0.08}$ & $0.58^{+0.01}_{-0.01}$ & $-73.1^{+0.9}_{-1.0}$ & $24.13^{+0.02}_{-0.02}$ & $25.17^{+0.02}_{-0.02}$ & $22.04^{+0.07}_{-0.07}$\\[+0.75em]
J1721+8842 & $4.02^{+0.01}_{-0.01}$ & $5.00^{+0.01}_{-0.01}$ & $0.86^{+0.01}_{-0.01}$ & $3.6^{+0.2}_{-0.2}$ & $23.91^{+0.01}_{-0.01}$ & $24.95^{+0.01}_{-0.01}$ & $20.88^{+0.01}_{-0.01}$\\[+0.75em]
J1817+2729 & $4.0$ & $2.00^{+0.05}_{-0.04}$ & $0.27^{+0.01}_{-0.01}$ & $30.2^{+0.1}_{-0.1}$ & $25.17^{+0.01}_{-0.01}$ & $26.21^{+0.01}_{-0.01}$ & $21.48^{+0.03}_{-0.02}$\\[+0.50em]
 & $1.0$ & $1.19^{+0.02}_{-0.03}$ & $0.25^{+0.01}_{-0.01}$ & $17.8^{+0.3}_{-0.3}$ & $25.95^{+0.02}_{-0.03}$ & $26.99^{+0.02}_{-0.03}$ & $-$ \\[+0.75em]
DES J2038-4008 & $4.0$ & $2.85^{+0.01}_{-0.01}$ & $0.63^{+0.01}_{-0.01}$ & $53.0^{+0.1}_{-0.1}$ & $24.25^{+0.01}_{-0.01}$ & $25.29^{+0.01}_{-0.01}$ & $20.16^{+0.01}_{-0.01}$\\[+0.50em]
 & $1.0$ & $3.17^{+0.05}_{-0.04}$ & $0.63^{+0.01}_{-0.01}$ & $-26.2^{+0.4}_{-0.5}$ & $24.94^{+0.01}_{-0.01}$ & $25.98^{+0.01}_{-0.01}$ & $22.14^{+0.05}_{-0.04}$\\[+0.75em]
WG2100-4452 & $4.0$ & $0.96^{+0.01}_{-0.01}$ & $0.61^{+0.01}_{-0.01}$ & $86.0^{+0.1}_{-0.1}$ & $24.29^{+0.01}_{-0.01}$ & $25.33^{+0.01}_{-0.01}$ & $18.95^{+0.01}_{-0.01}$\\[+0.50em]
 & $1.0$ & $2.96^{+0.06}_{-0.05}$ & $0.82^{+0.01}_{-0.01}$ & $-30.2^{+1.3}_{-1.3}$ & $24.66^{+0.01}_{-0.01}$ & $25.70^{+0.01}_{-0.01}$ & $22.14^{+0.05}_{-0.05}$\\[+0.75em]
J2145+6345 & $5.90^{+0.08}_{-0.15}$ & $1.00^{+0.09}_{-0.09}$ & $0.72^{+0.02}_{-0.02}$ & $-54.8^{+2.1}_{-2.6}$ & $23.91^{+0.04}_{-0.03}$ & $24.95^{+0.04}_{-0.03}$ & $21.00^{+0.12}_{-0.14}$\\[+0.75em]
J2205-3727 \textsuperscript{$\dagger$} & $6.60^{+0.16}_{-0.13}$ & $0.78^{+0.03}_{-0.03}$ & $0.74^{+0.01}_{-0.01}$ & $-76.7^{+0.6}_{-0.6}$ & $23.81^{+0.01}_{-0.01}$ & $24.85^{+0.01}_{-0.01}$ & $20.84^{+0.07}_{-0.07}$\\[+0.75em]
ATLAS J2344-3056 \textsuperscript{$\dagger$} & $3.60^{+0.21}_{-0.19}$ & $1.45^{+0.10}_{-0.09}$ & $0.84^{+0.01}_{-0.01}$ & $-22.9^{+0.8}_{-1.0}$ & $24.00^{+0.02}_{-0.03}$ & $25.04^{+0.02}_{-0.03}$ & $21.38^{+0.13}_{-0.12}$\\[+0.75em]
    \hline
    \multicolumn{8}{l}{\textsuperscript{$\dagger$}\scriptsize{Lenses model reconstructed by pipeline before restricting S\'ersic index of main deflector's light profile to 6.0.}}
  \end{tabular}
\end{table*}

\begin{table*}
  \caption{Astrometric positions of the main deflector's light profile centroid and lensed  QSO images. The total uncertainty on relative astrometry is dominated by 
  systematic errors associated with the reconstruction of the PSF on sub-pixel scale. We estimate it to be 6 mas by comparison with \textit{Gaia} (\S~\ref{sssec:systematic_astro}). Formal random uncertainties are negligible in comparison and therefore not listed.
  }
  \label{tab:astro_pos_results}
%   \footnotesize
  \scriptsize
%   \tiny
  \begin{tabular}{lcccccccccccc}
    \hline
    Name of & \multicolumn{2}{c}{Location} & \multicolumn{2}{c}{Main Deflector} & \multicolumn{2}{c}{Image A}  & \multicolumn{2}{c}{Image B} & \multicolumn{2}{c}{Image C} & \multicolumn{2}{c}{Image D} \\
    Lens System  & RA & DEC & $\Delta$RA & $\Delta$DEC & $\Delta$RA & $\Delta$DEC & $\Delta$RA & $\Delta$DEC & $\Delta$RA & $\Delta$DEC & $\Delta$RA & $\Delta$DEC\\
    & (degree) & (degree) & (arcsec) & (arcsec) & (arcsec) & (arcsec) & (arcsec) & (arcsec) & (arcsec) & (arcsec) & (arcsec) & (arcsec)\\ 
    \hline
    \\[-0.75em]
    J0029-3814 & $7.419298$ & $-38.240600$ & $-0.134$ & $-0.155$ & $1.156$ & $-0.513$ & $-0.592$ & $-0.490$ & $-0.975$ & $0.071$ & $-0.321$ & $0.384$\\[+0.75em]
PS J0030-1525 & $7.563492$ & $-15.417800$ & $-0.098$ & $0.020$ & $0.746$ & $-0.908$ & $-0.868$ & $0.002$ & $0.778$ & $0.887$ & $1.005$ & $0.608$\\[+0.75em]
DES J0053-2012 & $13.435033$ & $-20.209147$ & $-0.422$ & $0.429$ & $-0.529$ & $1.036$ & $1.414$ & $0.012$ & $0.691$ & $-0.731$ & $-1.424$ & $-1.111$\\[+0.75em]
PS J0147+4630 & $26.792372$ & $46.511872$ & $-0.145$ & $-1.137$ & $-0.317$ & $-2.296$ & $-1.218$ & $0.836$ & $0.029$ & $0.936$ & $1.191$ & $0.526$\\[+0.75em]
WG0214-2105 & $33.568175$ & $-21.093137$ & $0.071$ & $-0.015$ & $0.556$ & $-0.868$ & $-0.706$ & $-0.136$ & $-0.260$ & $0.790$ & $0.633$ & $0.517$\\[+0.75em]
SDSS J0248+1913 & $42.203067$ & $19.225228$ & $0.105$ & $0.071$ & $0.451$ & $-0.748$ & $-0.549$ & $-0.135$ & $-0.404$ & $0.699$ & $0.503$ & $0.661$\\[+0.75em]
WISE J0259-1635 & $44.928533$ & $-16.595370$ & $0.058$ & $-0.039$ & $0.035$ & $-0.730$ & $-0.727$ & $0.216$ & $0.434$ & $0.537$ & $0.752$ & $-0.342$\\[+0.75em]
J0343-2828 & $55.797650$ & $-28.477948$ & $-0.869$ & $0.894$ & $-1.251$ & $1.352$ & $1.959$ & $1.349$ & $1.364$ & $-0.746$ & $-1.329$ & $-1.952$\\[+0.75em]
DES J0405-3308 & $61.498960$ & $-33.147410$ & $-0.014$ & $-0.045$ & $0.691$ & $-0.279$ & $-0.374$ & $-0.605$ & $-0.529$ & $0.416$ & $0.349$ & $0.556$\\[+0.75em]
DES J0420-4037 & $65.194823$ & $-40.624087$ & $0.113$ & $-0.001$ & $0.821$ & $-0.579$ & $-0.586$ & $-0.349$ & $-0.339$ & $0.675$ & $0.287$ & $0.796$\\[+0.75em]
DES J0530-3730 & $82.654075$ & $-37.503113$ & $0.325$ & $-0.325$ & $0.597$ & $-0.517$ & $-0.384$ & $-0.237$ & $-0.143$ & $0.243$ & $-0.018$ & $0.311$\\[+0.75em]
PS J0630-1201 & $97.537708$ & $-12.022081$ & $-0.502$ & $0.275$ & $-1.370$ & $1.157$ & $1.151$ & $0.416$ & $0.822$ & $-0.704$ & $0.302$ & $-1.268$\\[+0.75em]
J0659+1629 & $104.766545$ & $16.485908$ & $0.390$ & $-0.066$ & $2.192$ & $-0.928$ & $-2.481$ & $-1.265$ & $1.205$ & $1.953$ & $2.274$ & $0.974$\\[+0.75em]
J0818-2613 & $124.617817$ & $-26.223740$ & $-0.882$ & $1.184$ & $-2.488$ & $-2.982$ & $-2.042$ & $2.603$ & $2.010$ & $1.366$ & $1.854$ & $0.695$\\[+0.75em]
W2M J1042+1641 & $160.592005$ & $16.687614$ & $-0.052$ & $0.034$ & $-0.858$ & $0.687$ & $0.694$ & $0.111$ & $0.526$ & $-0.468$ & $-0.087$ & $-0.803$\\[+0.75em]
J1131-4419 & $172.750079$ & $-44.333469$ & $-0.012$ & $0.050$ & $0.072$ & $-0.742$ & $-0.881$ & $0.349$ & $0.411$ & $0.793$ & $0.754$ & $0.446$\\[+0.75em]
2M1134-2103 & $173.668952$ & $-21.056299$ & $-0.154$ & $0.174$ & $1.326$ & $1.150$ & $0.593$ & $-0.609$ & $-1.356$ & $-1.384$ & $-0.660$ & $0.765$\\[+0.75em]
SDSS J1251+2935 & $192.781367$ & $29.594673$ & $0.185$ & $-0.050$ & $-0.885$ & $0.278$ & $0.829$ & $0.286$ & $0.895$ & $-0.304$ & $0.533$ & $-0.678$\\[+0.75em]
2M1310-1714 & $197.583583$ & $-17.249381$ & $-0.191$ & $0.522$ & $-1.093$ & $2.656$ & $2.790$ & $0.165$ & $1.508$ & $-2.447$ & $-2.295$ & $-2.222$\\[+0.75em]
SDSS J1330+1810 & $202.577718$ & $18.175763$ & $0.170$ & $0.068$ & $0.629$ & $0.700$ & $0.384$ & $-0.902$ & $-0.023$ & $-0.910$ & $-0.860$ & $0.256$\\[+0.75em]
SDSS J1433+6007 & $218.345150$ & $60.120839$ & $0.451$ & $-0.218$ & $1.558$ & $-0.338$ & $-0.486$ & $-1.913$ & $-1.243$ & $-0.288$ & $-0.480$ & $1.842$\\[+0.75em]
J1537-3010 & $234.355668$ & $-30.171336$ & $0.095$ & $-0.030$ & $1.491$ & $-0.815$ & $-0.471$ & $-1.126$ & $-1.352$ & $0.822$ & $0.772$ & $0.970$\\[+0.75em]
PS J1606-2333 & $241.500980$ & $-23.556122$ & $0.035$ & $0.025$ & $0.852$ & $0.414$ & $0.065$ & $-0.490$ & $-0.765$ & $-0.173$ & $-0.272$ & $0.567$\\[+0.75em]
J1721+8842 & $260.432958$ & $88.705847$ & $-0.287$ & $0.261$ & $-1.694$ & $-0.094$ & $0.312$ & $2.378$ & $1.589$ & $0.965$ & $1.242$ & $-1.361$\\[+0.75em]
J1817+2729 & $274.378603$ & $27.494383$ & $0.096$ & $0.105$ & $0.655$ & $-0.732$ & $-0.692$ & $-0.256$ & $-0.607$ & $0.553$ & $0.671$ & $0.709$\\[+0.75em]
DES J2038-4008 & $309.511333$ & $-40.137050$ & $0.117$ & $0.074$ & $0.819$ & $0.938$ & $0.945$ & $-1.141$ & $-0.572$ & $-1.112$ & $-1.367$ & $0.569$\\[+0.75em]
WG2100-4452 & $315.062075$ & $-44.868438$ & $-0.078$ & $0.072$ & $-0.286$ & $1.128$ & $1.318$ & $-0.401$ & $-0.703$ & $-1.083$ & $-1.125$ & $-0.749$\\[+0.75em]
J2145+6345 & $326.271094$ & $63.761447$ & $-0.174$ & $0.394$ & $-0.955$ & $-0.650$ & $-0.487$ & $1.020$ & $0.890$ & $0.280$ & $0.571$ & $-0.298$\\[+0.75em]
J2205-3727 & $331.434422$ & $-37.450361$ & $-0.046$ & $0.088$ & $0.859$ & $0.189$ & $-0.342$ & $-0.536$ & $-0.780$ & $0.066$ & $-0.491$ & $0.624$\\[+0.75em]
ATLAS J2344-3056 & $356.070733$ & $-30.940611$ & $0.055$ & $-0.115$ & $-0.431$ & $0.073$ & $0.150$ & $0.425$ & $0.442$ & $-0.245$ & $-0.191$ & $-0.583$\\[+0.75em]
    
    \hline
    \multicolumn{11}{l}{\textsuperscript{$\dagger$}\scriptsize{Astrometric position of 5th image for J0343-282: $\Delta$RA = $-0\arcsec.273$, $\Delta$DEC = $0\arcsec.473$; and for 2M1310-1714: $\Delta$RA = $-0\arcsec.181$, $\Delta$DEC = $0\arcsec.070$.}}
  \end{tabular}
\end{table*}

This section provides details on the lens systems that have been successfully processed by the automated pipeline. For each lens, we give a description of the deflector's mass profile parameters as well as details on the corresponding light profile components. 
For the system that cannot be successfully reconstructed by the framework, we list the reasons in Appendix \ref{Failure_modes} and discuss necessary modification that could be implemented in future iterations of the pipeline in order to achieve a fully automated reconstruction. We further show predicted time delays for flux variations between the quasar images, based on measured or assumed redshifts for the main deflector and lensed quasar.

\subsection{Lens models}

\begin{figure}
 \includegraphics[width=0.5\textwidth]{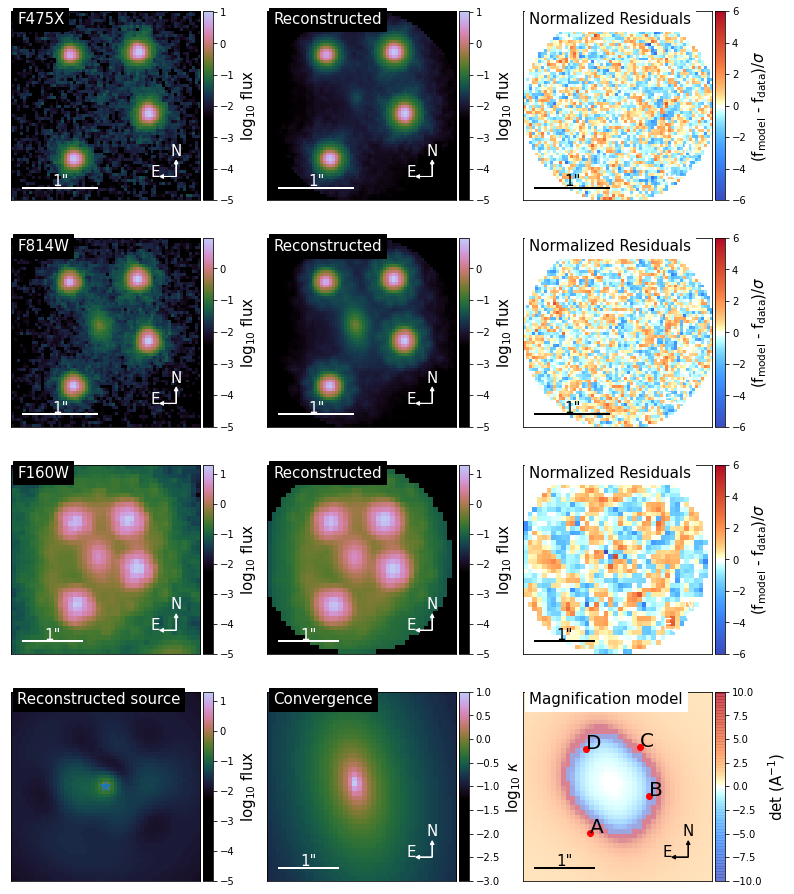}
 \caption{Comparison of observations with the reconstructed model for SDSS J0248+1913 in {\it{HST}} bands F475X (first row), F814W (second row), and F160W (third row). Also shown are the respective normalized residual for each band, after the subtraction of the data from the model. The last row shows the reconstructed source using information from the F160W band (column 1), a plot of the unitless convergence, $\kappa(\theta)$ (column 2), and a model plotting the magnification as well as the position of the lensed quasar images (column 3).}
 \label{fig:J0248}
\end{figure}

\begin{figure}
 \includegraphics[width=0.5\textwidth]{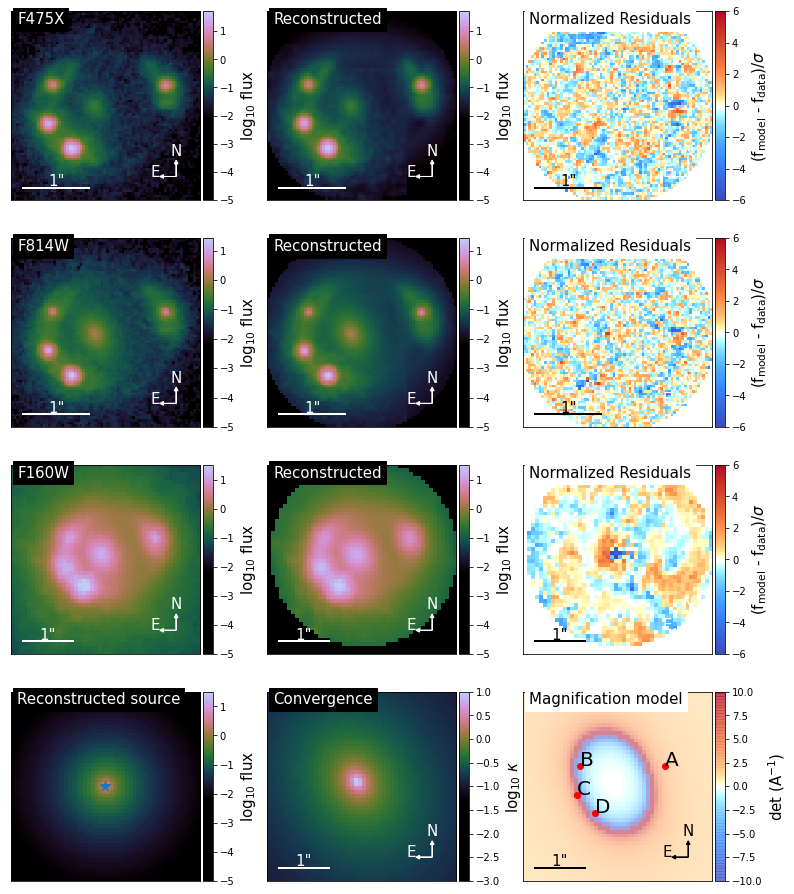}
 \caption{Comparison of observations with the reconstructed model for SDSSJ 1251+2935 in {\it{HST}} bands F475X (first row), F814W (second row), and F160W (third row). Also shown are the respective normalized residual for each band, after the subtraction of the data from the model. The last row shows the reconstructed source using information from the F160W band (column 1), a plot of the unitless convergence, $\kappa(\theta)$ (column 2), and a model plotting the magnification as well as the position of the lensed quasar images (column 3).}
 \label{fig:J1251}
\end{figure}

For 30 out of 31 lenses (97\%), our automated pipeline is able to reconstruct models based on the observational data. As an example, for two of the systems in our sample, we show in Figures~\ref{fig:J0248} and~\ref{fig:J1251} a comparison between the {\it HST} observations in each filter (column 1) and the corresponding reconstructed lens model (column 2). To demonstrate how well our models match the data, we include (in column 3) the normalized residuals after the subtraction of observational data from the reconstructed model. Also shown, in the 4th row for each figure, is a reconstruction of the lensed galaxy's light in {\it{HST}} band F160W (column 1) and the convergence, $\kappa(\theta)$, for the respective lens configuration (column 2). Lastly, the figures include a magnification model (column 3 in 4th row), indicating the position of the lensed quasar images. The corresponding convergence and external shear strength at the image positions can be found in Table~\ref{tab:conv_and_shear}, along with the magnification for each QSO image. For our estimates of the stellar convergences, $\kappa_{\star}$, at the image positions we use the lens light flux in the F160W band and assume a constant mass-to-light ratio. The normalization factor has been chosen such that within an area of 1/2 of the effective radius, the integrated stellar convergence is 2/3, or less, of the integrated convergence \citep[see][]{Auger10b}.\\

\subsection{Lens model parameters} \label{lensmodels}

\begin{table*}
  \caption{Model parameters for source light distributions, which are median values. The associated uncertainties are statistical in nature and were computed using 84th and 16th percentiles.}
  \label{tab:source_light_results}
%   \scriptsize
  \begin{tabular}{lcccccccccc}
    \hline
    Name of &   &  & \multicolumn{2}{c}{Centroid} & \multicolumn{2}{c}{F814W} & \multicolumn{2}{c}{F475X} & \multicolumn{2}{c}{F160W}\\
    Lens System &  $n_{\textrm{S\'ersic}}$ & $\theta_{\rm{e}}$ & $\Delta$RA & $\Delta$DEC & $n_{\rm{max}}$ & $\beta$& $n_{\rm{max}}$ & $\beta$& $n_{\rm{max}}$ & $\beta$\\
     &    & (arcsec)  &  (arcsec)  & (arcsec)  & & (arcsec) & & (arcsec) & & (arcsec) \\
    \hline
    \\[-0.75em]
J0029-3814 & $2.82^{+0.23}_{-0.21}$ & $1.05^{+0.03}_{-0.06}$ & $0.043^{+0.004}_{-0.005}$ & $-0.243^{+0.004}_{-0.003}$ & $-$  & $-$  & $-$  & $-$  & $-$  & $-$ \\[+0.75em]
PS J0030-1525 & $0.53^{+0.02}_{-0.01}$ & $0.44^{+0.01}_{-0.01}$ & $0.327^{+0.010}_{-0.012}$ & $0.045^{+0.003}_{-0.003}$ & $-$  & $-$  & $-$  & $-$  & $-$  & $-$ \\[+0.75em]
DES J0053-2012 & $0.50^{+0.01}_{-0.01}$ & $0.77^{+0.03}_{-0.03}$ & $-0.151^{+0.004}_{-0.005}$ & $0.022^{+0.015}_{-0.013}$ & $-$  & $-$  & $-$  & $-$  & $-$  & $-$ \\[+0.75em]
PS J0147+4630 & $1.30^{+0.23}_{-0.19}$ & $0.01^{+0.01}_{-0.01}$ & $-0.157^{+0.002}_{-0.002}$ & $-0.720^{+0.014}_{-0.015}$ & $-$  & $-$  & $-$  & $-$  & $-$  & $-$ \\[+0.75em]
WG0214-2105 & $0.92^{+0.26}_{-0.21}$ & $0.06^{+0.01}_{-0.01}$ & $0.096^{+0.001}_{-0.001}$ & $-0.052^{+0.001}_{-0.001}$ & $-$  & $-$  & $-$  & $-$  & $-$  & $-$ \\[+0.75em]
SDSS J0248+1913 & $3.62^{+0.12}_{-0.13}$ & $1.91^{+0.06}_{-0.12}$ & $0.112^{+0.001}_{-0.001}$ & $0.031^{+0.002}_{-0.002}$ & $-$  & $-$  & $0$ & $0.100^{+0.015}_{-0.018}$ & $5$ & $0.187^{+0.010}_{-0.010}$\\[+0.75em]
WISE J0259-1635 & $3.96^{+0.03}_{-0.06}$ & $1.09^{+0.01}_{-0.01}$ & $0.016^{+0.001}_{-0.001}$ & $-0.024^{+0.001}_{-0.001}$ & $3$ & $0.075^{+0.002}_{-0.002}$ & $4$ & $0.063^{+0.002}_{-0.001}$ & $1$ & $0.145^{+0.004}_{-0.004}$\\[+0.75em]
J0343-2828 & $3.99^{+0.01}_{-0.01}$ & $0.06^{+0.01}_{-0.01}$ & $-0.122^{+0.003}_{-0.003}$ & $0.149^{+0.003}_{-0.003}$ & $8$ & $0.161^{+0.009}_{-0.007}$ & $3$ & $0.031^{+0.003}_{-0.005}$ & $-$  & $-$ \\[+0.75em]
DES J0405-3308 & $3.96^{+0.03}_{-0.06}$ & $0.17^{+0.01}_{-0.01}$ & $0.017^{+0.001}_{-0.001}$ & $-0.039^{+0.001}_{-0.001}$ & $-$  & $-$  & $-$  & $-$  & $-$  & $-$ \\[+0.75em]
DES J0420-4037 & $0.59^{+0.39}_{-0.07}$ & $0.91^{+0.11}_{-0.12}$ & $0.166^{+0.001}_{-0.001}$ & $-0.033^{+0.001}_{-0.001}$ & $15$ & $0.037^{+0.001}_{-0.001}$ & $15$ & $0.037^{+0.001}_{-0.001}$ & $15$ & $0.141^{+0.005}_{-0.005}$\\[+0.75em]
DES J0530-3730 & $1.27^{+1.02}_{-0.51}$ & $1.09^{+0.01}_{-0.02}$ & $0.172^{+0.019}_{-0.008}$ & $-0.215^{+0.006}_{-0.005}$ & $5$ & $0.083^{+0.011}_{-0.012}$ & $5$ & $0.062^{+0.011}_{-0.016}$ & $-$  & $-$ \\[+0.75em]
PS J0630-1201 & $1.70^{+0.08}_{-0.07}$ & $1.10^{+0.01}_{-0.01}$ & $-0.275^{+0.007}_{-0.007}$ & $0.217^{+0.005}_{-0.005}$ & $-$  & $-$  & $-$  & $-$  & $-$  & $-$ \\[+0.75em]
J0659+1629 & $2.18^{+0.14}_{-0.11}$ & $1.10^{+0.01}_{-0.01}$ & $0.028^{+0.011}_{-0.011}$ & $-0.215^{+0.002}_{-0.002}$ & $-$  & $-$  & $-$  & $-$  & $-$  & $-$ \\[+0.75em]
J0818-2613 & $0.50^{+0.01}_{-0.01}$ & $0.01^{+0.01}_{-0.01}$ & $-1.064^{+0.002}_{-0.001}$ & $0.538^{+0.003}_{-0.002}$ & $-$  & $-$  & $-$  & $-$  & $6$ & $0.724^{+0.008}_{-0.001}$\\[+0.75em]
W2M J1042+1641 & $3.80^{+0.14}_{-0.30}$ & $0.54^{+0.08}_{-0.09}$ & $-0.194^{+0.002}_{-0.002}$ & $0.105^{+0.002}_{-0.002}$ & $5$ & $0.010^{+0.001}_{-0.001}$ & $5$ & $0.011^{+0.001}_{-0.001}$ & $-$  & $-$ \\[+0.75em]
J1131-4419 & $2.05^{+0.06}_{-0.05}$ & $1.09^{+0.01}_{-0.02}$ & $-0.052^{+0.002}_{-0.002}$ & $0.078^{+0.003}_{-0.002}$ & $7$ & $0.075^{+0.002}_{-0.002}$ & $3$ & $0.123^{+0.004}_{-0.003}$ & $5$ & $0.471^{+0.007}_{-0.008}$\\[+0.75em]
2M1134-2103 & $3.99^{+0.01}_{-0.02}$ & $1.10^{+0.01}_{-0.01}$ & $-0.097^{+0.002}_{-0.002}$ & $0.010^{+0.003}_{-0.002}$ & $-$  & $-$  & $-$  & $-$  & $-$  & $-$ \\[+0.75em]
SDSS J1251+2935 & $3.27^{+0.17}_{-0.14}$ & $0.24^{+0.01}_{-0.01}$ & $-0.004^{+0.002}_{-0.002}$ & $0.007^{+0.001}_{-0.001}$ & $5$ & $0.068^{+0.001}_{-0.001}$ & $5$ & $0.068^{+0.001}_{-0.001}$ & $-$  & $-$ \\[+0.75em]
2M1310-1714 & $4.00^{+0.01}_{-0.01}$ & $0.01^{+0.01}_{-0.01}$ & $-0.153^{+0.001}_{-0.001}$ & $-0.092^{+0.001}_{-0.001}$ & $20$ & $0.210^{+0.001}_{-0.002}$ & $8$ & $0.091^{+0.001}_{-0.001}$ & $1$ & $0.171^{+0.001}_{-0.001}$\\[+0.75em]
SDSS J1330+1810 & $1.90^{+0.09}_{-0.08}$ & $1.10^{+0.01}_{-0.01}$ & $0.045^{+0.006}_{-0.006}$ & $0.016^{+0.002}_{-0.002}$ & $2$ & $0.325^{+0.011}_{-0.011}$ & $2$ & $0.311^{+0.015}_{-0.017}$ & $-$  & $-$ \\[+0.75em]
SDSS J1433+6007 & $3.88^{+0.08}_{-0.15}$ & $0.64^{+0.05}_{-0.04}$ & $0.013^{+0.010}_{-0.010}$ & $-0.078^{+0.003}_{-0.003}$ & $-$  & $-$  & $-$  & $-$  & $-$  & $-$ \\[+0.75em]
J1537-3010 & $1.45^{+0.07}_{-0.07}$ & $0.16^{+0.01}_{-0.01}$ & $0.048^{+0.001}_{-0.001}$ & $-0.017^{+0.001}_{-0.001}$ & $5$ & $0.068^{+0.001}_{-0.001}$ & $8$ & $0.056^{+0.001}_{-0.001}$ & $-$  & $-$ \\[+0.75em]
PS J1606-2333 & $3.96^{+0.03}_{-0.07}$ & $0.67^{+0.03}_{-0.03}$ & $0.033^{+0.001}_{-0.001}$ & $0.025^{+0.005}_{-0.005}$ & $9$ & $0.108^{+0.003}_{-0.003}$ & $9$ & $0.105^{+0.002}_{-0.002}$ & $-$  & $-$ \\[+0.75em]
J1721+8842 & $1.73^{+0.22}_{-0.18}$ & $1.10^{+0.01}_{-0.01}$ & $-0.018^{+0.002}_{-0.002}$ & $0.288^{+0.001}_{-0.001}$ & $6$ & $1.075^{+0.017}_{-0.030}$ & $6$ & $1.061^{+0.027}_{-0.039}$ & $10$ & $0.117^{+0.001}_{-0.001}$\\[+0.75em]
J1817+2729 & $0.83^{+0.19}_{-0.13}$ & $0.46^{+0.06}_{-0.05}$ & $0.143^{+0.001}_{-0.001}$ & $0.027^{+0.001}_{-0.001}$ & $3$ & $0.012^{+0.002}_{-0.001}$ & $-$  & $-$  & $9$ & $0.643^{+0.019}_{-0.017}$\\[+0.75em]
DES J2038-4008 & $1.06^{+0.11}_{-0.10}$ & $0.11^{+0.01}_{-0.01}$ & $-0.027^{+0.001}_{-0.001}$ & $-0.031^{+0.001}_{-0.001}$ & $8$ & $0.144^{+0.002}_{-0.002}$ & $10$ & $0.126^{+0.001}_{-0.001}$ & $3$ & $0.257^{+0.002}_{-0.002}$\\[+0.75em]
WG2100-4452 & $3.96^{+0.03}_{-0.07}$ & $0.80^{+0.09}_{-0.08}$ & $0.025^{+0.003}_{-0.004}$ & $-0.075^{+0.006}_{-0.004}$ & $-$  & $-$  & $0$ & $0.179^{+0.009}_{-0.008}$ & $-$  & $-$ \\[+0.75em]
J2145+6345 & $3.94^{+0.04}_{-0.07}$ & $0.64^{+0.10}_{-0.08}$ & $-0.197^{+0.001}_{-0.001}$ & $0.214^{+0.005}_{-0.005}$ & $-$  & $-$  & $-$  & $-$  & $-$  & $-$ \\[+0.75em]
J2205-3727 & $1.74^{+0.15}_{-0.12}$ & $0.31^{+0.02}_{-0.02}$ & $0.046^{+0.003}_{-0.003}$ & $0.099^{+0.001}_{-0.001}$ & $-$  & $-$  & $-$  & $-$  & $-$  & $-$ \\[+0.75em]
ATLAS J2344-3056 & $0.57^{+0.03}_{-0.03}$ & $0.47^{+0.01}_{-0.01}$ & $0.005^{+0.001}_{-0.001}$ & $-0.085^{+0.001}_{-0.001}$ & $-$  & $-$  & $-$  & $-$  & $-$  & $-$ \\[+0.75em]
    \hline
    \end{tabular}
\end{table*}

The mean and associated uncertainties of the free model parameters for each lens are obtained from the MCMC chain. Therefore, the uncertainties listed do not account for systematic sources of error. In future analyses of this sample systematic errors will need to be estimated for each specific application. In some cases they can be dominant. We discuss some examples of systematic errors in the remainder of this paper and refer to the literature for additional examples.

A breakdown of the mass model components by attribute can be found in Table~\ref{tab:lens_mass_results}. These include the lens mass parameters of the main deflector, 
the attributes of the external shear profile associated with the combined impact of additional perturber along the line of sight, as well as the area enclosed by the inner caustics of the critical curve.

Table~\ref{tab:lens_light_results} details the lens light profile parameterization for each lensed system that is successfully processed by the pipeline. For lenses where the light profile of the main deflector is modeled by a double S\'ersic, we first list the parameters of profile with the S\'ersic index fixed at $4.0$, the de Vaucouleurs profile \citep{deVaucouleurs48}, and immediately below show the parameters of the light profile with the S\'ersic index fixed at $1.0$, the exponential profile.  

Table~\ref{tab:astro_pos_results} lists the astrometry of the point sources and galaxy centroid as inferred from our lens models.

Details on the reconstructed host galaxy of the lensed QSO can be found in Table~\ref{tab:source_light_results}. We note that many of the free parameters are highly correlated; the pairwise Pearson correlation coefficients are listed in Table~\ref{tab:cross_cor}.

\begin{table*}
  \caption{Cross-correlation strength between model parameters in our lens sample, as indicated by the pairwise Pearson correlation coefficient. Coefficients with an absolute value of 0.4 or greater are boldened to increase notability.}
  \label{tab:cross_cor}
  \tiny
%   \scriptsize
  \begin{tabular}{lccccccccccccccccccccc}
    \hline \\
    Parameter & \rotatebox[origin=c]{90}{$\theta_{\rm{E}}$}  & \rotatebox[origin=c]{90}{$\gamma$} & \rotatebox[origin=c]{90}{$q_{\rm{mass}}$} & \rotatebox[origin=c]{90}{$\Phi_{\rm{mass}}$} & \rotatebox[origin=c]{90}{$\gamma_{\rm{ext}}$} & \rotatebox[origin=c]{90}{$\phi_{\rm{ext}}$} & \rotatebox[origin=c]{90}{$n_{\textrm{S\'ersic}}$ (lens)} & \rotatebox[origin=c]{90}{$R_{\textrm{S\'ersic}}$ (lens)} &
    \rotatebox[origin=c]{90}{$q_{\rm{light}}$} & \rotatebox[origin=c]{90}{$\phi_{\rm{light}}$} & \rotatebox[origin=c]{90}{$n_{\textrm{S\'ersic}}$ (source)} & \rotatebox[origin=c]{90}{$R_{\textrm{S\'ersic}}$ (source)} &
    \rotatebox[origin=c]{90}{$n_{\rm{max}}$ (F814W)} &
    \rotatebox[origin=c]{90}{$\beta$ (F814W)} &     \rotatebox[origin=c]{90}{$n_{\rm{max}}$ (F475X)} &
    \rotatebox[origin=c]{90}{$\beta$ (F475X)} &
    \rotatebox[origin=c]{90}{$n_{\rm{max}}$ (F160W)} &
    \rotatebox[origin=c]{90}{$\beta$ (F160W)} & 
    \rotatebox[origin=c]{90}{z (lens)} & 
    \rotatebox[origin=c]{90}{z (source)} & 
    \rotatebox[origin=c]{90}{Caustic Area} \\ \\

    \hline \\
$\theta_{\rm{E}}$ & \bf{1.0} & -0.09 & 0.14 & -0.17 & 0.37 & -0.02 & -0.09 & \bf{0.44} & 0.36 & -0.16 & -0.22 & -0.17 & -0.12 & 0.19 & -0.14 & 0.21 & 0.15 & 0.33 & -0.22 & 0.07 & 0.16\\[+0.75em]
$\gamma$ & -0.09 & \bf{1.0} & -0.16 & 0.34 & -0.01 & 0.11 & -0.01 & 0.07 & -0.1 & 0.39 & 0.05 & -0.16 & -0.02 & -0.14 & 0.12 & -0.09 & -0.06 & 0.08 & 0.01 & -0.25 & 0.03\\[+0.75em]
$q_{\rm{mass}}$ & 0.14 & -0.16 & \bf{1.0} & -0.16 & -0.26 & \bf{-0.45} & 0.13 & 0.23 & 0.33 & -0.06 & -0.2 & -0.25 & -0.06 & -0.06 & -0.0 & -0.06 & 0.12 & -0.03 & 0.18 & 0.22 & -0.17\\[+0.75em]
$\phi_{\rm{mass}}$ & -0.17 & 0.34 & -0.16 & \bf{1.0} & -0.05 & -0.07 & -0.24 & 0.09 & -0.37 & \bf{0.9} & -0.03 & 0.24 & 0.07 & 0.11 & 0.26 & 0.18 & 0.34 & 0.33 & -0.05 & -0.3 & -0.3\\[+0.75em]
$\gamma_{\rm{ext}}$ & 0.37 & -0.01 & -0.26 & -0.05 & \bf{1.0} & -0.0 & 0.1 & -0.04 & 0.02 & -0.09 & -0.02 & 0.21 & -0.37 & -0.1 & -0.33 & -0.15 & -0.09 & 0.11 & 0.13 & 0.39 & 0.19\\[+0.75em]
$\phi_{\rm{ext}}$ & -0.02 & 0.11 & \bf{-0.45} & -0.07 & -0.0 & \bf{1.0} & 0.08 & -0.29 & 0.01 & -0.15 & 0.05 & -0.04 & 0.27 & -0.24 & 0.13 & -0.24 & -0.09 & 0.04 & 0.06 & -0.13 & 0.1\\[+0.75em]
$n_{\textrm{S\'ersic}}$ (lens) & -0.09 & -0.01 & 0.13 & -0.24 & 0.1 & 0.08 & \bf{1.0} & 0.11 & \bf{0.43} & -0.29 & 0.12 & -0.12 & -0.13 & -0.21 & -0.03 & -0.19 & -0.36 & -0.33 & 0.13 & 0.09 & 0.1\\[+0.75em]
$R_{\textrm{S\'ersic}}$ (lens) & \bf{0.44} & 0.07 & 0.23 & 0.09 & -0.04 & -0.29 & 0.11 & \bf{1.0} & 0.36 & 0.09 & -0.26 & -0.25 & 0.05 & \bf{0.62} & 0.19 & \bf{0.65} & 0.28 & 0.2 & \bf{-0.42} & -0.13 & 0.07\\[+0.75em]
$q_{\rm{light}}$ & 0.36 & -0.1 & 0.33 & -0.37 & 0.02 & 0.01 & \bf{0.43} & 0.36 & \bf{1.0} & -0.3 & -0.15 & -0.29 & -0.08 & 0.05 & -0.02 & 0.07 & -0.09 & -0.29 & -0.17 & -0.1 & 0.1\\[+0.75em]
$\phi_{\rm{light}}$ & -0.16 & 0.39 & -0.06 & \bf{0.9} & -0.09 & -0.15 & -0.29 & 0.09 & -0.3 & \bf{1.0} & -0.06 & 0.11 & -0.02 & 0.02 & 0.18 & 0.11 & 0.29 & 0.28 & 0.06 & -0.29 & -0.3\\[+0.75em]
$n_{\textrm{S\'ersic}}$ (source) & -0.22 & 0.05 & -0.2 & -0.03 & -0.02 & 0.05 & 0.12 & -0.26 & -0.15 & -0.06 & \bf{1.0} & 0.17 & 0.12 & 0.0 & 0.01 & -0.02 & -0.33 & -0.3 & 0.04 & -0.25 & 0.18\\[+0.75em]
$R_{\textrm{S\'ersic}}$ (source) & -0.17 & -0.16 & -0.25 & 0.24 & 0.21 & -0.04 & -0.12 & -0.25 & -0.29 & 0.11 & 0.17 & \bf{1.0} & -0.17 & 0.22 & -0.07 & 0.26 & 0.19 & -0.08 & 0.08 & 0.17 & -0.14\\[+0.75em]
$n_{\rm{max}}$ (F814W) & -0.12 & -0.02 & -0.06 & 0.07 & -0.37 & 0.27 & -0.13 & 0.05 & -0.08 & -0.02 & 0.12 & -0.17 & \bf{1.0} & 0.31 & \bf{0.87} & 0.21 & \bf{0.42} & 0.17 & -0.23 & -0.29 & 0.36\\[+0.75em]
$\beta$ (F814W) & 0.19 & -0.14 & -0.06 & 0.11 & -0.1 & -0.24 & -0.21 & \bf{0.62} & 0.05 & 0.02 & 0.0 & 0.22 & 0.31 & \bf{1.0} & 0.32 & \bf{0.97} & 0.4 & 0.03 & \bf{-0.47} & -0.09 & 0.13\\[+0.75em]
$n_{\rm{max}}$ (F475X) & -0.14 & 0.12 & -0.0 & 0.26 & -0.33 & 0.13 & -0.03 & 0.19 & -0.02 & 0.18 & 0.01 & -0.07 & \bf{0.87} & 0.32 & \bf{1.0} & 0.28 & \bf{0.44} & 0.04 & -0.19 & -0.37 & 0.14\\[+0.75em]
$\beta$ (F475X) & 0.21 & -0.09 & -0.06 & 0.18 & -0.15 & -0.24 & -0.19 & \bf{0.65} & 0.07 & 0.11 & -0.02 & 0.26 & 0.21 & \bf{0.97} & 0.28 & \bf{1.0} & 0.39 & 0.03 & \bf{-0.52} & -0.15 & 0.07\\[+0.75em]
$n_{\rm{max}}$ (F160W) & 0.15 & -0.06 & 0.12 & 0.34 & -0.09 & -0.09 & -0.36 & 0.28 & -0.09 & 0.29 & -0.33 & 0.19 & \bf{0.42} & 0.4 & \bf{0.44} & 0.39 & \bf{1.0} & \bf{0.64} & -0.16 & -0.03 & -0.14\\[+0.75em]
$\beta$ (F160W) & 0.33 & 0.08 & -0.03 & 0.33 & 0.11 & 0.04 & -0.33 & 0.2 & -0.29 & 0.28 & -0.3 & -0.08 & 0.17 & 0.03 & 0.04 & 0.03 & \bf{0.64} & \bf{1.0} & -0.01 & -0.08 & -0.1\\[+0.75em]
z (lens) & -0.22 & 0.01 & 0.18 & -0.05 & 0.13 & 0.06 & 0.13 & \bf{-0.42} & -0.17 & 0.06 & 0.04 & 0.08 & -0.23 & \bf{-0.47} & -0.19 & \bf{-0.52} & -0.16 & -0.01 & \bf{1.0} & 0.29 & -0.21\\[+0.75em]
z (source) & 0.07 & -0.25 & 0.22 & -0.3 & 0.39 & -0.13 & 0.09 & -0.13 & -0.1 & -0.29 & -0.25 & 0.17 & -0.29 & -0.09 & -0.37 & -0.15 & -0.03 & -0.08 & 0.29 & \bf{1.0} & 0.04\\[+0.75em]
Caustic Area & 0.16 & 0.03 & -0.17 & -0.3 & 0.19 & 0.1 & 0.1 & 0.07 & 0.1 & -0.3 & 0.18 & -0.14 & 0.36 & 0.13 & 0.14 & 0.07 & -0.14 & -0.1 & -0.21 & 0.04 & \bf{1.0}\\[+0.75em]
    \hline
    \end{tabular}
\end{table*}

\begin{figure}
 \includegraphics[width=0.5\textwidth]{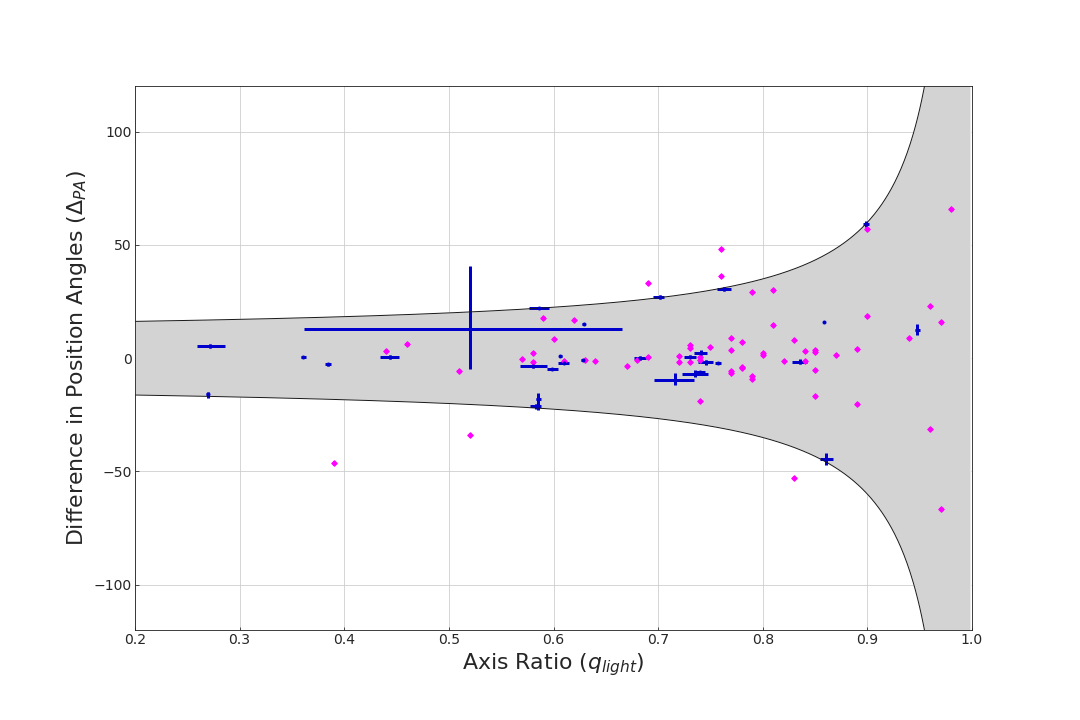}
 \caption{Difference in position angle between the main deflector's lens mass and lens light as function of the lens light's axis ratio for our lens sample (blue markers). The shaded area represents the prior set for the lens masses position angle while the magenta diamond markers represent the strong lenses in the SLACS sample.}
 \label{fig:PA_results}
\end{figure}

As further illustration, we briefly highlight some results for lens SDSS J0248+1913 and lens SDSS J1251+2935, which are shown in Figures~\ref{fig:J0248} and ~\ref{fig:J1251}, respectively. Analyzing the position angle (PA) of the lens mass distribution, we find that the convergence aligns well with orientation of the lens light profile for both systems, as can be seen in the corresponding UVIS filter F814W of the respective lens. For SDSS J0248+1913, the mass distribution's PA and the lens light distribution's PA are both 80 degrees North of East, while for SDSS J1251+2935 both PAs are 63 degrees North of East. To perform a similar analysis for all other systems, in Figure~\ref{fig:PA_results} we plot the difference between the PA of the main deflector's mass and primary lens light profile as a function of the light profile's axis ratio, with a resulting Pearson correlation coefficient of $0.9$ as shown in Table~\ref{tab:cross_cor}. The shaded area in Figure~\ref{fig:PA_results} represents the prior on the PA difference as discussed in Section~\ref{Priors}.

Even though our prior constraints allow for the convergence's axis ratio to be below the axis ratio of the light profile, we find that in both systems the lens mass distribution is more spherical compared to the respective lens light. We also find that due to a nearby galaxy, approximately $90$ degrees West of North, SDSS J0248+1913 experiences a stronger than average external shear, as reflected in the inferred value of $\gamma_{\rm{ext}} = 0.22$. Analogous evaluations can be performed for all remaining systems in our sample, using the results listed in the tables of this section. \\

\subsubsection{Systematic uncertainties on astrometry}
\label{sssec:systematic_astro}

We can estimate the systematic uncertainties on our astrometry by comparing the relative positions of the multiply imaged quasars with independent measurements based on the \textit{Gaia} satellite and with measurement based on the same {\it HST} images as analyzed in this paper, but with a different methodology \citep[see][]{Luhtaru21}.
We do not expect the measurements to agree perfectly since our positions are inferred from a forward modeling procedure taking into account the surface brightness of the quasar host galaxies and of the perturbers, while the comparison positions are measured in the image plane, without a lens model. However, we expect that this comparison should give us a robust upper limit to the systematic uncertainty on astrometry, which we expect is dominated by the uncertainty on our reconstruction of the PSF at subpixel scales \citep{Chen2021}.

\begin{figure}
 \includegraphics[width=0.5\textwidth]{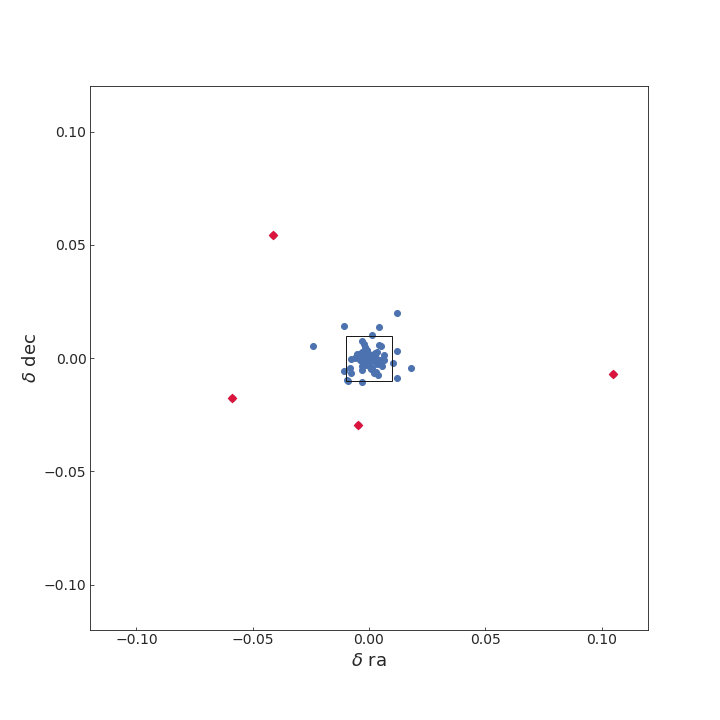}
  \includegraphics[width=0.5\textwidth]{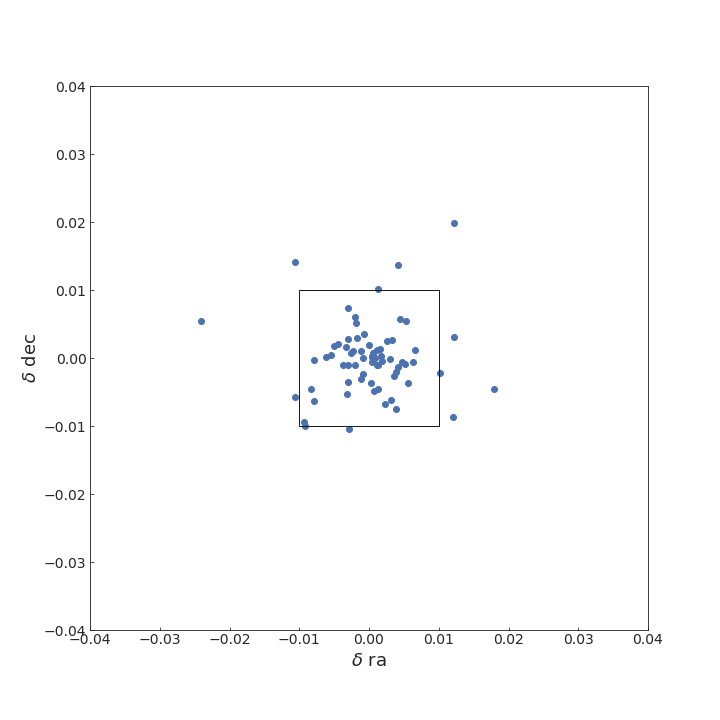}

 \caption{Comparison of the difference in quasar image positions that we inferred in this work by forward modeling {\it HST} images and the corresponding astrometry for systems with at least 3 detected images by the \textit{Gaia} satellite. The top panel shows the comparison for all the systems with the crimson diamonds representing J1721+8842 differences. The bottom panel zooms in, excluding the outlier J1721+8842}
 \label{fig:astrometry}
\end{figure}

Figure~\ref{fig:astrometry} shows the difference of the relative positions of the multiply image quasars measured in this work with respect to those measured from \textit{Gaia} data release 3. Only systems for which \textit{Gaia} measured at least three image positions are used for the comparison.  The lens J1721+8842 is a clear outlier in terms of astrometric precision. This is not surprising considering that our pipeline is not intended to deal with the complexity of the system, composed of two sets of multiple images. Excluding J1721+8842, the r.m.s. scatter is 6 mas and 5 mas respectively in RA and DEC. A comparison with the astrometry of \cite{Luhtaru21} yields very similar results, with r.m.s. scatter of 7 mas in both RA and DEC, excluding J1721+8842. Conservatively, assuming the \textit{Gaia} error to be negligible, we assign a systematic error of 6 mas on our relative astrometry listed in Table~\ref{tab:astro_pos_results}, with the exception of J1721+8842, for which a larger astrometry should be adopted until a more detailed model is developed.

The total astrometric uncertainty can be propagated into uncertainty in the estimated time delays and thus in the Hubble constant as described by \citet{Birrer19}. For the lenses analyzed in this work 6 mas will yield uncertainties on the H$_0$ well below the 5\% threshold and thus astrometric uncertainty is not a dominant contribution to the cosmographic error budget. For time delay cosmography, the dominant sources of errors are those arising from modeling choices, as discussed in the rest of the remainder of this paper, from the residual uncertainty on the PSF reconstruction \citep{Shajib22}, from the time delay measurements \citep{Millon20b}, and from the estimation of the effect of the mass along the line of sight \citep{Greene13}.

\begin{table}
  \caption{Median values for image magnification, unitless convergence, stellar convergence ($\kappa_{\star}$) estimated from the lens flux in the F160W band, and shear at the position of the quasar images. The associated uncertainties are statistical in nature and were computed using 84th and 16th percentiles.}
  \label{tab:conv_and_shear}
%   \scriptsize
  \tiny
  \begin{tabular}{lccccc}
    \hline

    Name of &  Image & $\kappa$ & $\kappa_{\star}$ & $\gamma$ & Image \\
    Lens System  &    &   &  &   & Magnification  \\
    \hline
 & A & $0.20^{+0.01}_{-0.01}$ & $0.027^{+0.001}_{-0.001}$ & $0.06^{+0.01}_{-0.01}$ & $1.56^{+0.05}_{-0.04}$\\
\\[-0.85em]
 & B & $0.84^{+0.02}_{-0.03}$ & $0.335^{+0.010}_{-0.010}$ & $0.99^{+0.03}_{-0.04}$ & $-1.05^{+0.07}_{-0.09}$\\
J0029-3814 \\[-0.85em]
 & C & $0.38^{+0.02}_{-0.01}$ & $0.067^{+0.003}_{-0.003}$ & $0.12^{+0.01}_{-0.01}$ & $2.70^{+0.14}_{-0.13}$\\
\\[-0.85em]
 & D & $0.88^{+0.02}_{-0.03}$ & $0.169^{+0.006}_{-0.007}$ & $1.04^{+0.04}_{-0.04}$ & $-0.93^{+0.06}_{-0.08}$\\
\hline
 & A & $0.45^{+0.01}_{-0.01}$ & $0.016^{+0.001}_{-0.001}$ & $0.28^{+0.01}_{-0.01}$ & $4.41^{+0.25}_{-0.20}$\\
\\[-0.85em]
 & B & $0.70^{+0.02}_{-0.02}$ & $0.411^{+0.016}_{-0.013}$ & $0.60^{+0.02}_{-0.02}$ & $-3.65^{+0.30}_{-0.35}$\\
PS J0030-1525 \\[-0.85em]
 & C & $0.55^{+0.01}_{-0.01}$ & $0.051^{+0.002}_{-0.002}$ & $0.42^{+0.01}_{-0.01}$ & $35.73^{+3.68}_{-3.05}$\\
\\[-0.85em]
 & D & $0.61^{+0.01}_{-0.01}$ & $0.114^{+0.004}_{-0.003}$ & $0.43^{+0.01}_{-0.01}$ & $-29.01^{+2.29}_{-2.56}$\\
\hline
 & A & $1.44^{+0.05}_{-0.05}$ & $0.345^{+0.012}_{-0.013}$ & $1.74^{+0.06}_{-0.08}$ & $-0.36^{+0.02}_{-0.03}$\\
\\[-0.85em]
 & B & $0.36^{+0.01}_{-0.01}$ & $0.018^{+0.001}_{-0.001}$ & $0.39^{+0.01}_{-0.01}$ & $3.95^{+0.21}_{-0.17}$\\
DES J0053-2012 \\[-0.85em]
 & C & $0.50^{+0.01}_{-0.01}$ & $0.034^{+0.001}_{-0.001}$ & $0.71^{+0.01}_{-0.02}$ & $-3.85^{+0.20}_{-0.26}$\\
\\[-0.85em]
 & D & $0.29^{+0.01}_{-0.01}$ & $0.028^{+0.001}_{-0.001}$ & $0.30^{+0.01}_{-0.01}$ & $2.43^{+0.11}_{-0.09}$\\
\hline
 & A & $1.03^{+0.02}_{-0.02}$ & $0.354^{+0.007}_{-0.009}$ & $1.32^{+0.03}_{-0.04}$ & $-0.57^{+0.03}_{-0.03}$\\
\\[-0.85em]
 & B & $0.39^{+0.01}_{-0.01}$ & $0.139^{+0.003}_{-0.003}$ & $0.51^{+0.01}_{-0.01}$ & $8.89^{+0.37}_{-0.34}$\\
PS J0147+4630 \\[-0.85em]
 & C & $0.42^{+0.01}_{-0.01}$ & $0.175^{+0.003}_{-0.004}$ & $0.63^{+0.01}_{-0.01}$ & $-15.70^{+0.63}_{-0.67}$\\
\\[-0.85em]
 & D & $0.37^{+0.01}_{-0.01}$ & $0.174^{+0.003}_{-0.004}$ & $0.54^{+0.01}_{-0.01}$ & $9.38^{+0.39}_{-0.36}$\\
\hline
 & A & $0.40^{+0.01}_{-0.01}$ & $0.137^{+0.001}_{-0.001}$ & $0.37^{+0.01}_{-0.01}$ & $4.46^{+0.26}_{-0.19}$\\
\\[-0.85em]
 & B & $0.52^{+0.01}_{-0.01}$ & $0.254^{+0.002}_{-0.002}$ & $0.66^{+0.02}_{-0.02}$ & $-4.66^{+0.23}_{-0.32}$\\
WG0214-2105 \\[-0.85em]
 & C & $0.43^{+0.01}_{-0.01}$ & $0.169^{+0.001}_{-0.001}$ & $0.43^{+0.01}_{-0.01}$ & $7.56^{+0.47}_{-0.34}$\\
\\[-0.85em]
 & D & $0.48^{+0.01}_{-0.01}$ & $0.261^{+0.002}_{-0.002}$ & $0.67^{+0.02}_{-0.02}$ & $-5.71^{+0.27}_{-0.37}$\\
\hline
 & A & $0.43^{+0.03}_{-0.03}$ & $0.016^{+0.004}_{-0.003}$ & $0.31^{+0.01}_{-0.02}$ & $4.42^{+0.65}_{-0.50}$\\
\\[-0.85em]
 & B & $0.45^{+0.04}_{-0.03}$ & $0.013^{+0.003}_{-0.003}$ & $0.66^{+0.04}_{-0.04}$ & $-7.17^{+0.82}_{-1.07}$\\
SDSS J0248+1913 \\[-0.85em]
 & C & $0.43^{+0.04}_{-0.03}$ & $0.011^{+0.003}_{-0.002}$ & $0.42^{+0.02}_{-0.02}$ & $6.67^{+1.04}_{-0.79}$\\
\\[-0.85em]
 & D & $0.60^{+0.03}_{-0.03}$ & $0.050^{+0.010}_{-0.008}$ & $0.60^{+0.04}_{-0.04}$ & $-4.83^{+0.63}_{-0.74}$\\
\hline
 & A & $0.49^{+0.01}_{-0.01}$ & $0.742^{+0.038}_{-0.041}$ & $0.73^{+0.02}_{-0.01}$ & $-3.52^{+0.15}_{-0.15}$\\
\\[-0.85em]
 & B & $0.30^{+0.01}_{-0.01}$ & $0.379^{+0.019}_{-0.020}$ & $0.43^{+0.01}_{-0.01}$ & $3.29^{+0.11}_{-0.12}$\\
WISE J0259-1635 \\[-0.85em]
 & C & $0.49^{+0.01}_{-0.01}$ & $0.749^{+0.043}_{-0.040}$ & $0.76^{+0.02}_{-0.02}$ & $-3.07^{+0.13}_{-0.13}$\\
\\[-0.85em]
 & D & $0.35^{+0.01}_{-0.01}$ & $0.839^{+0.045}_{-0.045}$ & $0.49^{+0.01}_{-0.01}$ & $5.25^{+0.20}_{-0.20}$\\
\hline
 & A & $1.25^{+0.01}_{-0.01}$ & $0.323^{+0.007}_{-0.008}$ & $1.37^{+0.01}_{-0.01}$ & $-0.55^{+0.01}_{-0.01}$\\
\\[-0.85em]
 & B & $0.39^{+0.01}_{-0.01}$ & $0.020^{+0.001}_{-0.001}$ & $0.30^{+0.01}_{-0.01}$ & $3.54^{+0.03}_{-0.03}$\\
J0343-2828 \\[-0.85em]
 & C & $0.57^{+0.01}_{-0.01}$ & $0.065^{+0.002}_{-0.002}$ & $0.71^{+0.01}_{-0.01}$ & $-3.06^{+0.04}_{-0.04}$\\
\\[-0.85em]
 & D & $0.36^{+0.01}_{-0.01}$ & $0.017^{+0.001}_{-0.001}$ & $0.24^{+0.01}_{-0.01}$ & $2.80^{+0.02}_{-0.02}$\\
\hline
 & A & $0.35^{+0.01}_{-0.01}$ & $0.114^{+0.003}_{-0.003}$ & $0.50^{+0.01}_{-0.01}$ & $5.84^{+0.31}_{-0.26}$\\
\\[-0.85em]
 & B & $0.55^{+0.01}_{-0.01}$ & $0.238^{+0.006}_{-0.006}$ & $0.68^{+0.02}_{-0.02}$ & $-3.82^{+0.22}_{-0.24}$\\
DES J0405-3308 \\[-0.85em]
 & C & $0.34^{+0.01}_{-0.01}$ & $0.127^{+0.004}_{-0.003}$ & $0.53^{+0.01}_{-0.01}$ & $6.74^{+0.34}_{-0.29}$\\
\\[-0.85em]
 & D & $0.51^{+0.01}_{-0.01}$ & $0.216^{+0.005}_{-0.005}$ & $0.64^{+0.02}_{-0.02}$ & $-5.78^{+0.34}_{-0.36}$\\
\hline
 & A & $0.40^{+0.01}_{-0.02}$ & $0.057^{+0.003}_{-0.003}$ & $0.43^{+0.01}_{-0.01}$ & $5.72^{+0.29}_{-0.27}$\\
\\[-0.85em]
 & B & $0.54^{+0.01}_{-0.01}$ & $0.122^{+0.006}_{-0.006}$ & $0.59^{+0.02}_{-0.02}$ & $-7.34^{+0.37}_{-0.39}$\\
DES J0420-4037 \\[-0.85em]
 & C & $0.46^{+0.01}_{-0.02}$ & $0.083^{+0.004}_{-0.004}$ & $0.47^{+0.01}_{-0.01}$ & $15.13^{+0.80}_{-0.78}$\\
\\[-0.85em]
 & D & $0.55^{+0.01}_{-0.01}$ & $0.136^{+0.006}_{-0.006}$ & $0.54^{+0.02}_{-0.01}$ & $-10.90^{+0.61}_{-0.64}$\\
\hline
 & A & $0.40^{+0.03}_{-0.05}$ & $2.362^{+0.869}_{-0.647}$ & $0.55^{+0.04}_{-0.02}$ & $12.50^{+19.34}_{-8.40}$\\
\\[-0.85em]
 & B & $0.43^{+0.02}_{-0.03}$ & $0.311^{+0.135}_{-0.099}$ & $0.58^{+0.03}_{-0.03}$ & $-51.51^{+75.11}_{-73.07}$\\
DES J0530-3730 \\[-0.85em]
 & C & $0.45^{+0.02}_{-0.04}$ & $0.273^{+0.114}_{-0.089}$ & $0.54^{+0.04}_{-0.02}$ & $92.64^{+189.15}_{-184.91}$\\
\\[-0.85em]
 & D & $0.49^{+0.03}_{-0.06}$ & $0.342^{+0.171}_{-0.134}$ & $0.53^{+0.04}_{-0.03}$ & $-66.41^{+149.10}_{-147.22}$\\
\hline
 & A & $0.62^{+0.02}_{-0.02}$ & $0.295^{+0.067}_{-0.049}$ & $0.88^{+0.03}_{-0.02}$ & $-1.59^{+0.11}_{-0.10}$\\
\\[-0.85em]
 & B & $0.33^{+0.01}_{-0.01}$ & $0.038^{+0.009}_{-0.007}$ & $0.64^{+0.01}_{-0.01}$ & $21.69^{+0.59}_{-0.41}$\\
PS J0630-1201 \\[-0.85em]
 & C & $0.41^{+0.01}_{-0.01}$ & $0.112^{+0.026}_{-0.019}$ & $0.60^{+0.01}_{-0.01}$ & $-63.78^{+2.04}_{-1.78}$\\
\\[-0.85em]
 & D & $0.48^{+0.01}_{-0.01}$ & $0.123^{+0.027}_{-0.019}$ & $0.49^{+0.01}_{-0.01}$ & $34.60^{+2.19}_{-1.99}$\\
\hline
 & A & $0.67^{+0.01}_{-0.01}$ & $0.080^{+0.003}_{-0.003}$ & $0.53^{+0.02}_{-0.02}$ & $-5.86^{+0.38}_{-0.44}$\\
\\[-0.85em]
 & B & $0.40^{+0.01}_{-0.01}$ & $0.029^{+0.001}_{-0.001}$ & $0.25^{+0.01}_{-0.01}$ & $3.40^{+0.18}_{-0.17}$\\
J0659+1629 \\[-0.85em]
 & C & $0.72^{+0.01}_{-0.01}$ & $0.068^{+0.003}_{-0.003}$ & $0.56^{+0.02}_{-0.02}$ & $-4.31^{+0.29}_{-0.34}$\\
\\[-0.85em]
 & D & $0.58^{+0.01}_{-0.01}$ & $0.065^{+0.003}_{-0.002}$ & $0.31^{+0.01}_{-0.01}$ & $12.24^{+0.83}_{-0.73}$\\
\hline
 & A & $0.40^{+0.01}_{-0.01}$ & $0.142^{+0.001}_{-0.001}$ & $0.18^{+0.01}_{-0.01}$ & $3.02^{+0.05}_{-0.03}$\\
\\[-0.85em]
 & B & $0.68^{+0.01}_{-0.02}$ & $0.396^{+0.003}_{-0.003}$ & $1.02^{+0.01}_{-0.01}$ & $-1.07^{+0.02}_{-0.03}$\\
J0818-2613 \\[-0.85em]
 & C & $0.35^{+0.01}_{-0.01}$ & $0.247^{+0.002}_{-0.004}$ & $0.60^{+0.01}_{-0.01}$ & $16.19^{+0.16}_{-0.11}$\\
\\[-0.85em]
 & D & $0.36^{+0.01}_{-0.01}$ & $0.234^{+0.002}_{-0.003}$ & $0.69^{+0.01}_{-0.01}$ & $-16.13^{+0.11}_{-0.17}$\\
\hline
 & A & $0.30^{+0.01}_{-0.01}$ & $0.133^{+0.003}_{-0.003}$ & $0.48^{+0.01}_{-0.01}$ & $3.93^{+0.16}_{-0.18}$\\
\\[-0.85em]
 & B & $0.40^{+0.01}_{-0.01}$ & $0.246^{+0.005}_{-0.004}$ & $0.65^{+0.01}_{-0.01}$ & $-15.26^{+0.66}_{-0.60}$\\
W2M J1042+1641 \\[-0.85em]
 & C & $0.38^{+0.01}_{-0.01}$ & $0.226^{+0.005}_{-0.004}$ & $0.59^{+0.01}_{-0.01}$ & $22.60^{+1.01}_{-1.04}$\\
\\[-0.85em]
 & D & $0.49^{+0.01}_{-0.01}$ & $0.284^{+0.005}_{-0.005}$ & $0.61^{+0.01}_{-0.01}$ & $-9.13^{+0.56}_{-0.50}$\\
    \hline
  \end{tabular}
  \columnbreak
\end{table}

\begin{table}
  \caption{Median values for image magnification, unitless convergence, stellar convergence ($\kappa_{\star}$) estimated from the lens flux in the F160W band, and shear at the position of the quasar images. The associated uncertainties are statistical in nature and were computed using 84th and 16th percentiles.}
  \label{tab:conv_and_shear_2}
%   \scriptsize
  \tiny
  \begin{tabular}{lccccc}
    \hline
    Name of &  Image & $\kappa$ & $\kappa_{st}$ & $\gamma$ & Image \\
    Lens System  &    &   &  &   & Magnification  \\
    \hline
 & A & $0.71^{+0.01}_{-0.01}$ & $0.087^{+0.005}_{-0.005}$ & $0.70^{+0.02}_{-0.02}$ & $-2.49^{+0.13}_{-0.12}$\\
\\[-0.85em]
 & B & $0.35^{+0.01}_{-0.01}$ & $0.021^{+0.002}_{-0.002}$ & $0.43^{+0.01}_{-0.01}$ & $4.08^{+0.15}_{-0.17}$\\
J1131-4419 \\[-0.85em]
 & C & $0.58^{+0.01}_{-0.01}$ & $0.149^{+0.007}_{-0.007}$ & $0.55^{+0.01}_{-0.01}$ & $-7.92^{+0.48}_{-0.42}$\\
\\[-0.85em]
 & D & $0.44^{+0.01}_{-0.01}$ & $0.091^{+0.005}_{-0.005}$ & $0.46^{+0.01}_{-0.01}$ & $9.82^{+0.42}_{-0.47}$\\
\hline
 & A & $0.22^{+0.01}_{-0.01}$ & $0.036^{+0.001}_{-0.002}$ & $0.14^{+0.01}_{-0.01}$ & $1.72^{+0.03}_{-0.03}$\\
\\[-0.85em]
 & B & $0.62^{+0.01}_{-0.02}$ & $0.178^{+0.004}_{-0.005}$ & $1.12^{+0.02}_{-0.02}$ & $-0.90^{+0.03}_{-0.03}$\\
2M1134-2103 \\[-0.85em]
 & C & $0.21^{+0.01}_{-0.01}$ & $0.030^{+0.001}_{-0.001}$ & $0.08^{+0.01}_{-0.01}$ & $1.63^{+0.03}_{-0.03}$\\
\\[-0.85em]
 & D & $0.96^{+0.02}_{-0.03}$ & $0.335^{+0.008}_{-0.014}$ & $1.54^{+0.03}_{-0.03}$ & $-0.42^{+0.02}_{-0.01}$\\
\hline
 & A & $0.30^{+0.01}_{-0.01}$ & $0.077^{+0.003}_{-0.003}$ & $0.28^{+0.01}_{-0.01}$ & $2.44^{+0.04}_{-0.04}$\\
\\[-0.85em]
 & B & $0.54^{+0.01}_{-0.01}$ & $0.380^{+0.005}_{-0.004}$ & $0.63^{+0.01}_{-0.01}$ & $-5.49^{+0.11}_{-0.10}$\\
SDSS J1251+2935 \\[-0.85em]
 & C & $0.45^{+0.01}_{-0.01}$ & $0.225^{+0.005}_{-0.005}$ & $0.47^{+0.01}_{-0.01}$ & $11.63^{+0.21}_{-0.21}$\\
\\[-0.85em]
 & D & $0.50^{+0.01}_{-0.01}$ & $0.278^{+0.008}_{-0.007}$ & $0.63^{+0.01}_{-0.01}$ & $-6.63^{+0.13}_{-0.13}$\\
\hline
 & A & $0.64^{+0.01}_{-0.01}$ & $0.089^{+0.001}_{-0.001}$ & $0.64^{+0.01}_{-0.01}$ & $-3.64^{+0.13}_{-0.05}$\\
\\[-0.85em]
 & B & $0.46^{+0.01}_{-0.01}$ & $0.034^{+0.001}_{-0.001}$ & $0.46^{+0.01}_{-0.01}$ & $12.09^{+0.07}_{-0.19}$\\
2M1310-1714 \\[-0.85em]
 & C & $0.54^{+0.01}_{-0.01}$ & $0.048^{+0.001}_{-0.001}$ & $0.54^{+0.01}_{-0.01}$ & $-13.28^{+0.34}_{-0.13}$\\
\\[-0.85em]
 & D & $0.46^{+0.01}_{-0.01}$ & $0.028^{+0.001}_{-0.001}$ & $0.44^{+0.01}_{-0.01}$ & $9.60^{+0.07}_{-0.15}$\\
\hline
 & A & $1.00^{+0.04}_{-0.04}$ & $0.404^{+0.009}_{-0.005}$ & $1.00^{+0.05}_{-0.05}$ & $-1.01^{+0.10}_{-0.11}$\\
\\[-0.85em]
 & B & $0.43^{+0.02}_{-0.02}$ & $0.095^{+0.002}_{-0.002}$ & $0.40^{+0.01}_{-0.01}$ & $5.93^{+0.49}_{-0.50}$\\
SDSS J1330+1810 \\[-0.85em]
 & C & $0.66^{+0.01}_{-0.01}$ & $0.230^{+0.005}_{-0.004}$ & $0.58^{+0.02}_{-0.02}$ & $-4.62^{+0.50}_{-0.53}$\\
\\[-0.85em]
 & D & $0.27^{+0.02}_{-0.02}$ & $0.038^{+0.001}_{-0.001}$ & $0.45^{+0.01}_{-0.01}$ & $2.99^{+0.19}_{-0.18}$\\
\hline
 & A & $0.74^{+0.01}_{-0.01}$ & $0.191^{+0.009}_{-0.008}$ & $0.75^{+0.02}_{-0.02}$ & $-2.04^{+0.12}_{-0.13}$\\
\\[-0.85em]
 & B & $0.49^{+0.01}_{-0.01}$ & $0.031^{+0.002}_{-0.001}$ & $0.32^{+0.01}_{-0.01}$ & $6.45^{+0.35}_{-0.31}$\\
SDSS J1433+6007 \\[-0.85em]
 & C & $0.75^{+0.01}_{-0.01}$ & $0.101^{+0.005}_{-0.005}$ & $0.65^{+0.02}_{-0.02}$ & $-2.81^{+0.14}_{-0.16}$\\
\\[-0.85em]
 & D & $0.44^{+0.01}_{-0.01}$ & $0.027^{+0.001}_{-0.001}$ & $0.21^{+0.01}_{-0.01}$ & $3.77^{+0.17}_{-0.16}$\\
\hline
 & A & $0.39^{+0.01}_{-0.01}$ & $0.094^{+0.001}_{-0.001}$ & $0.29^{+0.01}_{-0.01}$ & $3.51^{+0.11}_{-0.10}$\\
\\[-0.85em]
 & B & $0.62^{+0.01}_{-0.01}$ & $0.252^{+0.003}_{-0.003}$ & $0.77^{+0.01}_{-0.01}$ & $-2.24^{+0.08}_{-0.08}$\\
J1537-3010 \\[-0.85em]
 & C & $0.38^{+0.01}_{-0.01}$ & $0.086^{+0.001}_{-0.001}$ & $0.27^{+0.01}_{-0.01}$ & $3.19^{+0.10}_{-0.09}$\\
\\[-0.85em]
 & D & $0.61^{+0.01}_{-0.01}$ & $0.262^{+0.003}_{-0.003}$ & $0.75^{+0.01}_{-0.01}$ & $-2.39^{+0.08}_{-0.09}$\\
\hline
 & A & $0.33^{+0.01}_{-0.01}$ & $0.073^{+0.005}_{-0.005}$ & $0.19^{+0.01}_{-0.01}$ & $2.43^{+0.06}_{-0.06}$\\
\\[-0.85em]
 & B & $0.94^{+0.01}_{-0.01}$ & $0.478^{+0.018}_{-0.021}$ & $0.87^{+0.02}_{-0.02}$ & $-1.34^{+0.05}_{-0.06}$\\
PS J1606-2333 \\[-0.85em]
 & C & $0.39^{+0.01}_{-0.01}$ & $0.087^{+0.006}_{-0.006}$ & $0.21^{+0.01}_{-0.01}$ & $3.08^{+0.09}_{-0.08}$\\
\\[-0.85em]
 & D & $0.81^{+0.01}_{-0.01}$ & $0.352^{+0.014}_{-0.014}$ & $0.81^{+0.01}_{-0.02}$ & $-1.60^{+0.06}_{-0.07}$\\
\hline
 & A & $0.78^{+0.01}_{-0.01}$ & $0.475^{+0.001}_{-0.001}$ & $0.81^{+0.01}_{-0.01}$ & $-1.65^{+0.02}_{-0.02}$\\
\\[-0.85em]
 & B & $0.44^{+0.01}_{-0.01}$ & $0.207^{+0.001}_{-0.001}$ & $0.37^{+0.01}_{-0.01}$ & $5.54^{+0.07}_{-0.07}$\\
J1721+8842 \\[-0.85em]
 & C & $0.55^{+0.01}_{-0.01}$ & $0.292^{+0.001}_{-0.001}$ & $0.60^{+0.01}_{-0.01}$ & $-6.34^{+0.09}_{-0.09}$\\
\\[-0.85em]
 & D & $0.42^{+0.01}_{-0.01}$ & $0.217^{+0.001}_{-0.001}$ & $0.36^{+0.01}_{-0.01}$ & $4.75^{+0.06}_{-0.05}$\\
\hline
 & A & $0.43^{+0.01}_{-0.01}$ & $0.057^{+0.002}_{-0.001}$ & $0.46^{+0.01}_{-0.01}$ & $9.02^{+0.53}_{-0.48}$\\
\\[-0.85em]
 & B & $0.53^{+0.01}_{-0.01}$ & $0.497^{+0.011}_{-0.012}$ & $0.54^{+0.01}_{-0.01}$ & $-14.09^{+0.88}_{-0.93}$\\
J1817+2729 \\[-0.85em]
 & C & $0.47^{+0.01}_{-0.01}$ & $0.090^{+0.003}_{-0.003}$ & $0.47^{+0.01}_{-0.01}$ & $15.67^{+0.97}_{-0.88}$\\
\\[-0.85em]
 & D & $0.52^{+0.01}_{-0.01}$ & $0.504^{+0.009}_{-0.006}$ & $0.58^{+0.01}_{-0.01}$ & $-8.88^{+0.54}_{-0.58}$\\
\hline
 & A & $0.58^{+0.01}_{-0.01}$ & $0.458^{+0.001}_{-0.001}$ & $1.06^{+0.01}_{-0.01}$ & $-1.05^{+0.02}_{-0.02}$\\
\\[-0.85em]
 & B & $0.24^{+0.01}_{-0.01}$ & $0.178^{+0.001}_{-0.001}$ & $0.48^{+0.01}_{-0.01}$ & $2.88^{+0.04}_{-0.04}$\\
DES J2038-4008 \\[-0.85em]
 & C & $0.46^{+0.01}_{-0.01}$ & $0.330^{+0.001}_{-0.001}$ & $0.86^{+0.01}_{-0.01}$ & $-2.28^{+0.04}_{-0.04}$\\
\\[-0.85em]
 & D & $0.22^{+0.01}_{-0.01}$ & $0.168^{+0.001}_{-0.001}$ & $0.43^{+0.01}_{-0.01}$ & $2.38^{+0.03}_{-0.03}$\\
\hline
 & A & $0.76^{+0.03}_{-0.02}$ & $0.292^{+0.008}_{-0.006}$ & $1.00^{+0.02}_{-0.03}$ & $-1.07^{+0.04}_{-0.05}$\\
\\[-0.85em]
 & B & $0.24^{+0.01}_{-0.01}$ & $0.100^{+0.003}_{-0.002}$ & $0.40^{+0.01}_{-0.01}$ & $2.38^{+0.10}_{-0.06}$\\
WG2100-4452 \\[-0.85em]
 & C & $0.49^{+0.02}_{-0.01}$ & $0.181^{+0.005}_{-0.004}$ & $0.66^{+0.01}_{-0.02}$ & $-5.93^{+0.20}_{-0.28}$\\
\\[-0.85em]
 & D & $0.35^{+0.02}_{-0.01}$ & $0.132^{+0.004}_{-0.003}$ & $0.53^{+0.01}_{-0.02}$ & $7.25^{+0.31}_{-0.23}$\\
\hline
 & A & $0.34^{+0.01}_{-0.01}$ & $0.045^{+0.004}_{-0.004}$ & $0.29^{+0.01}_{-0.01}$ & $2.84^{+0.16}_{-0.13}$\\
\\[-0.85em]
 & B & $0.87^{+0.04}_{-0.04}$ & $0.280^{+0.022}_{-0.024}$ & $1.00^{+0.04}_{-0.04}$ & $-1.01^{+0.08}_{-0.09}$\\
J2145+6345 \\[-0.85em]
 & C & $0.40^{+0.01}_{-0.01}$ & $0.082^{+0.007}_{-0.007}$ & $0.43^{+0.01}_{-0.01}$ & $5.80^{+0.36}_{-0.30}$\\
\\[-0.85em]
 & D & $0.52^{+0.02}_{-0.02}$ & $0.135^{+0.011}_{-0.011}$ & $0.64^{+0.02}_{-0.02}$ & $-5.43^{+0.35}_{-0.43}$\\
\hline
 & A & $0.31^{+0.01}_{-0.01}$ & $0.052^{+0.002}_{-0.002}$ & $0.33^{+0.01}_{-0.01}$ & $2.77^{+0.12}_{-0.10}$\\
\\[-0.85em]
 & B & $0.59^{+0.01}_{-0.01}$ & $0.125^{+0.004}_{-0.004}$ & $0.65^{+0.02}_{-0.02}$ & $-3.82^{+0.19}_{-0.23}$\\
J2205-3727 \\[-0.85em]
 & C & $0.42^{+0.01}_{-0.01}$ & $0.082^{+0.003}_{-0.003}$ & $0.46^{+0.01}_{-0.01}$ & $8.13^{+0.41}_{-0.34}$\\
\\[-0.85em]
 & D & $0.58^{+0.01}_{-0.01}$ & $0.140^{+0.004}_{-0.004}$ & $0.63^{+0.02}_{-0.02}$ & $-4.53^{+0.21}_{-0.27}$\\
\hline
 & A & $0.61^{+0.01}_{-0.01}$ & $0.328^{+0.006}_{-0.005}$ & $0.65^{+0.02}_{-0.02}$ & $-3.80^{+0.19}_{-0.22}$\\
\\[-0.85em]
 & B & $0.40^{+0.01}_{-0.01}$ & $0.235^{+0.003}_{-0.004}$ & $0.39^{+0.01}_{-0.01}$ & $4.80^{+0.22}_{-0.19}$\\
ATLAS J2344-3056 \\[-0.85em]
 & C & $0.62^{+0.01}_{-0.01}$ & $0.459^{+0.007}_{-0.007}$ & $0.67^{+0.02}_{-0.02}$ & $-3.31^{+0.17}_{-0.19}$\\
\\[-0.85em]
 & D & $0.40^{+0.01}_{-0.01}$ & $0.247^{+0.004}_{-0.003}$ & $0.40^{+0.01}_{-0.01}$ & $4.79^{+0.23}_{-0.20}$\\
    \hline
  \end{tabular}
\end{table}

\subsubsection{Comparison to published mass models}

For several systems, mass models based on ground-based imaging exist in the literature \citep[e.g.][]{rusu2018, Lemon18, Lemon19, lemon2020, lemon2022}. Given the difference in data resolution and depth, modeling approaches, treatment of perturbers, and parameterization it is difficult to perform a detailed quantitative comparison. Overall, quantities such as Einstein radius, axis ratios, and position angle, are in agreement within the uncertainties. The external shear depends crucially on the choice of mass components and precision of the main galaxy position, which is often uncertain in ground based data. A more detailed comparison will have to be based on the same data, a common parameterization, and choice of mass model components.

Comparing our results for system DES J2038-4008 to those obtained via the cosmography-grade lens model of \citet{Shajib22}, we find excellent agreement for the power-law slope, Einstein radius, axis ratio and PA of the mass profile, shear strength and shear PA. We further find that our predicted time delays match very well the predictions by \citet{Shajib22}, with the largest difference of 0.6 days resulting from the greatest time delay prediction of 25.7 days, between images A and D, corresponding to a $2.3\%$ difference.

For J1721+8842, we compare our results with those by \citet{lemon2022} and find good agreement for the power-law slope (\citet{lemon2022} used a singular isothermal ellipsoid or SIE, which is PEMD with a fixed slope of $2.0$), the Einstein radius, axis ratio and PA of the mass profile, as well as for the shear strength. For the shear direction, we find a discrepancy of nearly 80 degrees, however, in our model we mask out the second image pair, which \citet{lemon2022} use as additional constraint. 
A comparison of magnification values, shear, and convergence at the image positions with the best-fit model of \citet{lemon2022} shows agreement within a few percent, which is remarkable given the complexity of the system and the assumption of an SIE in \citet{lemon2022} versus the power law used in our model.
We further compare our predicted time delays and find excellent agreement, with a largest difference of $0.6$ days and the highest predicted time delay in \citet{lemon2022} showing no difference to our result. Additionally, we compare our results for J1721+8842 with those by \citet{Mangat21} and, again, find reasonable agreement for the power-law slope (\citet{Mangat21} use an SIE), the Einstein radius, axis ratio and PA of the mass profile, and shear. After rescaling to the cosmology assumed by \citet{Mangat21}, we further find that our predicted time delays and image magnifications agree within a few
percent.

\subsection{Predicted time delays}

\begin{table*}
  \caption{Median values for Fermat potential differences between quasar images and associated predicted time delays using listed measured or assumed redshifts. The associated uncertainties are statistical in nature and were computed using 84th and 16th percentiles. Our calculations assume a flat $\Lambda$CDM cosmology with $\Omega_{\rm{m,0}} = 0.3$, $\Omega_{\rm{\Lambda,0}} = 0.7$, and $H_0 = 70$ km s$^{-1}$ Mpc$^{-1}$. For unmeasured deflector redshifts, we adopt a fiducial $z_{\rm{d}} = 0.5$.}
%   For unmeasured redshifts, we adopt a fiducial $z_{\rm{d}} = 0.5$ and of $z_{\rm{s}} = 2.0$, for the deflector and the source, respectively.}
  \label{tab:time_delays}
  \begin{tabular}{lcccccccc}
    \hline
    Name of &  z$_{\rm{d}}$ & z$_{\rm{s}}$ & $\Delta \Phi_{\rm{AB}}$ & $\Delta \Phi_{\rm{AC}}$ & $\Delta \Phi_{\rm{AD}}$ & $\Delta t_{\rm{AB}}$ & $\Delta t_{\rm{AC}}$ & $\Delta t_{\rm{AD}}$  \\
    Lens System  &    &   &  & & &  (days)  & (days) & (days)   \\
    \hline
    \\[-0.75em]
    J0029-3814& $0.863$ & $2.821$ & $-0.6295^{+0.0295}_{-0.0188}$ & $-0.5754^{+0.0274}_{-0.0172}$ & $-0.6436^{+0.0297}_{-0.0190}$ & $-100.0^{+4.7}_{-3.0}$ & $-91.4^{+4.4}_{-2.7}$ & $-102.2^{+4.7}_{-3.0}$\\[+0.75em]
PS J0030-1525\textsuperscript{$\dagger$} & $0.5$ & $3.36$ & $-0.2685^{+0.0142}_{-0.0136}$ & $-0.1151^{+0.0048}_{-0.0045}$ & $-0.1158^{+0.0048}_{-0.0046}$ & $-19.8^{+1.0}_{-1.0}$ & $-8.5^{+0.4}_{-0.3}$ & $-8.5^{+0.4}_{-0.3}$\\[+0.75em]
DES J0053-2012\textsuperscript{$\dagger$} & $0.5$ & $3.8$ & $1.3895^{+0.0263}_{-0.0391}$ & $1.3460^{+0.0249}_{-0.0369}$ & $1.6476^{+0.0323}_{-0.0501}$ & $100.5^{+1.9}_{-2.8}$ & $97.3^{+1.8}_{-2.7}$ & $119.2^{+2.3}_{-3.6}$\\[+0.75em]
PS J0147+4630& $0.678$ & $2.377$ & $2.5538^{+0.0434}_{-0.0479}$ & $2.5329^{+0.0430}_{-0.0474}$ & $2.5520^{+0.0434}_{-0.0478}$ & $306.1^{+5.2}_{-5.7}$ & $303.6^{+5.1}_{-5.7}$ & $305.9^{+5.2}_{-5.7}$\\[+0.75em]
WG0214-2105& $0.22$ & $3.229$ & $-0.1206^{+0.0040}_{-0.0039}$ & $-0.0800^{+0.0027}_{-0.0026}$ & $-0.1046^{+0.0035}_{-0.0034}$ & $-3.5^{+0.1}_{-0.1}$ & $-2.3^{+0.1}_{-0.1}$ & $-3.0^{+0.1}_{-0.1}$\\[+0.75em]
SDSS J0248+1913\textsuperscript{$\dagger$} & $0.5$ & $2.44$ & $-0.0737^{+0.0051}_{-0.0051}$ & $-0.0557^{+0.0039}_{-0.0040}$ & $-0.1047^{+0.0078}_{-0.0080}$ & $-5.8^{+0.4}_{-0.4}$ & $-4.4^{+0.3}_{-0.3}$ & $-8.2^{+0.6}_{-0.6}$\\[+0.75em]
WISE J0259-1635& $0.905$ & $2.16$ & $0.1022^{+0.0023}_{-0.0023}$ & $-0.0128^{+0.0003}_{-0.0003}$ & $0.0279^{+0.0006}_{-0.0006}$ & $20.3^{+0.5}_{-0.4}$ & $-2.5^{+0.1}_{-0.1}$ & $5.5^{+0.1}_{-0.1}$\\[+0.75em]
J0343-2828& $0.385$ & $1.655$ & $1.8668^{+0.0071}_{-0.0081}$ & $1.5985^{+0.0062}_{-0.0064}$ & $2.2440^{+0.0092}_{-0.0106}$ & $115.8^{+0.4}_{-0.5}$ & $99.1^{+0.4}_{-0.4}$ & $139.2^{+0.6}_{-0.7}$\\[+0.75em]
DES J0405-3308\textsuperscript{$\dagger$} & $0.5$ & $1.713$ & $-0.0587^{+0.0020}_{-0.0022}$ & $-0.0074^{+0.0003}_{-0.0004}$ & $-0.0308^{+0.0012}_{-0.0013}$ & $-5.2^{+0.2}_{-0.2}$ & $-0.7^{+0.1}_{-0.1}$ & $-2.7^{+0.1}_{-0.1}$\\[+0.75em]
DES J0420-4037& $0.358$ & $2.4$ & $-0.1101^{+0.0030}_{-0.0032}$ & $-0.0870^{+0.0024}_{-0.0027}$ & $-0.0930^{+0.0026}_{-0.0029}$ & $-5.7^{+0.2}_{-0.2}$ & $-4.5^{+0.1}_{-0.1}$ & $-4.8^{+0.1}_{-0.1}$\\[+0.75em]
DES J0530-3730\textsuperscript{$\dagger$} & $0.5$ & $2.838$ & $-0.0159^{+0.0138}_{-0.0238}$ & $-0.0152^{+0.0129}_{-0.0234}$ & $-0.0153^{+0.0129}_{-0.0234}$ & $-1.2^{+1.1}_{-1.8}$ & $-1.2^{+1.0}_{-1.8}$ & $-1.2^{+1.0}_{-1.8}$\\[+0.75em]
PS J0630-1201\textsuperscript{$\dagger$} & $0.5$ & $3.34$ & $0.9574^{+0.0334}_{-0.0324}$ & $0.9520^{+0.0335}_{-0.0324}$ & $0.9539^{+0.0335}_{-0.0325}$ & $70.7^{+2.5}_{-2.4}$ & $70.3^{+2.5}_{-2.4}$ & $70.4^{+2.5}_{-2.4}$\\[+0.75em]
J0659+1629& $0.766$ & $3.083$ & $2.3321^{+0.0751}_{-0.0674}$ & $0.0404^{+0.0015}_{-0.0014}$ & $0.1064^{+0.0047}_{-0.0043}$ & $302.6^{+9.7}_{-8.7}$ & $5.2^{+0.2}_{-0.2}$ & $13.8^{+0.6}_{-0.6}$\\[+0.75em]
J0818-2613\textsuperscript{$\dagger$} & $0.5$ & $2.164$ & $-4.7682^{+0.0857}_{-0.0506}$ & $-1.5446^{+0.0178}_{-0.0122}$ & $-1.5484^{+0.0178}_{-0.0122}$ & $-387.5^{+7.0}_{-4.1}$ & $-125.5^{+1.4}_{-1.0}$ & $-125.8^{+1.4}_{-1.0}$\\[+0.75em]
W2M J1042+1641& $0.599$ & $2.5$ & $-0.1777^{+0.0066}_{-0.0078}$ & $-0.1748^{+0.0065}_{-0.0077}$ & $-0.1807^{+0.0066}_{-0.0079}$ & $-17.6^{+0.7}_{-0.8}$ & $-17.3^{+0.6}_{-0.8}$ & $-17.9^{+0.7}_{-0.8}$\\[+0.75em]
J1131-4419\textsuperscript{$\dagger$} & $0.5$ & $1.09$ & $0.2117^{+0.0060}_{-0.0053}$ & $0.1023^{+0.0021}_{-0.0023}$ & $0.1072^{+0.0023}_{-0.0024}$ & $24.2^{+0.7}_{-0.6}$ & $11.7^{+0.2}_{-0.3}$ & $12.2^{+0.3}_{-0.3}$\\[+0.75em]
2M1134-2103\textsuperscript{$\dagger$} & $0.5$ & $2.77$ & $-0.4707^{+0.0077}_{-0.0085}$ & $0.1825^{+0.0024}_{-0.0032}$ & $-0.9526^{+0.0122}_{-0.0117}$ & $-36.0^{+0.6}_{-0.7}$ & $14.0^{+0.2}_{-0.2}$ & $-72.8^{+0.9}_{-0.9}$\\[+0.75em]
SDSS J1251+2935& $0.41$ & $0.802$ & $-0.3271^{+0.0031}_{-0.0031}$ & $-0.3188^{+0.0030}_{-0.0030}$ & $-0.3242^{+0.0031}_{-0.0030}$ & $-33.6^{+0.3}_{-0.3}$ & $-32.8^{+0.3}_{-0.3}$ & $-33.3^{+0.3}_{-0.3}$\\[+0.75em]
2M1310-1714& $0.293$ & $1.975$ & $1.0199^{+0.0214}_{-0.0085}$ & $0.8912^{+0.0184}_{-0.0076}$ & $1.2109^{+0.0259}_{-0.0103}$ & $43.2^{+0.9}_{-0.4}$ & $37.7^{+0.8}_{-0.3}$ & $51.3^{+1.1}_{-0.4}$\\[+0.75em]
SDSS J1330+1810& $0.373$ & $1.393$ & $0.2867^{+0.0126}_{-0.0127}$ & $0.2808^{+0.0121}_{-0.0122}$ & $0.4553^{+0.0251}_{-0.0242}$ & $18.0^{+0.8}_{-0.8}$ & $17.6^{+0.8}_{-0.8}$ & $28.6^{+1.6}_{-1.5}$\\[+0.75em]
SDSS J1433+6007& $0.407$ & $2.737$ & $0.6681^{+0.0202}_{-0.0184}$ & $0.5380^{+0.0186}_{-0.0153}$ & $1.0323^{+0.0257}_{-0.0265}$ & $39.8^{+1.2}_{-1.1}$ & $32.0^{+1.1}_{-0.9}$ & $61.4^{+1.5}_{-1.6}$\\[+0.75em]
J1537-3010& $0.592$ & $1.721$ & $-0.3265^{+0.0049}_{-0.0050}$ & $0.0975^{+0.0016}_{-0.0017}$ & $-0.2925^{+0.0046}_{-0.0042}$ & $-36.9^{+0.6}_{-0.6}$ & $11.0^{+0.2}_{-0.2}$ & $-33.0^{+0.5}_{-0.5}$\\[+0.75em]
PS J1606-2333\textsuperscript{$\dagger$} & $0.5$ & $1.69$ & $-0.2161^{+0.0041}_{-0.0042}$ & $-0.1102^{+0.0024}_{-0.0023}$ & $-0.1868^{+0.0038}_{-0.0037}$ & $-19.1^{+0.4}_{-0.4}$ & $-9.7^{+0.2}_{-0.2}$ & $-16.5^{+0.3}_{-0.3}$\\[+0.75em]
J1721+8842& $0.184$ & $2.37$ & $1.1885^{+0.0068}_{-0.0071}$ & $1.0683^{+0.0059}_{-0.0063}$ & $1.2809^{+0.0074}_{-0.0079}$ & $29.0^{+0.2}_{-0.2}$ & $26.0^{+0.1}_{-0.2}$ & $31.2^{+0.2}_{-0.2}$\\[+0.75em]
J1817+2729\textsuperscript{$\dagger$} & $0.5$ & $3.07$ & $-0.0474^{+0.0020}_{-0.0024}$ & $-0.0385^{+0.0016}_{-0.0020}$ & $-0.0724^{+0.0031}_{-0.0036}$ & $-3.6^{+0.1}_{-0.2}$ & $-2.9^{+0.1}_{-0.2}$ & $-5.4^{+0.2}_{-0.3}$\\[+0.75em]
DES J2038-4008& $0.228$ & $0.777$ & $0.5230^{+0.0043}_{-0.0038}$ & $0.3863^{+0.0031}_{-0.0026}$ & $0.6563^{+0.0059}_{-0.0050}$ & $20.5^{+0.2}_{-0.1}$ & $15.1^{+0.1}_{-0.1}$ & $25.7^{+0.2}_{-0.2}$\\[+0.75em]
WG2100-4452& $0.203$ & $0.92$ & $0.8140^{+0.0093}_{-0.0099}$ & $0.4141^{+0.0040}_{-0.0051}$ & $0.4206^{+0.0041}_{-0.0051}$ & $25.8^{+0.3}_{-0.3}$ & $13.1^{+0.1}_{-0.2}$ & $13.3^{+0.1}_{-0.2}$\\[+0.75em]
J2145+6345\textsuperscript{$\dagger$} & $0.5$ & $1.56$ & $-0.5187^{+0.0242}_{-0.0221}$ & $-0.1477^{+0.0086}_{-0.0084}$ & $-0.1604^{+0.0092}_{-0.0090}$ & $-47.5^{+2.2}_{-2.0}$ & $-13.5^{+0.8}_{-0.8}$ & $-14.7^{+0.8}_{-0.8}$\\[+0.75em]
J2205-3727\textsuperscript{$\dagger$} & $0.5$ & $1.848$ & $-0.2200^{+0.0075}_{-0.0058}$ & $-0.2006^{+0.0069}_{-0.0054}$ & $-0.2126^{+0.0072}_{-0.0056}$ & $-18.8^{+0.6}_{-0.5}$ & $-17.1^{+0.6}_{-0.5}$ & $-18.2^{+0.6}_{-0.5}$\\[+0.75em]
ATLAS J2344-3056\textsuperscript{$\dagger$} & $0.5$ & $1.298$ & $0.0260^{+0.0007}_{-0.0008}$ & $-0.0066^{+0.0002}_{-0.0002}$ & $0.0264^{+0.0007}_{-0.0008}$ & $2.6^{+0.1}_{-0.1}$ & $-0.7^{+0.1}_{-0.1}$ & $2.7^{+0.1}_{-0.1}$\\[+0.75em]
    
    \hline
    \multicolumn{9}{l}{\textsuperscript{$\dagger$}\scriptsize{System with a fiducial deflector redshift of $z_{\rm{d}} = 0.5$.}}
  \end{tabular}
\end{table*}

For each system, we predict the time delay, $\Delta t$, between images of the lensed quasar. These predictions can be used to determine for which system high-cadence observations are viable and to give guidance on the duration of long-term monitoring campaigns, as well as when to expect observed variations to appear in other images for the purpose of scheduling follow-up observations. To predict the time delays, we adopt a flat $\Lambda$CDM cosmology with standard values for present matter density, radiation, and the cosmological constant, at $\Omega_{\rm{m,0}} = 0.3$, $\Omega_{\rm{r,0}} = 0.0$, and $\Omega_{\rm{\Lambda,0}} = 0.7$, respectively, and the Hubble constant at $H_0 = 70$ km s$^{-1}$ Mpc$^{-1}$. For calculations where a component's redshift, due to lack of measurements, is currently unknown, we assume typical values of $z_{\rm{d}} = 0.5$, for the deflector, and of $z_{\rm{s}} = 2.0$, for the source. The predicted time delays for each successfully reconstructed lens model are summarized in Table~\ref{tab:time_delays}.\\

\subsection{Efficiency of the uniform framework}

To give an estimate on the time savings introduced by modeling strong lenses using our automated pipeline, we provide the total processing time for two systems, SDSS J0248+1913 and SDSS J1251+2935, broken down between the time needed for the PSO steps and the time required to probe the posterior distributions through an MCMC. In the case of SDSS J0248+1913, the PSO fitting time is 5 hours and 56 mins, while the run-time of the MCMC is 5 hours and 10 mins, giving a total reconstruction time of 11 hours and 6 mins. The PSO fitting time corresponds to using 19 threads on a machine with a hyper-threaded Intel(R) Core(TM) i9-9820X CPU clocked at 3.30 GHz, while the MCMC run-time corresponds to a computation using 20 threads on the same architecture. 
For SDSS J1251+2935, the PSO fitting time is 8 hours, 55 mins, and the associated MCMC time to find convergence is 8 hours, 12 mins, for a total computation time of 17 hours and 7 mins. The MCMC run-time corresponds to the same resource level as used for the modeling of SDSS J0248+1913, however, the PSO fitting is associated with 20 threads on a machine hosting an Intel(R) Core(TM) i9-10980XE CPU clocked at 3.00 GHz. By comparison, these processing times are much shorter than traditional lens modeling times, which can require up to 1 million CPU hours for extremely complex lens configurations. Furthermore, our pipeline's speed is comparable to the processing times of \citet{Shajib19}'s framework, but it requires no human input and intervention along the reconstruction process. 

The time required to set up the pipeline is conservatively estimated at 1 hour per lens. Additionally, we approximate another 6 hours of required investigator time per lens to reduce the data, prior to the pipeline processing, and 3 hours per system to review results and move data between machines, arriving at a conservative total of 10 hours of investigator overhead. This overhead represents a minimal level of investigator time and would be required for most types of analyses, considering the necessity of data preparation and quality control. This time is much smaller than the typical amount of investigator time adopted by previous non-automated studies, and comparable with the amount of time per system invested by \citet{Shajib19}. \\

\subsection{Difference in Fermat potential of the quasar images as a metric for cosmography} \label{stability_of_fermat_pot}

\begin{figure}
 \includegraphics[width=0.45\textwidth]{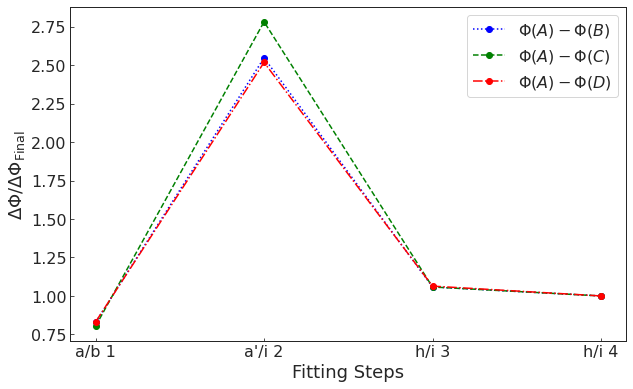}
 \caption{Difference in the Fermat potential between the image positions for SDSS J0248+1913, as a function of modeling steps, from the initial setup of the reconstruction through the final PSO fitting after adding additional source complexity to the model.}
 \label{fig:fermat_PSO_J0248}
\end{figure}

To assess the stability in our models and their utility for cosmography, we introduce a new metric that tracks the changes in the Fermat potential at the position of the quasar images for each step or modeling choice in our pipeline. We then compute the absolute difference in the Fermat potential at the image positions and normalize it with the results of the final model.

We expect the stability of the Fermat potential difference to depend on three factors: i) the information content of the multiple images of the extended source; ii) the overall symmetry and configuration of the multiple images of the quasars (highly symmetric crosses will have fairly similar potential at the location of the images); iii) complexity of the mass distribution of the deflector and presence of perturbers. 

This metric allows us to visualize the impact of each model decision along the reconstruction. It also  gives us a way to use the metric as a tool to evaluate systematic uncertainties resulting from modeling choices by applying it to a large sample as we will demonstrate for the case study SDSS J0248+1913 in Section~\ref{Systematics}. Ultimately, this new metric also gives us a way to assess how close our reconstructed models come to the quality required for cosmography. 

In this section we first discuss in some detail a case study and then present some statistics about the performance of the pipeline with respect to this metric across the sample.

\subsubsection{Case study}

To demonstrate the described tracking mechanism for modeling choices, we show in Figure~\ref{fig:fermat_PSO_J0248} the evolution of the difference in Fermat potential at the image positions for lens SDSS J0248+1913, starting with the first model setup up to the initialization of the MCMC run. Stepping through the decision framework, Figure~\ref{fig:fermat_PSO_J0248} illustrates the resulting changes in Fermat potential differences from an initial configuration  (step a.), used for fitting the most informative band (step b), through the first simultaneous fitting sequence of all bands (step a'/i), up to adding source complexity via increase in shapelet order in the final steps h/i 3 and h/i 4.

The conclusion of Figure~\ref{fig:fermat_PSO_J0248} is that modeling choices can alter the difference in Fermat potential at a level that is significant w.r.t. our target precision of 3-5\% and therefore need to be properly addressed.  
In contrast, the statistical uncertainties for a fixed model choice, as explored by the MCMC, are tiny ($\sim10^{-13}$), and absolutely negligible. 
%much smaller than our target uncertainty of 3-5\%, and 
Therefore they do not contribute significantly to the cosmological error budget.\\

\subsubsection{Assessment of Fermat potential stability on the sample }

Table~\ref{tab:Fermat_stability} summarizes the performance of our pipeline on the sample, in terms of Fermat potential stability. We focus on the absolute value of the Fermat potential difference between the images with the largest difference, since those will have the longest time delay. The longest time delay is the easiest one to measure precisely and therefore usually the dominant contributor to the cosmographic measurement. The first row describes the stability with respect to the last two steps for model changes within the PSO process, while the second row describes stability during the MCMC chain, which we derive from the comparison of the final MCMC result to the results for the final model of the PSO. As in the case above, it is clear that modeling choices, and not residual statistical errors once a model has been defined, are the dominant source of uncertainty in Fermat potential difference, so we will focus our discussion on the first row. 

For convenience, we bin the sample into four general categories: i) cosmography grade models, which we define as lenses with an uncertainty of less than $3\%$; ii) nearly cosmography grade, defined with errors of between $3-5\%$, which will require a likely modest amount of additional effort to achieve accurate Hubble constant measurements; iii) models with uncertainties in the range 5-10\%, which may already be useful for other applications but require substantial work to reach cosmography grade; iv) models with uncertainties larger than 10\%, which should be used only as a starting point for further investigation. We caution the reader that this assessment is indicative only and caveats apply. For example, we have used an informative prior on the mass density profile slope, which could bias the inference if not removed or properly accounted for. Furthermore, some systems may have good enough lens models but may still yield low precision cosmology if the sources do not vary enough to measure the time delay or if the line of sight is overly complicated.

Remarkably, the pipeline yields cosmography-grade or nearly cosmography grade models (as defined above) for 10/30 systems (WG0214-2105, J0343-2828, DES J0420-4037, J0659+1629, 2M1310-1714, J1721+8842, J1817+2729, WG2100-4452 are cosmography grade; PS J0147+4630 and J1537-3010 are nearly cosmography grade); however, we urge caution with regards to the results of J1721+8842, as this system is highly complex and, as shown in Figure~\ref{fig:astrometry}, we find small variances between our predicted image positions and the corresponding \textit{Gaia} measurements.
Six lenses are in the 5-10\% range, but require substantial work (SDSS J0248+1913, WISE J0259-1635, J0818-2613, 2M1134-2103, DES J2038-4008, SDSS J1251+2935). The remaining fourteen systems we assess as currently far from cosmography grade given the stability in Fermat potential differences observed during the steps of model reconstruction. 

The fact that half of the lenses are far from cosmography grade is not surprising, considering the quality of the data, the lensing configurations, and the simplifications inherent to our modeling procedure.  Table~\ref{tab:time_delays} and Figures~\ref{fig:td_plots_1} to~\ref{fig:td_plots_4} shed some light on the causes of instability. In certain cases (e.g. J2145+6345) the quasars are so bright that little extended source light is visible. In others (e.g. ATLAS J233-3056) the image separation is so small and the system so symmetric (resulting in very short predicted time delays) that it is difficult to imagine getting a cosmography grade model with current technology. From the modeling point of view, some cases seem easily improvable, while others will require a more flexible modeling scheme than the one applied here. In the first category are all those systems for which there was a substantial jump in the Fermat potential difference between the one band and three band model. In many of those cases the three band model was considered acceptable but the metric still recorded the jump (e.g. J0029-3814). For some of those cases, further exploration may reveal that cosmography grade models are achievable within the present modeling assumptions. The latter category includes systems such as PS J0630-1201, where there is clearly more than one deflector and therefore will require more complex models. Additional work, left for future studies, is needed to find out if those systems can be made cosmography grade with additional modeling efforts and/or better data, or whether it is more cost effective to focus on the low hanging fruits.  

The fact that our automated pipeline yields "cosmography grade" models for a 1/3 of the sample is an important result, although some caveats must be kept in mind. The most important caveat is that our pipeline uses informative priors on the slope of the mass density profile, ellipticity, and alignment of mass and light profiles, to avoid non-physical solutions. However, such informative priors could have artificially reduced the uncertainty on the Fermat potential. For example, J1817+2729 and WG0214-2105 are labelled "cosmography grade", but the offset between the PA of mass and light is close to the boundary of the prior. It is thus possible that with a less informative prior the uncertainty would have been larger. Analyzing these types of issues is beyond the scope of our automated pipeline and is left for future work.  

In conclusion, our finding that a third of the sample is "cosmography grade" -- if confirmed by a more detailed analysis of a subset of the systems -- would imply that investigator time can be cut significantly shorter than in previous state-of-the-art studies, paving the way for studies of much larger future samples. 
 
\begin{table}
  \caption{Stability in Fermat Potential. Listed are the number of systems for which a change in the Fermat potential difference is within the table's thresholds. The table includes both, model changes within the PSO and changes during the MCMC chain. }
  \label{tab:Fermat_stability}
  \begin{tabular}{lcccc}
    \hline
    Stability &  $<3\%$ & $3-5\%$ & $5-10\%$ & $\geq10\%$  \\
    \hline
PSO & 8 & 2 & 6 & 14\\
MCMC & 24 & 3 & 2 & 1 \\

    \hline
  \end{tabular}
\end{table}

\section{Discussion} \label{Discussion}

\subsection{Systematics in source complexity} \label{Systematics}

Even small increases in source complexity, represented by the maximum shapelet order, $n_{\rm{max}}$, have a substantial impact on the computational time in the reconstruction of the lens models. Additionally, unnecessary complexity would only result in "fitting the noise". Therefore, the models produced by our automated pipeline should have a sufficient complexity to accurately reflect the  data, but not more.

To assess the impact of systematic uncertainties in a model's source light complexity, we use the metric established in Section~\ref{stability_of_fermat_pot} and introduce a small perturbation in the shapelet order, $n_{\rm{max}}$, of a successfully converged lens model. 
We first increase $n_{\rm{max}}$ by 2 and then probe the model's parameter space with a narrow search region using a PSO with 150 particles for 500 iterations. If the converged lens model does not include any shapelets for a particular band, we add the shapelet profile to the existing S\'ersic light profile of the source. The particle number and iterations for the PSO routine are set high enough that any deviation in the stability of the Fermat potential will become apparent after the PSO's execution. We then re-evaluate the difference in the Fermat potential at the image positions, using the result of the perturbation, and compare it to the previously converged model result. This test in stability is repeated with a decrease in the model's source complexity. For this, we decrease $n_{\rm{max}}$ by 2, unless no shapelets were used to model additional source complexity. 
We repeat probing the parameter space with a PSO routine, using the same particle number, at the same same uncertainty level, and the same number of iterations as selected for investigating the impact of a higher source complexity. We then compare the Fermat potential difference of the quasar images between the perturbed and the baseline model.

\begin{figure}
 \includegraphics[width=0.5\textwidth]{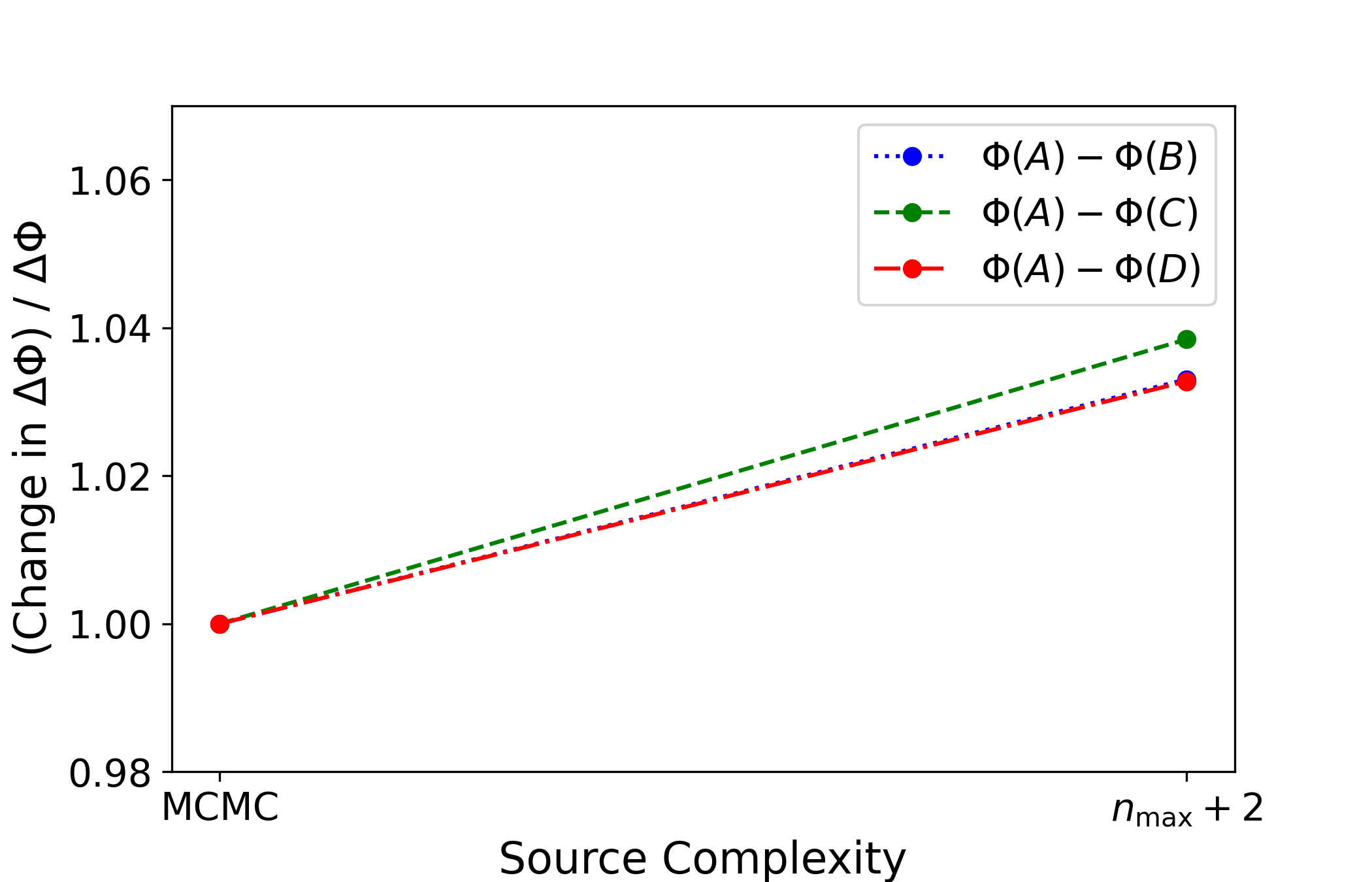}
 \caption{Stability of the difference in Fermat potential at the image positions, w.r.t. an increase in source complexity for SDSS J0248-1913. The increase of the highest shapelet order parameter, $n_{\rm{max}}$, by 2 results in a less than 4\% change for all Fermat potential differences.}
 \label{fig:fermat_increase_J0248}
\end{figure}

\begin{figure}
 \includegraphics[width=0.5\textwidth]{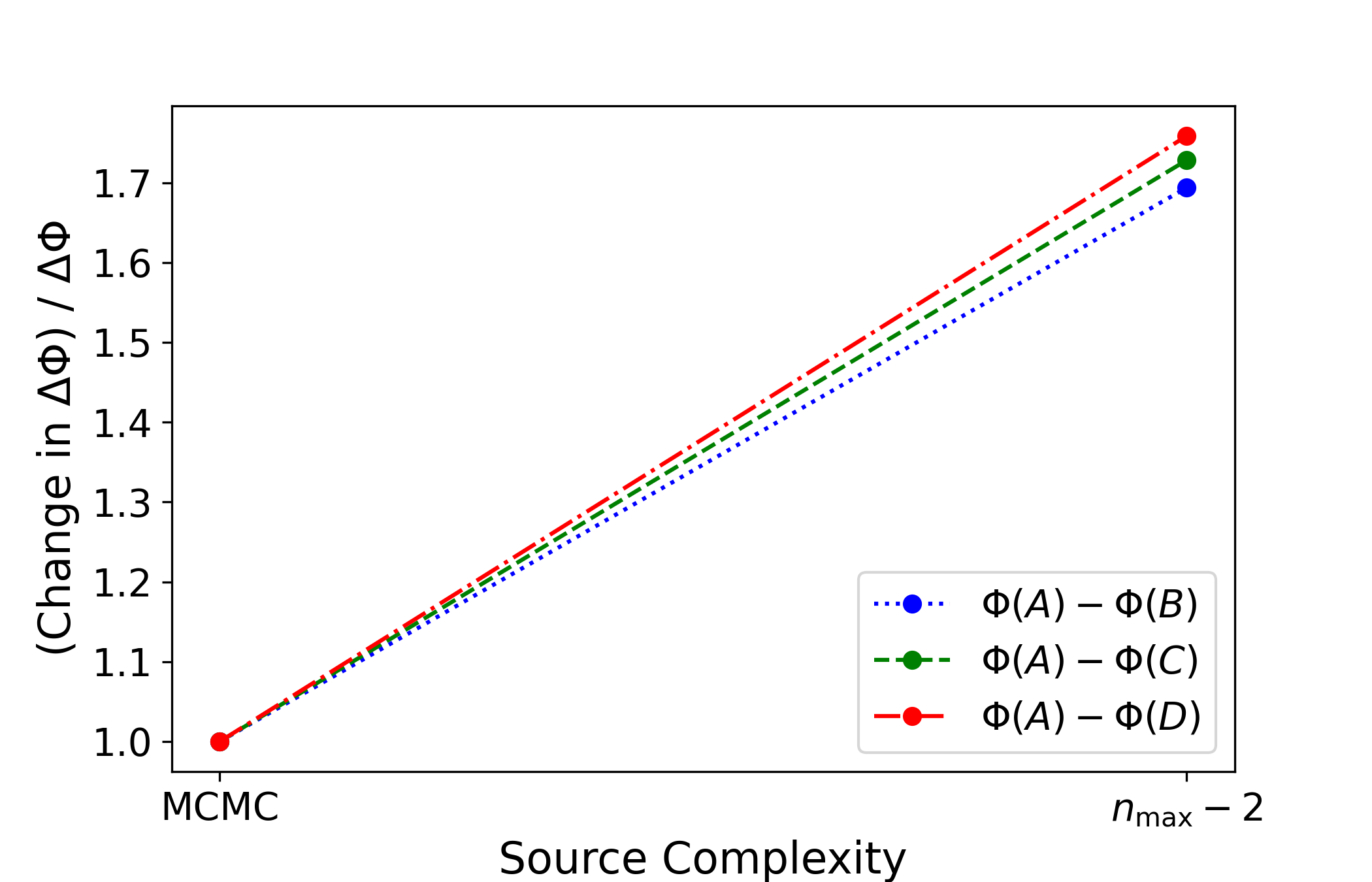}
 \caption{Stability of the difference in Fermat potential at the image positions, w.r.t. a decrease in source complexity for SDSS J0248-1913. The decrease in the maximum shapelet order, $n_{\rm{max}}$, by 2 drive the model towards changes that end in nearly doubling the images' Fermat potential difference.}
 \label{fig:fermat_decrease_J0248}
\end{figure}

For lens system SDSSJ0248+1913 we find no significant impact on the Fermat potential difference at the image positions for an increase in source complexity, as shown in Figure~\ref{fig:fermat_increase_J0248}. Additionally, the PSO converged after 191 iterations, reflecting the model's stability. However, as illustrated in Figure~\ref{fig:fermat_decrease_J0248}, we find the opposite for a decrease in $n_{\rm{max}}$, where we observe a noteworthy change in the Fermat potential difference. These findings are reassuring, as they demonstrate the reconstructed model's stability, and further infer that adding unnecessary complexity does not change the derived model parameters. In contrast, lowering complexity enters a regime in which the model does not sufficiently describe the data and with it signals a change in the Fermat potential for the system. As Figures~\ref{fig:fermat_increase_J0248} and~\ref{fig:fermat_decrease_J0248} demonstrate, the complexity increase results in a change of less than 4\%, which is below our required uncertainty threshold, while a decrease in source complexity pushes the stable model towards a less precise description and with it results in a change far above our target of a 5\% error budget.\\

\subsection{Future improvements} \label{improvements}
\label{ssec:future}

The pipeline presented here is a major step forward.  Future work should be able to further improve on our result by carrying out the following additional steps:

\begin{itemize}

\item Assess the impact of other modeling choices on the reconstruction of strongly lensed systems. Utilizing our pipeline on large samples will enable us to test how much models are affected by variations in parameters such as PSF symmetry, cut-out size, or sub-grid resolution.

\item
Expand and build up the capabilities of our pipeline to be able to reconstruct more complex lens model scenarios. For instance, our pipeline is limited to one main deflector. In cases where observations show two lensing galaxies (i.e. J0343-282 and 2M1310-1714) we currently designate one of the deflectors as primary, using a PEMD mass profile, and model the second lens galaxy as satellite with an SIS mass profile. This simplified approach, however, limits us when other significant perturbers, such as satellites to the main deflector, are present and their impact should not be ignored in the reconstruction. Another aspect that could be generalized is the choice of a satellite's mass and light profile. Not every main deflector companion is sufficiently represented by our current modeling choice of an SIS  with a circular S\'ersic light profile, leaving room for the inclusion of other possible options, for instance a Singular Isothermal Ellipsoid.\\

\item
Develop the capability to model multi-lens-plane systems and massive perturbers. In most cases a single lens plane approximation proves sufficient for the model.  However, if masssive perturbers are found outside the lens plane, these deflectors should be modeled at their correct respective distances to insure accurate computations of the Fermat potential difference between quasar images. Since we have observations in multiple bands available, we have the necessary information to incorporate photometric redshift estimates for all model components, and with it could potentially facilitate a multi-plane lens model reconstructions, if determined preferential.\\

\item
Given that our sample includes lenses for which a double S\'ersic provides an improved but not yet complete description of the main deflector's light profile (i.e. DES J2038-4008), future versions of the pipeline would benefit from the inclusion of other lens light descriptors for highly complex systems, both, in addition or in lieu of the S\'ersic functions used for this work. 

\end{itemize}

\section{Summary} \label{Summary}

We developed a lens modeling pipeline aimed at minimizing an investigator's time. We then applied it to a sample of 30 quadruply imaged quasars and one lensed compact galaxy. Out of these 31 lenses, $30$ systems can be processed successfully by our pipeline, while  the remaining system is too complex for the current capabilities of the pipeline. Explicit details on specific model parameters for each lens system are shown in the tables of Section~\ref{Results}. 
Our main results can be summarized as follows:

\begin{itemize}

\item
Our pipeline produces lens models using typically 10 hours of investigator time and 100 hours of CPU time. This is an improvement of many orders of magnitude with respect to studies of individual lenses and comparable to what was achieved by \citet{Shajib19b}, but for a larger sample and with less human intervention during the process. 

\item Based on the pipeline output, we provide for each lens an extensive set of lens model parameters and forecasted properties such as time delays, convergence, and magnification at the location of the images. 

\item 
We introduce a new metric to assess the quality of our models, i.e. the stability of the difference in the Fermat potential between multiple images. We demonstrate the usefulness of the metric in assessing the impact of modeling choices and recommend it as convergence/stability indicator in future studies. The factors contributing to the instability of the Fermat potential differences are i) overall information content in the multiple images of the extended sources, ii) symmetry and configuration of the multiple images of the quasars; iii) complexity of the lensing gravitational potential.

\item
We show that in terms of Fermat potential stability statistical errors are always subdominant with respect to those induced by modeling choices.
\item
For a third of the sample (10/30), our pipeline produces models that are cosmography or nearly cosmography grade (i.e. stability in Fermat potential $<3$\% or 3-5\%). For 6/30 quads, the models have Fermat potential differences stable within 5-10\% and could therefore become cosmography grade with some additional effort. In the remaining 14/30 models, the Fermat potential differences are larger than 10\%.  Further investigations are needed to establish which of those systems could become cosmography grade with additional work based on the current pipeline, which ones will require extensions of the pipeline, and which ones are instead intrinsically limited by the quality of the data and the lensing configuration.

\item
We apply small perturbations, both upwards and downwards, to the source complexity of a converged model and find that, as long as the level of source light is sufficiently well represented in our models, a perturbation will not significantly affect the Fermat potential difference between quasar image positions.

\end{itemize}

Our modeling of an unprecedented large sample of quads is a major step forward in time-delay cosmography.  Although further analysis and verification is needed before they can be used for cosmography, these results pave the way to the uniform modeling of large samples of quads (100 or more) that are expected to be discovered in the near future \citep[e.g.][]{Oguri10}. Further improvements are possible by, both, running the existing pipeline more extensively, or carrying out the steps outlined in Section~\ref{ssec:future}.

Going forward, the strategic question that needs to be answered for time-delay cosmography is whether to focus on the fraction of lenses that require less work to be modeled or whether it is necessary to tackle more complex systems at the cost of expanding and customizing the modeling pipeline. Up until today, painstaking work on individual lenses has been carried out, owing to a combination of small samples and the invaluable lessons learned from the study of the first few objects. As the samples of lenses increase by orders of magnitude, the strategy will have to adapt to harness the power of large samples \citep[e.g.,][]{Sonnenfeld21a,Sonnenfeld21b} and exploit the insights from detailed studies of smaller samples, while keeping the investigator time investment manageable and optimizing the observational resources needed for follow-up. For example, given large samples of lenses it may be preferable to prioritize those systems with expected long time delays (therefore more easily measurable at high precision) and with the main deflector galaxy not completely overwhelmed by the lensed quasars light (and thus easier to measure spatially resolved kinematics).

\section*{Acknowledgements}

This research is based on observations made with the NASA/ESA Hubble Space Telescope obtained from the Space Telescope Science Institute, which is operated by the Association of Universities for Research in Astronomy, Inc., under NASA contract NAS 5-26555. These observations are associated with programs HST-GO-15320 and HST-GO-15652. Support for the two programs was provided by NASA through a grant from the Space Telescope Science Institute, which is operated by the Association of Universities for Research in Astronomy, Inc., under NASA contract NAS 5-26555.

TS TT CDF acknowledge support by the the National Science Foundation through grant NSF-AST-1906976 and NSF-AST-1907396 "Collaborative Research: Toward a 1\% measurement of the Hubble Constant with gravitational time delays". TT acknowledges support by the Packard Foundation through a Packard Research Fellowship. Support for this work was provided by NASA through the NASA Hubble Fellowship grant
HST-HF2-51492 awarded to AJS by the Space Telescope Science Institute, which is operated by the Association of Universities for Research in Astronomy, Inc., for NASA, under contract NAS5-26555. 

This programme is in part supported by the Swiss National Science Foundation (SNSF) and by the European Research Council (ERC) under the European Union's Horizon 2020 research and innovation programme (COSMICLENS: grant agreement No 787886).
This project has received funding from the European Research
Council (ERC) under the European Union's Horizon
2020 research and innovation programme (COSMICLENS :
grant agreement No 787886).

The authors acknowledge data and feedback provided by E. Glikman and C.E.Rusu.
AA's work is funded by Villum Experiment Grant \textit{Cosmic Beacons} (project number 36225). 
TA acknowledges support from FONDECYT Regular 1190335, the Millennium Science Initiative ICN12\_009  and the ANID BASAL project FB210003.
MWA-W acknowledges support from the Kavli Foundation.
CS is supported by an `Hintze Fellow' at the Oxford Centre for Astrophysical Surveys, which is funded through generous support from the Hintze Family Charitable  Foundation. 
RGM would like to acknowledge the support of the UK Science and Technology Facilities Council (STFC). 
IK is supported by JSPS KAKENHI Grant Number JP20K04016.
VM acknowledges support from project "Fortalecimiento del Sistema de Investigaci\'on e Innovaci\'on de la Universidad de Valpara\'{\i}so (UVA20993)".
SE, SS and SHS thank the Max Planck Society for support through the Max Planck Research Group for SHS.  
This research is supported in part by the Excellence Cluster ORIGINS which is funded by the Deutsche Forschungsgemeinschaft (DFG, German Research Foundation) under Germany's Excellence Strategy -- EXC-2094 -- 390783311.

Funding for the DES Projects has been provided by the U.S. Department of Energy, the U.S. National Science Foundation, the Ministry of Science and Education of Spain, 
the Science and Technology Facilities Council of the United Kingdom, the Higher Education Funding Council for England, the National Center for Supercomputing 
Applications at the University of Illinois at Urbana-Champaign, the Kavli Institute of Cosmological Physics at the University of Chicago, 
the Center for Cosmology and Astro-Particle Physics at the Ohio State University,
the Mitchell Institute for Fundamental Physics and Astronomy at Texas A\&M University, Financiadora de Estudos e Projetos, 
Funda{\c c}{\~a}o Carlos Chagas Filho de Amparo {\`a} Pesquisa do Estado do Rio de Janeiro, Conselho Nacional de Desenvolvimento Cient{\'i}fico e Tecnol{\'o}gico and 
the Minist{\'e}rio da Ci{\^e}ncia, Tecnologia e Inova{\c c}{\~a}o, the Deutsche Forschungsgemeinschaft and the Collaborating Institutions in the Dark Energy Survey. 

The Collaborating Institutions are Argonne National Laboratory, the University of California at Santa Cruz, the University of Cambridge, Centro de Investigaciones Energ{\'e}ticas, 
Medioambientales y Tecnol{\'o}gicas-Madrid, the University of Chicago, University College London, the DES-Brazil Consortium, the University of Edinburgh, 
the Eidgen{\"o}ssische Technische Hochschule (ETH) Z{\"u}rich, 
Fermi National Accelerator Laboratory, the University of Illinois at Urbana-Champaign, the Institut de Ci{\`e}ncies de l'Espai (IEEC/CSIC), 
the Institut de F{\'i}sica d'Altes Energies, Lawrence Berkeley National Laboratory, the Ludwig-Maximilians Universit{\"a}t M{\"u}nchen and the associated Excellence Cluster Universe, 
the University of Michigan, NSF's NOIRLab, the University of Nottingham, The Ohio State University, the University of Pennsylvania, the University of Portsmouth, 
SLAC National Accelerator Laboratory, Stanford University, the University of Sussex, Texas A\&M University, and the OzDES Membership Consortium.

Based in part on observations at Cerro Tololo Inter-American Observatory at NSF's NOIRLab (NOIRLab Prop. ID 2012B-0001; PI: J. Frieman), which is managed by the Association of Universities for Research in Astronomy (AURA) under a cooperative agreement with the National Science Foundation.

The DES data management system is supported by the National Science Foundation under Grant Numbers AST-1138766 and AST-1536171.
The DES participants from Spanish institutions are partially supported by MICINN under grants ESP2017-89838, PGC2018-094773, PGC2018-102021, SEV-2016-0588, SEV-2016-0597, and MDM-2015-0509, some of which include ERDF funds from the European Union. IFAE is partially funded by the CERCA program of the Generalitat de Catalunya.
Research leading to these results has received funding from the European Research
Council under the European Union's Seventh Framework Program (FP7/2007-2013) including ERC grant agreements 240672, 291329, and 306478.
We  acknowledge support from the Brazilian Instituto Nacional de Ci\^encia
e Tecnologia (INCT) do e-Universo (CNPq grant 465376/2014-2).

This manuscript has been authored by Fermi Research Alliance, LLC under Contract No. DE-AC02-07CH11359 with the U.S. Department of Energy, Office of Science, Office of High Energy Physics.

This research made use of \textsc{Lenstronomy} \citep{Birrer18}, \textsc{emcee} \citep{Foreman-Mackey13}, \textsc{fastell} \citep{Barkana98}, \textsc{sextractor} \citep{Bertin96}, \textsc{NumPy} \citep{Oliphant15}, \textsc{SciPy} \citep{Jones01}, \textsc{Astropy} \citep{AstropyCollaboration18}, \textsc{Jupyter} \citep{Kluyver16}, \textsc{Matplotlib} \citep{Hunter07}, \textsc{seaborn} \citep{Waskom21}, and \textsc{draw.io} at {\href{https://www.draw.io}{https://www.draw.io}}.

\section*{Data availability}

All data underlying this article are publicly available through the {\it HST} archive.

\section*{Affiliations}
\noindent
{\it
% List of institutions
$\dagger$ Packard Fellow\\
$\ddagger$ NHFP Einstein Fellow\\
$^{1}$ Department of Physics and Astronomy, University of California, Los Angeles, CA 90095, USA\\
$^{2}$ Kavli Institute for Particle Astrophysics \& Cosmology, P. O. Box 2450, Stanford University, Stanford, CA 94305, USA\\
$^{3}$ SLAC National Accelerator Laboratory, Menlo Park, CA 94025, USA\\
$^{4}$ Department of Physics and Astronomy, Stony Brook University, Stony Brook, NY 11794, USA\\
$^{5}$ Kavli Institute for Cosmological Physics, University of Chicago, Chicago, IL 60637, USA\\
$^{6}$ Institute of Physics, Laboratoire d'Astrophysique, Ecole Polytechnique F\'ed\'erale de Lausanne (EPFL), Observatoire de Sauverny, CH-1290 Versoix, Switzerland\\
$^{7}$ STAR Institute, Quartier Agora - All\'ee du six Ao\^ut, 19c B-4000 Li\`ege, Belgium\\
$^{8}$ DARK, Niels Bohr Institute, Jagtvej 128, 2200 Copenhagen, Denmark\\
$^{9}$ Departamento de Ciencias Fisicas, Universidad Andres Bello Fernandez Concha 700, Las Condes, Santiago, Chile\\
$^{10}$ Millennium Institute of Astrophysics, Nuncio Monse{\~{n}}or S{\'{o}}tero Sanz 100, Of 104, Providencia, Santiago, Chile\\
$^{11}$ Institute of Astronomy, University of Cambridge, Madingley Road, Cambridge CB3 0HA, UK\\
$^{12}$ Kavli Institute for Cosmology, University of Cambridge, Madingley Road, Cambridge CB3 0HA, UK\\
$^{13}$ Instituto de F\'isica y Astronom\'ia, Facultad de Ciencias, Universidad de Valpara\'iso, Avda. Gran Breta\~na 1111, Valpara\'iso, Chile\\
$^{14}$ MIT Kavli Institute for Astrophysics and Space Research, Cambridge, MA 02139, USA\\
$^{15}$ Sub-Department of Astrophysics, Department of Physics, University of Oxford, Denys Wilkinson Building, Keble Road, Oxford OX1 3RH, United Kingdom\\
$^{16}$ INAF - Osservatorio Astronomico di Capodimonte, Salita Moiariello, 16, I-80131 Napoli, Italy\\
$^{17}$ Department of Liberal Arts, Tokyo University of Technology, Ota-ku, Tokyo 144-8650, Japan\\
$^{18}$ Max-Planck-Institut f\"ur Astrophysik, Karl-Schwarzschild Str.~1, 85748 Garching, Germany\\
$^{19}$ Technische Universit\"at M\"unchen, Physik Department, James-Franck Str.~1, 85748 Garching, Germany\\
$^{20}$ Department of Physics and Astronomy, University of California, Davis, 1 Shields Ave., Davis, CA 95161, USA\\
$^{21}$ Fermi National Accelerator Laboratory, P. O. Box 500, Batavia, IL 60510, USA\\
$^{22}$ The Inter-University Centre for Astronomy and Astrophysics (IUCAA), Post Bag 4, Ganeshkhind, Pune 411007, India\\
$^{23}$ Kavli Institute for the Physics and Mathematics of the Universe (IPMU), 5-1-5 Kashiwanoha, Kashiwa-shi, Chiba 277-8583, Japan\\
$^{24}$ Institute of Astronomy and Astrophysics, Academia Sinica, 11F of ASMAB, No.1, Section 4, Roosevelt Road, Taipei 10617, Taiwan\\
$^{25}$ Laborat\'orio Interinstitucional de e-Astronomia - LIneA, Rua Gal. Jos\'e Cristino 77, Rio de Janeiro, RJ - 20921-400, Brazil\\
$^{26}$ Department of Physics, University of Michigan, Ann Arbor, MI 48109, USA\\
$^{27}$ Institute of Cosmology and Gravitation, University of Portsmouth, Portsmouth, PO1 3FX, UK\\
$^{28}$ CNRS, UMR 7095, Institut d'Astrophysique de Paris, F-75014, Paris, France\\
$^{29}$ Sorbonne Universit\'es, UPMC Univ Paris 06, UMR 7095, Institut d'Astrophysique de Paris, F-75014, Paris, France\\
$^{30}$ Department of Physics \& Astronomy, University College London, Gower Street, London, WC1E 6BT, UK\\
$^{31}$ Instituto de Astrofisica de Canarias, E-38205 La Laguna, Tenerife, Spain\\
$^{32}$ Universidad de La Laguna, Dpto. Astrofísica, E-38206 La Laguna, Tenerife, Spain\\
$^{33}$ Center for Astrophysical Surveys, National Center for Supercomputing Applications, 1205 West Clark St., Urbana, IL 61801, USA\\
$^{34}$ Department of Astronomy, University of Illinois at Urbana-Champaign, 1002 W. Green Street, Urbana, IL 61801, USA\\
$^{35}$ Institut de F\'{\i}sica d'Altes Energies (IFAE), The Barcelona Institute of Science and Technology, Campus UAB, 08193 Bellaterra (Barcelona) Spain\\
$^{36}$ Jodrell Bank Center for Astrophysics, School of Physics and Astronomy, University of Manchester, Oxford Road, Manchester, M13 9PL, UK\\
$^{37}$ University of Nottingham, School of Physics and Astronomy, Nottingham NG7 2RD, UK\\
$^{38}$ Astronomy Unit, Department of Physics, University of Trieste, via Tiepolo 11, I-34131 Trieste, Italy\\
$^{39}$ INAF-Osservatorio Astronomico di Trieste, via G. B. Tiepolo 11, I-34143 Trieste, Italy\\
$^{40}$ Institute for Fundamental Physics of the Universe, Via Beirut 2, 34014 Trieste, Italy\\
$^{41}$ Observat\'orio Nacional, Rua Gal. Jos\'e Cristino 77, Rio de Janeiro, RJ - 20921-400, Brazil\\
$^{42}$ Hamburger Sternwarte, Universit\"{a}t Hamburg, Gojenbergsweg 112, 21029 Hamburg, Germany\\
$^{43}$ Centro de Investigaciones Energ\'eticas, Medioambientales y Tecnol\'ogicas (CIEMAT), Madrid, Spain\\
$^{44}$ Department of Physics, IIT Hyderabad, Kandi, Telangana 502285, India\\
$^{45}$ Santa Cruz Institute for Particle Physics, Santa Cruz, CA 95064, USA\\
$^{46}$ Institute of Theoretical Astrophysics, University of Oslo. P.O. Box 1029 Blindern, NO-0315 Oslo, Norway\\
$^{47}$ Instituto de Fisica Teorica UAM/CSIC, Universidad Autonoma de Madrid, 28049 Madrid, Spain\\
$^{48}$ Institut d'Estudis Espacials de Catalunya (IEEC), 08034 Barcelona, Spain\\
$^{49}$ Institute of Space Sciences (ICE, CSIC),  Campus UAB, Carrer de Can Magrans, s/n,  08193 Barcelona, Spain\\
$^{50}$ University Observatory, Faculty of Physics, Ludwig-Maximilians-Universit\"at, Scheinerstr. 1, 81679 Munich, Germany\\
$^{51}$ School of Mathematics and Physics, University of Queensland,  Brisbane, QLD 4072, Australia\\
$^{52}$ Center for Cosmology and Astro-Particle Physics, The Ohio State University, Columbus, OH 43210, USA\\
$^{53}$ Department of Physics, The Ohio State University, Columbus, OH 43210, USA\\
$^{54}$ Center for Astrophysics $\vert$ Harvard \& Smithsonian, 60 Garden Street, Cambridge, MA 02138, USA\\
$^{55}$ Australian Astronomical Optics, Macquarie University, North Ryde, NSW 2113, Australia\\
$^{56}$ Lowell Observatory, 1400 Mars Hill Rd, Flagstaff, AZ 86001, USA\\
$^{57}$ Instituci\'o Catalana de Recerca i Estudis Avan\c{c}ats, E-08010 Barcelona, Spain\\
$^{58}$ Department of Astronomy, University of California, Berkeley,  501 Campbell Hall, Berkeley, CA 94720, USA\\
$^{59}$ Department of Astrophysical Sciences, Princeton University, Peyton Hall, Princeton, NJ 08544, USA\\
$^{60}$ Department of Astronomy and Astrophysics, University of Chicago, Chicago, IL 60637, USA\\
$^{61}$ \\
$^{62}$ Department of Physics and Astronomy, Pevensey Building, University of Sussex, Brighton, BN1 9QH, UK\\
$^{63}$ School of Physics and Astronomy, University of Southampton,  Southampton, SO17 1BJ, UK\\
$^{64}$ Computer Science and Mathematics Division, Oak Ridge National Laboratory, Oak Ridge, TN 37831\\
$^{65}$ Excellence Cluster Origins, Boltzmannstr.\ 2, 85748 Garching, Germany\\
$^{66}$ Max Planck Institute for Extraterrestrial Physics, Giessenbachstrasse, 85748 Garching, Germany\\
$^{67}$ Universit\"ats-Sternwarte, Fakult\"at f\"ur Physik, Ludwig-Maximilians Universit\"at M\"unchen, Scheinerstr. 1, 81679 M\"unchen, Germany\\
}

% Entry for the table of contents, for this guide only
% \addcontentsline{toc}{section}{Acknowledgements}

%%%%%%%%%%%%%%%%%%%%%%%%%%%%%%%%%%%%%%%%%%%%%%%%%%



\bibliographystyle{mnras}
\bibliography{ajshajib} % if your bibtex file is called example.bib

% \begin{thebibliography}{99}
% \bibitem[\protect\citeauthoryear{Author}{2013}]{author2013}
% Author A.~N., 2013, Journal of Improbable Astronomy, 1, 1
% \bibitem[\protect\citeauthoryear{Jones}{2015}]{jones2015}
% Jones C.~D., 2015, Journal of Interesting Stuff, 17, 198
% \bibitem[\protect\citeauthoryear{Smith}{2014}]{smith2014}
% Smith A.~B., 2014, The Example Journal, 12, 345 (Paper I)
% \end{thebibliography}

%%%%%%%%%%%%%%%%%%%%%%%%%%%%%%%%%%%%%%%%%%%%%%%%%%

%%%%%%%%%%%%%%%%% APPENDICES %%%%%%%%%%%%%%%%%%%%%

\appendix

\section{Photometry of Quasar Images}

In Tables~\ref{tab:QSO_photometry} and~\ref{tab:QSO_photometry_2}, we report the QSO image magnitudes for each successfully modeled lens using the AB system.

\begin{table*}
  \caption{Median values for quasar (and, in case of J0343-2828, compact galaxy) image magnitudes in the AB system. The associated uncertainties are statistical in nature and were computed using 84th and 16th percentiles.}
  \label{tab:QSO_photometry}
  \scriptsize
%   \tiny
  \begin{tabular}{lccccc}
    \hline
    Name of Lens &  Filter & A & B & C & D \\
    \hline
    \\[-0.75em]
 & F475X & $21.601^{+0.004}_{-0.004}$ & $22.032^{+0.005}_{-0.005}$ & $20.992^{+0.005}_{-0.006}$ & $21.829^{+0.003}_{-0.004}$\\[+0.75em]
J0029-3814 & F814W & $21.397^{+0.004}_{-0.004}$ & $21.816^{+0.002}_{-0.003}$ & $20.628^{+0.003}_{-0.003}$ & $21.533^{+0.003}_{-0.003}$\\[+0.75em]
 & F160W & $21.255^{+0.007}_{-0.005}$ & $21.491^{+0.007}_{-0.007}$ & $20.373^{+0.007}_{-0.005}$ & $21.203^{+0.005}_{-0.005}$\\[+0.75em]\hline
     \\[-0.75em]
 & F475X & $22.066^{+0.010}_{-0.009}$ & $22.699^{+0.024}_{-0.025}$ & $21.836^{+0.009}_{-0.009}$ & $22.872^{+0.019}_{-0.017}$\\[+0.75em]
PS J0030-1525 & F814W & $20.779^{+0.007}_{-0.008}$ & $20.855^{+0.015}_{-0.016}$ & $20.586^{+0.008}_{-0.009}$ & $20.661^{+0.012}_{-0.013}$\\[+0.75em]
 & F160W & $19.299^{+0.008}_{-0.009}$ & $20.142^{+0.031}_{-0.028}$ & $19.344^{+0.014}_{-0.015}$ & $18.799^{+0.017}_{-0.014}$\\[+0.75em]\hline
     \\[-0.75em]
 & F475X & $22.600^{+0.009}_{-0.009}$ & $20.600^{+0.002}_{-0.002}$ & $20.474^{+0.002}_{-0.002}$ & $20.431^{+0.002}_{-0.002}$\\[+0.75em]
DES J0053-2012 & F814W & $21.473^{+0.013}_{-0.011}$ & $19.767^{+0.004}_{-0.004}$ & $19.651^{+0.004}_{-0.003}$ & $19.589^{+0.004}_{-0.004}$\\[+0.75em]
 & F160W & $21.415^{+0.010}_{-0.010}$ & $19.048^{+0.004}_{-0.003}$ & $18.918^{+0.003}_{-0.003}$ & $19.016^{+0.004}_{-0.004}$\\[+0.75em]\hline
     \\[-0.75em]
 & F475X & $18.681^{+0.001}_{-0.001}$ & $17.256^{+0.001}_{-0.001}$ & $16.477^{+0.001}_{-0.001}$ & $16.712^{+0.001}_{-0.001}$\\[+0.75em]
PS J0147+4630 & F814W & $18.219^{+0.002}_{-0.001}$ & $16.495^{+0.001}_{-0.001}$ & $15.824^{+0.001}_{-0.001}$ & $16.073^{+0.001}_{-0.001}$\\[+0.75em]
 & F160W & $18.068^{+0.007}_{-0.006}$ & $16.102^{+0.002}_{-0.002}$ & $15.376^{+0.001}_{-0.001}$ & $15.756^{+0.002}_{-0.002}$\\[+0.75em]\hline
     \\[-0.75em]
 & F475X & $20.664^{+0.003}_{-0.003}$ & $20.603^{+0.004}_{-0.003}$ & $20.617^{+0.004}_{-0.004}$ & $21.405^{+0.006}_{-0.005}$\\[+0.75em]
WG0214-2105 & F814W & $20.390^{+0.002}_{-0.002}$ & $20.339^{+0.003}_{-0.003}$ & $20.517^{+0.006}_{-0.005}$ & $21.260^{+0.008}_{-0.006}$\\[+0.75em]
 & F160W & $19.874^{+0.004}_{-0.003}$ & $19.837^{+0.003}_{-0.003}$ & $19.920^{+0.006}_{-0.005}$ & $20.601^{+0.006}_{-0.006}$\\[+0.75em]\hline
     \\[-0.75em]
 & F475X & $21.255^{+0.002}_{-0.002}$ & $20.981^{+0.002}_{-0.002}$ & $21.005^{+0.002}_{-0.002}$ & $21.792^{+0.003}_{-0.004}$\\[+0.75em]
SDSS J0248+1913 & F814W & $20.405^{+0.002}_{-0.002}$ & $20.174^{+0.002}_{-0.001}$ & $20.172^{+0.002}_{-0.002}$ & $20.715^{+0.006}_{-0.002}$\\[+0.75em]
 & F160W & $20.276^{+0.006}_{-0.005}$ & $20.056^{+0.008}_{-0.008}$ & $20.067^{+0.007}_{-0.007}$ & $20.502^{+0.007}_{-0.007}$\\[+0.75em]\hline
     \\[-0.75em]
 & F475X & $21.191^{+0.002}_{-0.002}$ & $21.083^{+0.003}_{-0.003}$ & $20.493^{+0.003}_{-0.002}$ & $20.331^{+0.003}_{-0.003}$\\[+0.75em]
WISE J0259-1635 & F814W & $19.731^{+0.004}_{-0.004}$ & $19.424^{+0.003}_{-0.003}$ & $19.024^{+0.003}_{-0.003}$ & $18.747^{+0.003}_{-0.003}$\\[+0.75em]
 & F160W & $18.948^{+0.005}_{-0.005}$ & $18.756^{+0.003}_{-0.003}$ & $18.331^{+0.003}_{-0.003}$ & $18.059^{+0.003}_{-0.003}$\\[+0.75em]\hline
     \\[-0.75em]
 & F475X & $24.532^{+0.078}_{-0.057}$ & $-$  & $-$  & $-$ \\[+0.75em]
J0343-2828 & F814W & $24.166^{+0.100}_{-0.069}$ & $21.808^{+0.052}_{-0.045}$ & $22.118^{+0.056}_{-0.046}$ & $22.452^{+0.069}_{-0.053}$\\[+0.75em]
 & F160W & $-$  & $-$  & $-$  & $-$ \\[+0.75em]\hline
     \\[-0.75em]
 & F475X & $21.754^{+0.004}_{-0.004}$ & $22.473^{+0.005}_{-0.005}$ & $21.588^{+0.004}_{-0.004}$ & $21.753^{+0.005}_{-0.005}$\\[+0.75em]
DES J0405-3308 & F814W & $19.975^{+0.002}_{-0.002}$ & $20.482^{+0.003}_{-0.003}$ & $19.863^{+0.003}_{-0.003}$ & $20.083^{+0.002}_{-0.003}$\\[+0.75em]
 & F160W & $19.675^{+0.012}_{-0.010}$ & $19.798^{+0.008}_{-0.009}$ & $19.480^{+0.013}_{-0.012}$ & $19.883^{+0.013}_{-0.014}$\\[+0.75em]\hline
     \\[-0.75em]
 & F475X & $21.779^{+0.006}_{-0.005}$ & $19.921^{+0.004}_{-0.004}$ & $20.537^{+0.005}_{-0.004}$ & $21.562^{+0.005}_{-0.005}$\\[+0.75em]
DES J0420-4037 & F814W & $21.759^{+0.007}_{-0.008}$ & $19.917^{+0.004}_{-0.004}$ & $20.577^{+0.007}_{-0.008}$ & $21.393^{+0.007}_{-0.007}$\\[+0.75em]
 & F160W & $21.781^{+0.022}_{-0.019}$ & $20.352^{+0.017}_{-0.015}$ & $20.738^{+0.025}_{-0.022}$ & $22.493^{+0.092}_{-0.066}$\\[+0.75em]\hline
     \\[-0.75em]
 & F475X & $21.170^{+0.074}_{-0.030}$ & $19.076^{+0.010}_{-0.008}$ & $20.016^{+0.049}_{-0.056}$ & $18.618^{+0.018}_{-0.013}$\\[+0.75em]
DES J0530-3730 & F814W & $21.220^{+0.132}_{-0.135}$ & $18.814^{+0.032}_{-0.020}$ & $19.030^{+0.041}_{-0.092}$ & $18.147^{+0.038}_{-0.013}$\\[+0.75em]
 & F160W & $-$  & $19.464^{+0.128}_{-0.093}$ & $18.827^{+0.057}_{-0.058}$ & $18.671^{+0.060}_{-0.067}$\\[+0.75em]\hline
     \\[-0.75em]
 & F475X & $23.477^{+0.023}_{-0.014}$ & $21.053^{+0.003}_{-0.003}$ & $21.162^{+0.004}_{-0.004}$ & $21.218^{+0.005}_{-0.004}$\\[+0.75em]
PS J0630-1201 & F814W & $22.270^{+0.008}_{-0.008}$ & $19.960^{+0.002}_{-0.002}$ & $20.056^{+0.003}_{-0.003}$ & $20.059^{+0.004}_{-0.003}$\\[+0.75em]
 & F160W & $21.020^{+0.009}_{-0.007}$ & $18.778^{+0.002}_{-0.002}$ & $18.692^{+0.002}_{-0.002}$ & $18.825^{+0.002}_{-0.002}$\\[+0.75em]\hline
     \\[-0.75em]
 & F475X & $20.757^{+0.004}_{-0.004}$ & $20.576^{+0.002}_{-0.002}$ & $20.930^{+0.004}_{-0.004}$ & $19.273^{+0.002}_{-0.002}$\\[+0.75em]
J0659+1629 & F814W & $20.280^{+0.003}_{-0.003}$ & $19.985^{+0.002}_{-0.002}$ & $20.383^{+0.004}_{-0.003}$ & $18.735^{+0.002}_{-0.002}$\\[+0.75em]
 & F160W & $19.021^{+0.003}_{-0.003}$ & $18.889^{+0.003}_{-0.002}$ & $19.276^{+0.004}_{-0.004}$ & $17.578^{+0.003}_{-0.002}$\\[+0.75em]\hline
     \\[-0.75em]
 & F475X & $20.876^{+0.003}_{-0.003}$ & $21.021^{+0.003}_{-0.003}$ & $19.075^{+0.002}_{-0.002}$ & $19.254^{+0.002}_{-0.002}$\\[+0.75em]
J0818-2613 & F814W & $19.688^{+0.002}_{-0.003}$ & $19.890^{+0.003}_{-0.002}$ & $17.889^{+0.002}_{-0.003}$ & $17.964^{+0.002}_{-0.002}$\\[+0.75em]
 & F160W & $18.688^{+0.002}_{-0.002}$ & $18.811^{+0.005}_{-0.008}$ & $16.887^{+0.002}_{-0.002}$ & $16.985^{+0.003}_{-0.004}$\\[+0.75em]\hline
     \\[-0.75em]
 & F475X & $-$  & $22.526^{+0.036}_{-0.031}$ & $23.709^{+0.045}_{-0.038}$ & $-$ \\[+0.75em]
W2M J1042+1641 & F814W & $23.213^{+0.044}_{-0.039}$ & $21.611^{+0.015}_{-0.016}$ & $23.454^{+0.086}_{-0.074}$ & $22.806^{+0.038}_{-0.036}$\\[+0.75em]
 & F160W & $21.291^{+0.016}_{-0.014}$ & $17.706^{+0.006}_{-0.006}$ & $20.344^{+0.017}_{-0.017}$ & $20.549^{+0.011}_{-0.013}$\\[+0.75em]\hline
  \end{tabular}
\end{table*}

\begin{table*}
  \caption{Median values for quasar (and, in case of J0343-2828, compact galaxy) image magnitudes in the AB system. The associated uncertainties are statistical in nature and were computed using 84th and 16th percentiles.}
  \label{tab:QSO_photometry_2}
  \scriptsize
%   \tiny
  \begin{tabular}{lccccc}
    \hline
    Name of Lens &  Filter & A & B & C & D \\
    \hline
    \\[-0.75em]
 & F475X & $21.252^{+0.006}_{-0.006}$ & $21.086^{+0.007}_{-0.007}$ & $20.456^{+0.012}_{-0.012}$ & $20.242^{+0.005}_{-0.007}$\\[+0.75em]
J1131-4419 & F814W & $20.629^{+0.010}_{-0.008}$ & $20.782^{+0.023}_{-0.009}$ & $20.035^{+0.007}_{-0.008}$ & $19.890^{+0.008}_{-0.008}$\\[+0.75em]
 & F160W & $20.317^{+0.012}_{-0.014}$ & $19.993^{+0.012}_{-0.009}$ & $19.420^{+0.012}_{-0.011}$ & $19.235^{+0.011}_{-0.012}$\\[+0.75em]\hline
     \\[-0.75em]
 & F475X & $18.554^{+0.001}_{-0.002}$ & $18.577^{+0.001}_{-0.001}$ & $18.631^{+0.002}_{-0.002}$ & $20.187^{+0.005}_{-0.003}$\\[+0.75em]
2M1134-2103 & F814W & $17.784^{+0.001}_{-0.001}$ & $17.943^{+0.003}_{-0.001}$ & $17.826^{+0.001}_{-0.001}$ & $19.679^{+0.006}_{-0.013}$\\[+0.75em]
 & F160W & $17.116^{+0.002}_{-0.003}$ & $17.346^{+0.007}_{-0.003}$ & $17.101^{+0.004}_{-0.003}$ & $19.238^{+0.015}_{-0.014}$\\[+0.75em]\hline
     \\[-0.75em]
 & F475X & $21.466^{+0.007}_{-0.007}$ & $21.864^{+0.009}_{-0.008}$ & $20.169^{+0.007}_{-0.007}$ & $19.183^{+0.002}_{-0.002}$\\[+0.75em]
SDSS J1251+2935 & F814W & $21.439^{+0.004}_{-0.004}$ & $21.797^{+0.012}_{-0.012}$ & $20.117^{+0.008}_{-0.009}$ & $19.143^{+0.004}_{-0.004}$\\[+0.75em]
 & F160W & $21.796^{+0.079}_{-0.075}$ & $22.027^{+0.165}_{-0.140}$ & $20.147^{+0.043}_{-0.045}$ & $19.305^{+0.018}_{-0.020}$\\[+0.75em]\hline
     \\[-0.75em]
 & F475X & $21.719^{+0.026}_{-0.013}$ & $20.982^{+0.014}_{-0.024}$ & $20.612^{+0.004}_{-0.003}$ & $20.670^{+0.012}_{-0.003}$\\[+0.75em]
2M1310-1714 & F814W & $21.162^{+0.015}_{-0.013}$ & $19.748^{+0.003}_{-0.003}$ & $19.671^{+0.003}_{-0.003}$ & $19.922^{+0.007}_{-0.005}$\\[+0.75em]
 & F160W & $21.211^{+0.029}_{-0.016}$ & $19.952^{+0.019}_{-0.012}$ & $19.422^{+0.008}_{-0.006}$ & $20.053^{+0.025}_{-0.011}$\\[+0.75em]\hline
     \\[-0.75em]
 & F475X & $22.812^{+0.015}_{-0.015}$ & $20.890^{+0.013}_{-0.012}$ & $21.410^{+0.011}_{-0.010}$ & $21.974^{+0.028}_{-0.026}$\\[+0.75em]
SDSS J1330+1810 & F814W & $21.497^{+0.015}_{-0.015}$ & $19.880^{+0.010}_{-0.010}$ & $20.169^{+0.009}_{-0.010}$ & $20.933^{+0.014}_{-0.015}$\\[+0.75em]
 & F160W & $20.266^{+0.017}_{-0.017}$ & $19.332^{+0.009}_{-0.009}$ & $19.517^{+0.011}_{-0.011}$ & $20.235^{+0.017}_{-0.015}$\\[+0.75em]\hline
     \\[-0.75em]
 & F475X & $21.972^{+0.004}_{-0.005}$ & $20.095^{+0.001}_{-0.001}$ & $20.468^{+0.002}_{-0.002}$ & $20.265^{+0.001}_{-0.001}$\\[+0.75em]
SDSS J1433+6007 & F814W & $21.782^{+0.008}_{-0.011}$ & $20.048^{+0.003}_{-0.003}$ & $20.365^{+0.004}_{-0.004}$ & $20.175^{+0.003}_{-0.003}$\\[+0.75em]
 & F160W & $21.793^{+0.012}_{-0.011}$ & $20.369^{+0.004}_{-0.004}$ & $20.455^{+0.007}_{-0.007}$ & $20.518^{+0.004}_{-0.005}$\\[+0.75em]\hline
     \\[-0.75em]
 & F475X & $21.203^{+0.014}_{-0.014}$ & $20.954^{+0.007}_{-0.006}$ & $21.137^{+0.013}_{-0.014}$ & $22.462^{+0.033}_{-0.033}$\\[+0.75em]
J1537-3010 & F814W & $21.049^{+0.016}_{-0.015}$ & $20.359^{+0.008}_{-0.007}$ & $20.743^{+0.014}_{-0.014}$ & $21.835^{+0.026}_{-0.023}$\\[+0.75em]
 & F160W & $20.263^{+0.026}_{-0.027}$ & $20.191^{+0.016}_{-0.014}$ & $20.206^{+0.022}_{-0.021}$ & $21.551^{+0.051}_{-0.049}$\\[+0.75em]\hline
     \\[-0.75em]
 & F475X & $19.524^{+0.004}_{-0.004}$ & $19.938^{+0.002}_{-0.002}$ & $19.712^{+0.005}_{-0.006}$ & $20.249^{+0.004}_{-0.004}$\\[+0.75em]
PS J1606-2333 & F814W & $18.920^{+0.002}_{-0.003}$ & $19.410^{+0.004}_{-0.004}$ & $19.118^{+0.004}_{-0.004}$ & $19.483^{+0.003}_{-0.003}$\\[+0.75em]
 & F160W & $19.450^{+0.011}_{-0.010}$ & $20.364^{+0.024}_{-0.023}$ & $19.830^{+0.014}_{-0.015}$ & $19.537^{+0.013}_{-0.013}$\\[+0.75em]\hline
     \\[-0.75em]
 & F475X & $20.837^{+0.004}_{-0.004}$ & $20.051^{+0.003}_{-0.003}$ & $19.349^{+0.002}_{-0.002}$ & $19.951^{+0.003}_{-0.003}$\\[+0.75em]
J1721+8842 & F814W & $20.476^{+0.007}_{-0.007}$ & $19.196^{+0.003}_{-0.002}$ & $18.621^{+0.002}_{-0.002}$ & $19.303^{+0.003}_{-0.003}$\\[+0.75em]
 & F160W & $20.099^{+0.017}_{-0.017}$ & $18.317^{+0.004}_{-0.004}$ & $17.783^{+0.005}_{-0.005}$ & $18.440^{+0.006}_{-0.007}$\\[+0.75em]\hline
     \\[-0.75em]
 & F475X & $20.796^{+0.005}_{-0.004}$ & $22.176^{+0.015}_{-0.014}$ & $19.836^{+0.002}_{-0.002}$ & $21.492^{+0.007}_{-0.007}$\\[+0.75em]
J1817+2729 & F814W & $20.305^{+0.009}_{-0.011}$ & $21.133^{+0.015}_{-0.014}$ & $19.229^{+0.008}_{-0.008}$ & $20.881^{+0.009}_{-0.009}$\\[+0.75em]
 & F160W & $18.936^{+0.010}_{-0.009}$ & $19.792^{+0.008}_{-0.007}$ & $18.111^{+0.008}_{-0.007}$ & $19.238^{+0.004}_{-0.004}$\\[+0.75em]\hline
     \\[-0.75em]
 & F475X & $21.368^{+0.005}_{-0.005}$ & $20.562^{+0.004}_{-0.004}$ & $20.524^{+0.003}_{-0.003}$ & $20.753^{+0.003}_{-0.003}$\\[+0.75em]
DES J2038-4008 & F814W & $20.567^{+0.010}_{-0.010}$ & $19.571^{+0.008}_{-0.009}$ & $19.632^{+0.008}_{-0.009}$ & $19.770^{+0.007}_{-0.007}$\\[+0.75em]
 & F160W & $19.640^{+0.022}_{-0.022}$ & $18.559^{+0.018}_{-0.017}$ & $18.508^{+0.017}_{-0.016}$ & $18.566^{+0.016}_{-0.014}$\\[+0.75em]\hline
     \\[-0.75em]
 & F475X & $22.158^{+0.008}_{-0.008}$ & $21.757^{+0.009}_{-0.009}$ & $20.103^{+0.003}_{-0.003}$ & $21.028^{+0.008}_{-0.008}$\\[+0.75em]
WG2100-4452 & F814W & $21.581^{+0.002}_{-0.002}$ & $21.240^{+0.003}_{-0.003}$ & $19.211^{+0.002}_{-0.002}$ & $20.292^{+0.003}_{-0.003}$\\[+0.75em]
 & F160W & $22.548^{+0.031}_{-0.028}$ & $21.534^{+0.022}_{-0.019}$ & $19.485^{+0.007}_{-0.006}$ & $20.359^{+0.017}_{-0.018}$\\[+0.75em]\hline
     \\[-0.75em]
 & F475X & $19.384^{+0.002}_{-0.002}$ & $19.717^{+0.002}_{-0.003}$ & $18.336^{+0.003}_{-0.007}$ & $17.872^{+0.001}_{-0.001}$\\[+0.75em]
J2145+6345 & F814W & $18.177^{+0.006}_{-0.005}$ & $18.551^{+0.006}_{-0.006}$ & $17.126^{+0.003}_{-0.003}$ & $16.746^{+0.002}_{-0.002}$\\[+0.75em]
 & F160W & $17.478^{+0.023}_{-0.026}$ & $17.768^{+0.014}_{-0.014}$ & $16.314^{+0.010}_{-0.010}$ & $15.914^{+0.004}_{-0.004}$\\[+0.75em]\hline
     \\[-0.75em]
 & F475X & $22.513^{+0.013}_{-0.010}$ & $22.130^{+0.010}_{-0.010}$ & $21.588^{+0.008}_{-0.007}$ & $23.103^{+0.015}_{-0.015}$\\[+0.75em]
J2205-3727 & F814W & $21.902^{+0.007}_{-0.007}$ & $21.554^{+0.007}_{-0.007}$ & $21.035^{+0.005}_{-0.005}$ & $22.285^{+0.009}_{-0.008}$\\[+0.75em]
 & F160W & $22.183^{+0.009}_{-0.008}$ & $21.643^{+0.010}_{-0.010}$ & $21.222^{+0.011}_{-0.010}$ & $22.494^{+0.021}_{-0.018}$\\[+0.75em]\hline
     \\[-0.75em]
 & F475X & $22.912^{+0.003}_{-0.004}$ & $21.684^{+0.001}_{-0.001}$ & $21.031^{+0.001}_{-0.001}$ & $21.623^{+0.002}_{-0.002}$\\[+0.75em]
ATLAS J2344-3056 & F814W & $21.835^{+0.007}_{-0.006}$ & $21.214^{+0.005}_{-0.004}$ & $20.785^{+0.003}_{-0.003}$ & $21.281^{+0.004}_{-0.004}$\\[+0.75em]
 & F160W & $21.253^{+0.023}_{-0.021}$ & $21.097^{+0.016}_{-0.015}$ & $20.421^{+0.020}_{-0.018}$ & $21.303^{+0.021}_{-0.019}$\\[+0.75em]\hline
  \end{tabular}
\end{table*}

\section{Fermat Potential Plots}

\begin{figure*}
 \includegraphics[width=\textwidth]{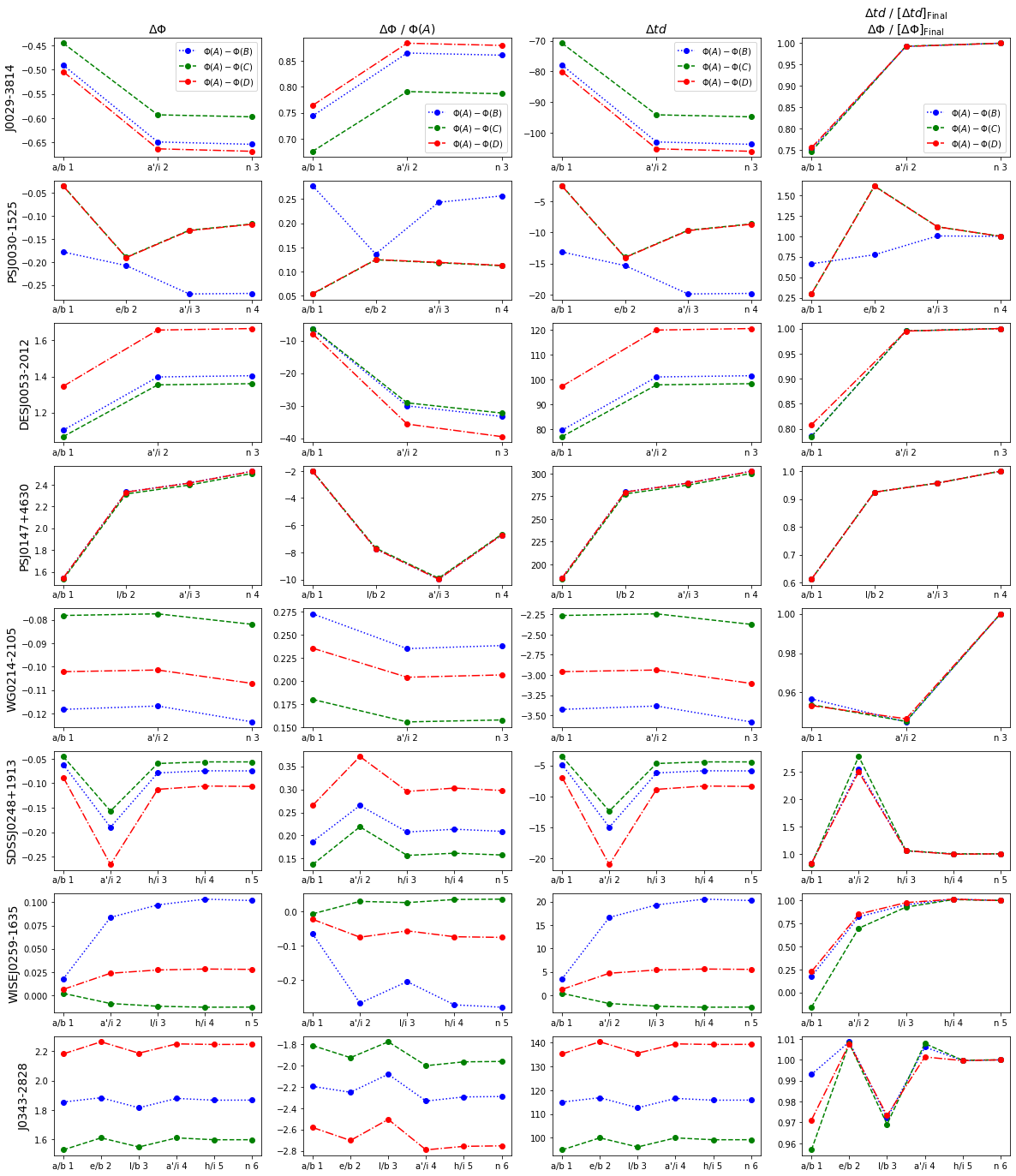}
 \caption{Difference in Fermat potential between image positions (column 1) and difference normalized by Fermat potential at image position A (column 2) for lens systems 1 - 8. Also shown are the differences in the predicted time delays between image positions associated with the Fermat potential differences (column 3). Column 4 shows the Fermat potential/time delay differences normalized by the final step in the reconstruction chain. In each plot, the dotted blue lines represent the difference between image A and B, the dashed green lines the difference between image A and C, and the dash-dotted red line the difference between image A and D.}
 \label{fig:td_plots_1}
\end{figure*}

\begin{figure*}
 \includegraphics[width=\textwidth]{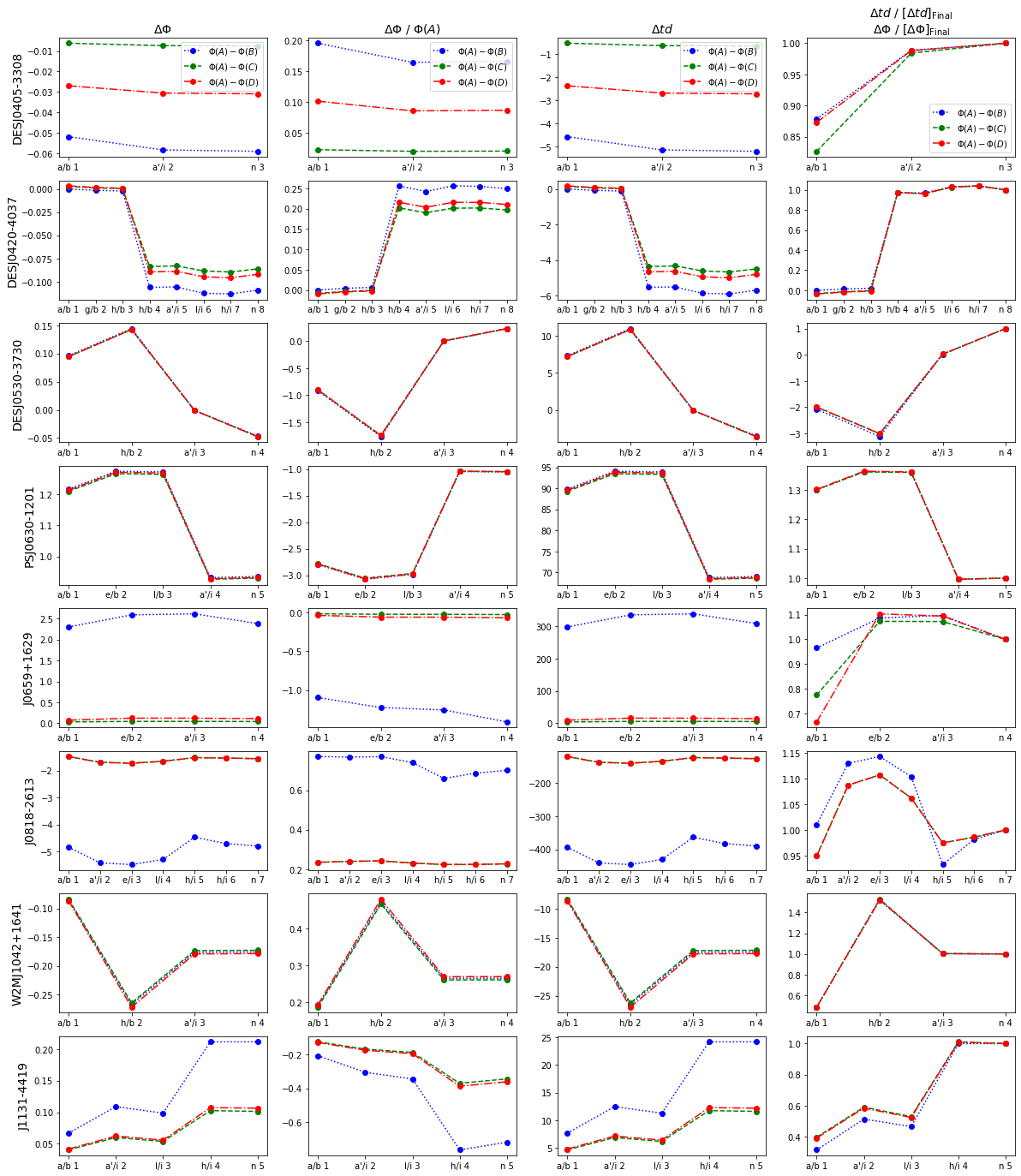}
 \caption{Difference in Fermat potential between image positions (column 1) and difference normalized by Fermat potential at image position A (column 2) for lens systems 9 - 16. Also shown are the differences in the predicted time delays between image positions associated with the Fermat potential differences (column 3). Column 4 shows the Fermat potential/time delay differences normalized by the final step in the reconstruction chain. In each plot, the dotted blue lines represent the difference between image A and B, the dashed green lines the difference between image A and C, and the dash-dotted red line the difference between image A and D.}
 \label{fig:td_plots_2}
\end{figure*}

\begin{figure*}
 \includegraphics[width=\textwidth]{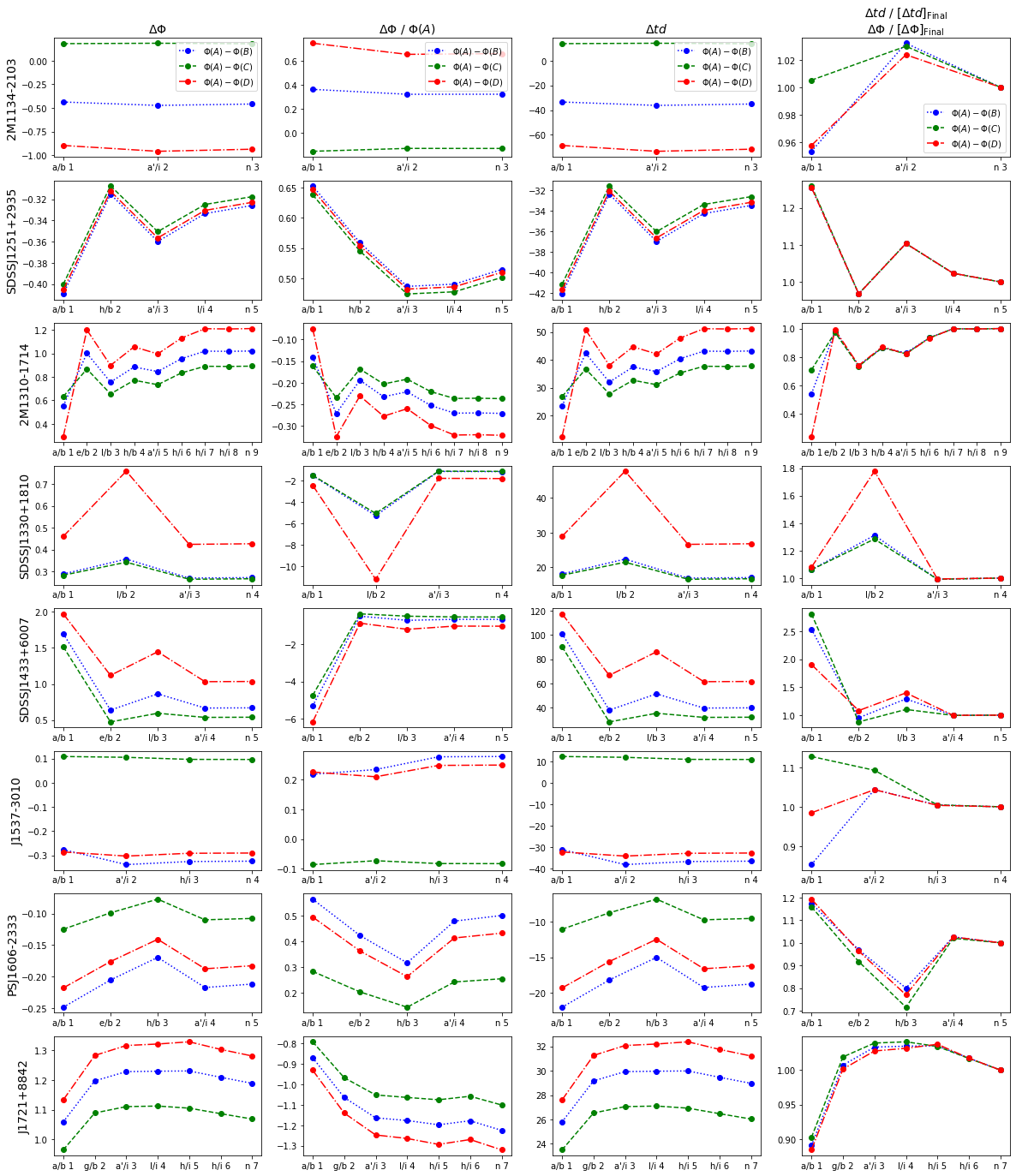}
 \caption{Difference in Fermat potential between image positions (column 1) and difference normalized by Fermat potential at image position A (column 2) for lens systems 17 - 24. Also shown are the differences in the predicted time delays between image positions associated with the Fermat potential differences (column 3). Column 4 shows the Fermat potential/time delay differences normalized by the final step in the reconstruction chain. In each plot, the dotted blue lines represent the difference between image A and B, the dashed green lines the difference between image A and C, and the dash-dotted red line the difference between image A and D.}
 \label{fig:td_plots_3}
\end{figure*}

\begin{figure*}
 \includegraphics[width=\textwidth]{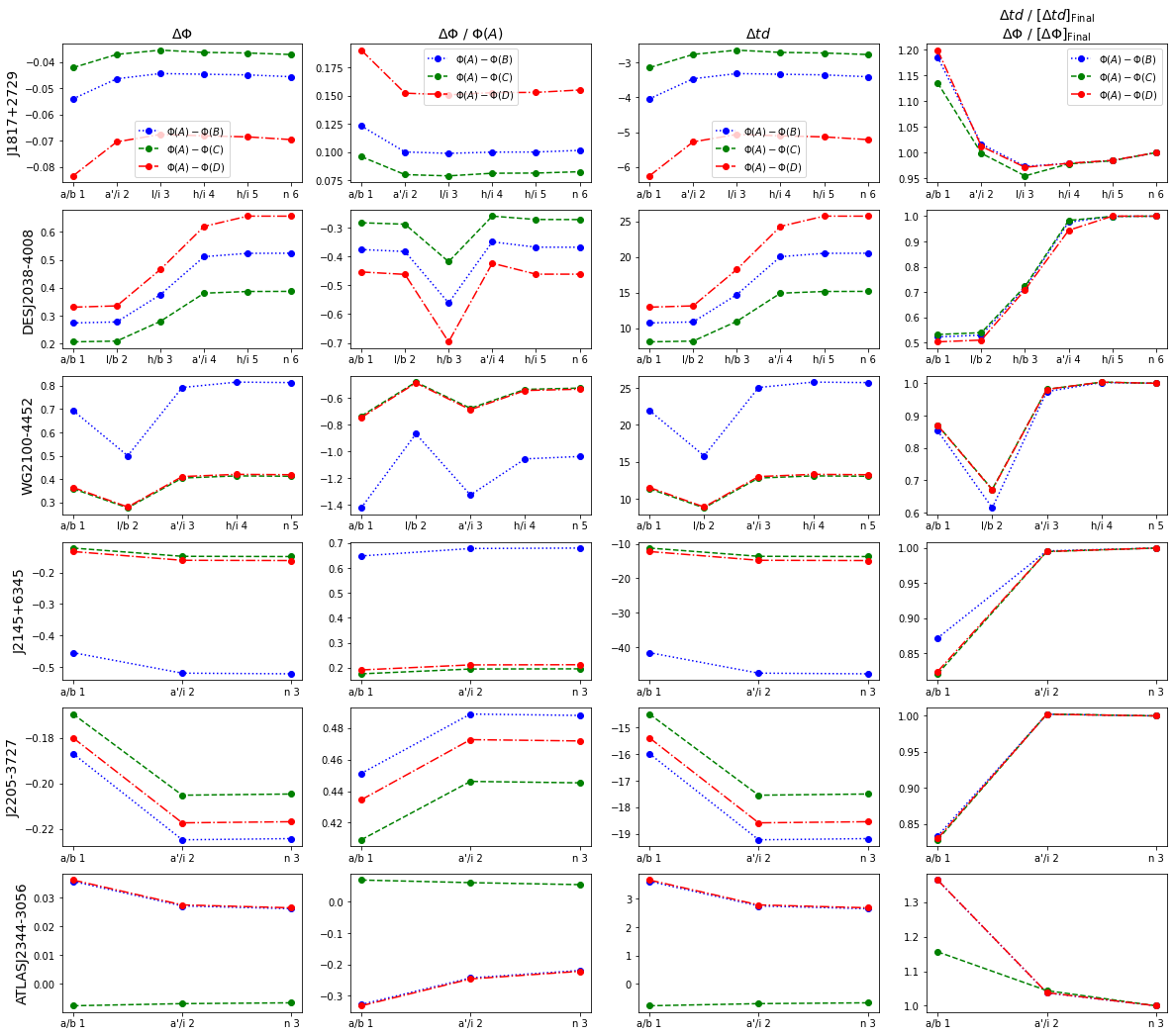}
 \caption{Difference in Fermat potential between image positions (column 1) and difference normalized by Fermat potential at image position A (column 2) for lens systems 25 - 30. Also shown are the differences in the predicted time delays between image positions associated with the Fermat potential differences (column 3). Column 4 shows the Fermat potential/time delay differences normalized by the final step in the reconstruction chain. In each plot, the dotted blue lines represent the difference between image A and B, the dashed green lines the difference between image A and C, and the dash-dotted red line the difference between image A and D.}
 \label{fig:td_plots_4}
\end{figure*}

In this section, we provide in Figures~\ref{fig:td_plots_1},~\ref{fig:td_plots_2},~\ref{fig:td_plots_3}, and~\ref{fig:td_plots_4} the evolution of the difference in Fermat potential between image positions throughout the lens modeling process, along with the associated evolution of the predicted time delay differences.

\section{Lens Models} \label{remaining_lens_models}

\begin{figure*}
 \includegraphics[width=\textwidth]{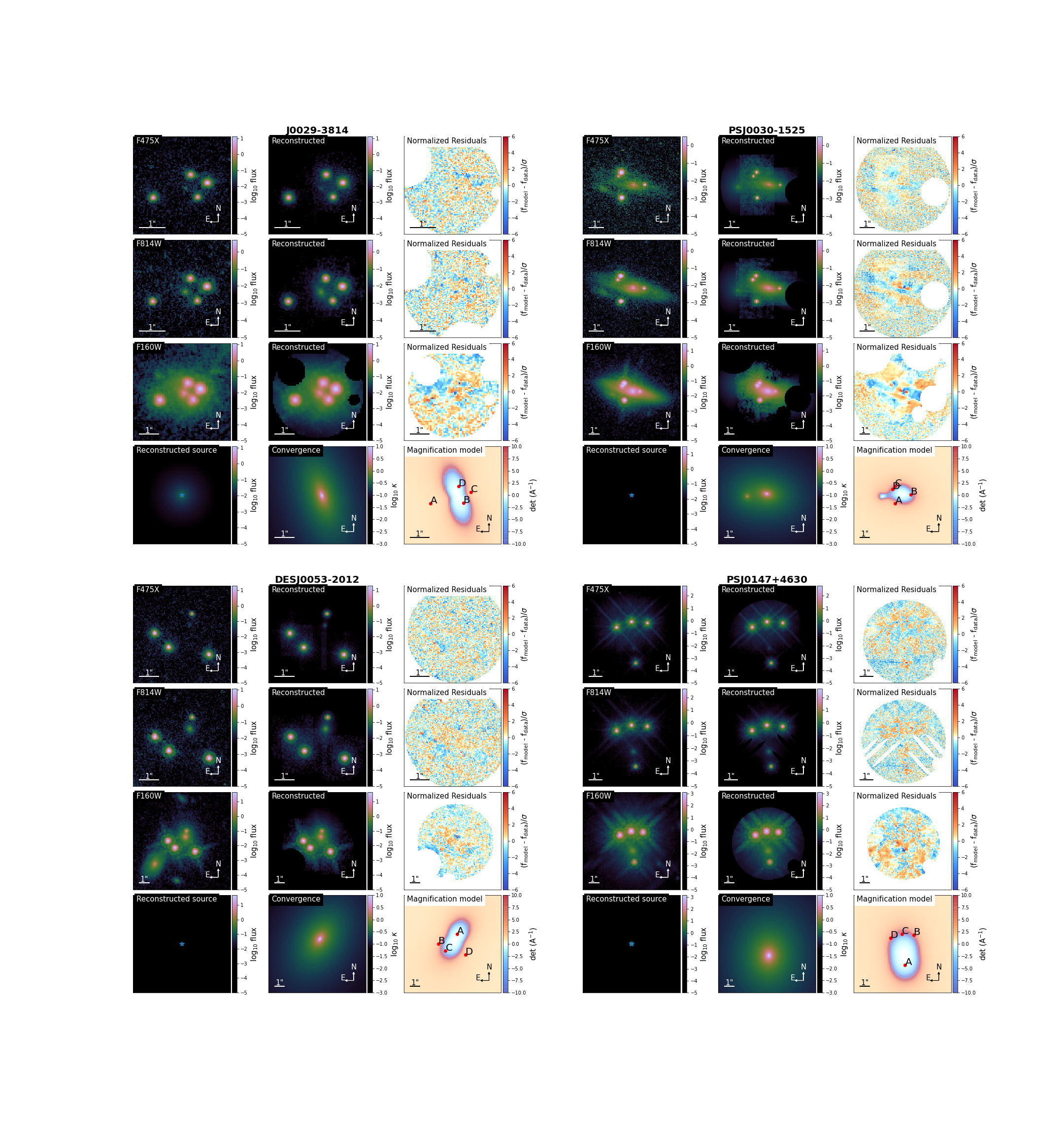}
 \caption{Comparison of observations with the reconstructed model for J0029-3814 (top left), PS J0030-1525 (top right), DES J0053-2012 (bottom left), and PS J0147+4630 (bottom right), in {\it{HST}} bands F475X (first row), F814W (second row), and F160W (third row). Also shown are the respective normalized residual for each band, after the subtraction of the data from the model. The last row shows the reconstructed source using information from the F160W band (column 1), a plot of the unitless convergence, $\kappa(\theta)$ (column 2), and a model plotting the magnification as well as the position of the lensed quasar images (column 3).}
 \label{fig:model_plots_1}
\end{figure*}

\begin{figure*}
 \includegraphics[width=\textwidth]{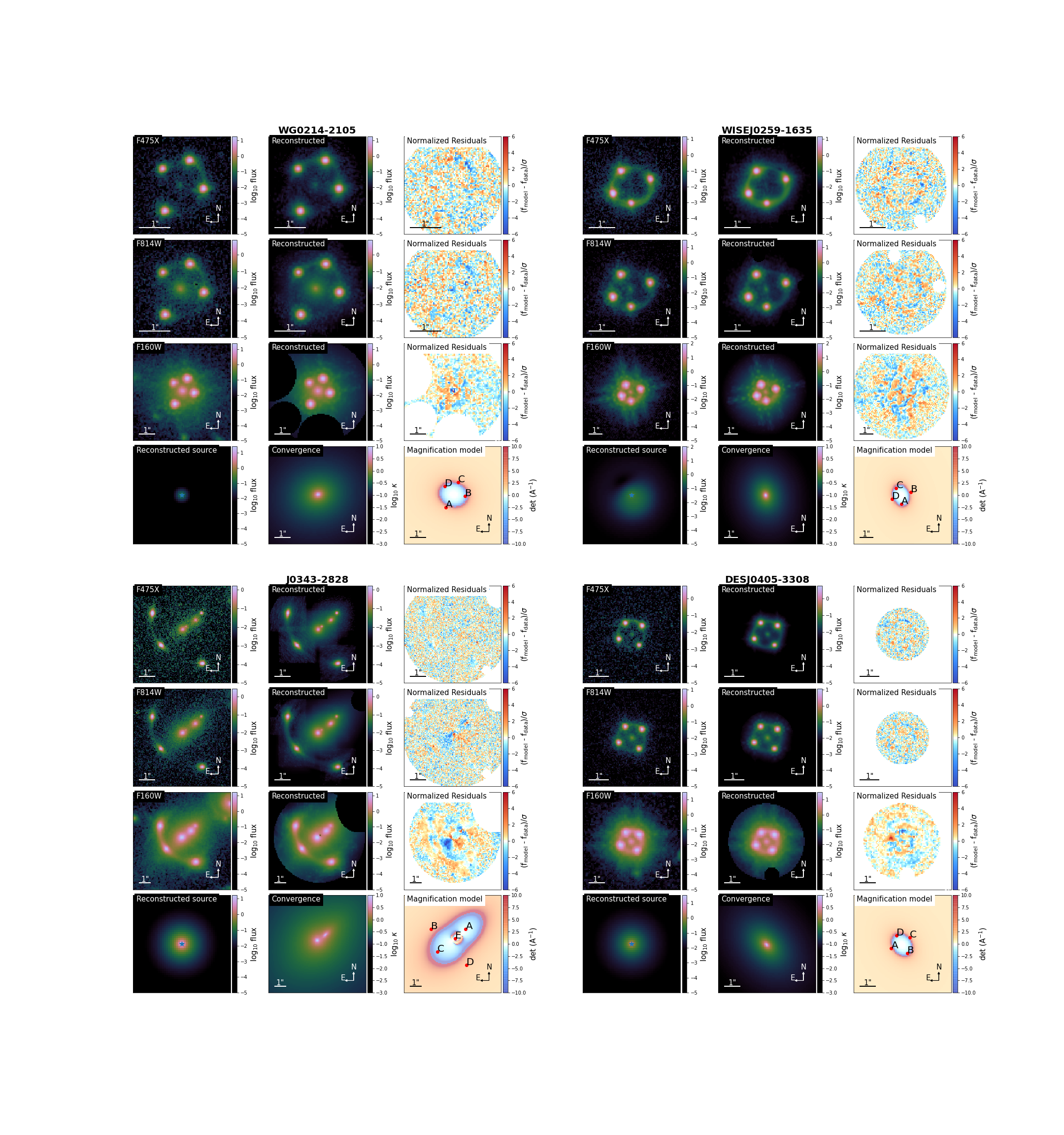}
 \caption{Comparison of observations with the reconstructed model for WG0214-2105 (top left), WISE J0259-1635 (top right), J0343-2828 (bottom left), and DES J0405-3308 (bottom right), in {\it{HST}} bands F475X (first row), F814W (second row), and F160W (third row). Also shown are the respective normalized residual for each band, after the subtraction of the data from the model. The last row shows the reconstructed source using information from the F160W band (column 1), a plot of the unitless convergence, $\kappa(\theta)$ (column 2), and a model plotting the magnification as well as the position of the lensed quasar images (column 3).}
 \label{fig:model_plots_2}
\end{figure*}

\begin{figure*}
 \includegraphics[width=\textwidth]{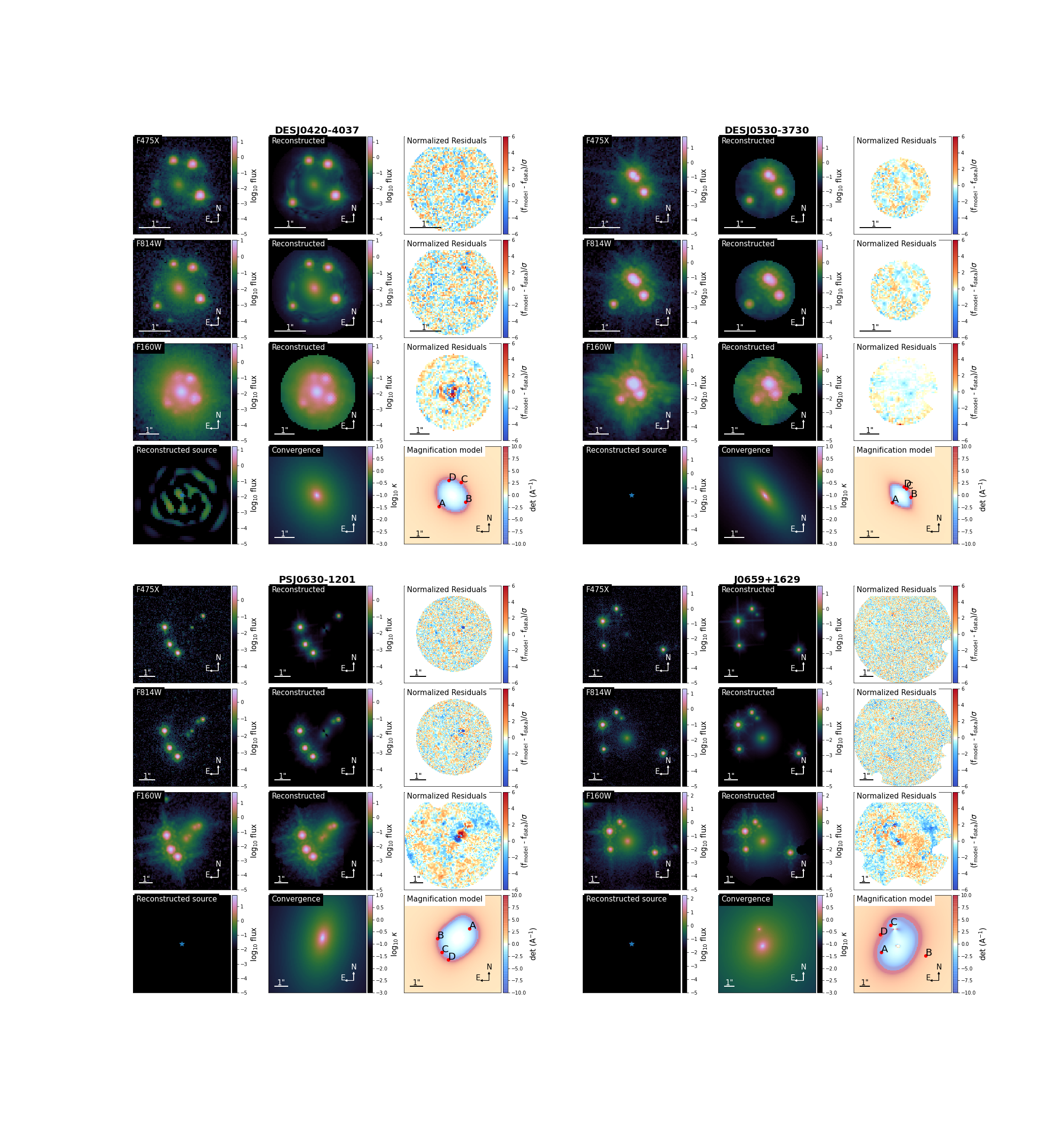}
 \caption{Comparison of observations with the reconstructed model for DES J0420-4037 (top left), DES J0530-3730 (top right), PS J0630-1201 (bottom left), and J0659+1629 (bottom right), in {\it{HST}} bands F475X (first row), F814W (second row), and F160W (third row). Also shown are the respective normalized residual for each band, after the subtraction of the data from the model. The last row shows the reconstructed source using information from the F160W band (column 1), a plot of the unitless convergence, $\kappa(\theta)$ (column 2), and a model plotting the magnification as well as the position of the lensed quasar images (column 3).}
 \label{fig:model_plots_3}
\end{figure*}

\begin{figure*}
 \includegraphics[width=\textwidth]{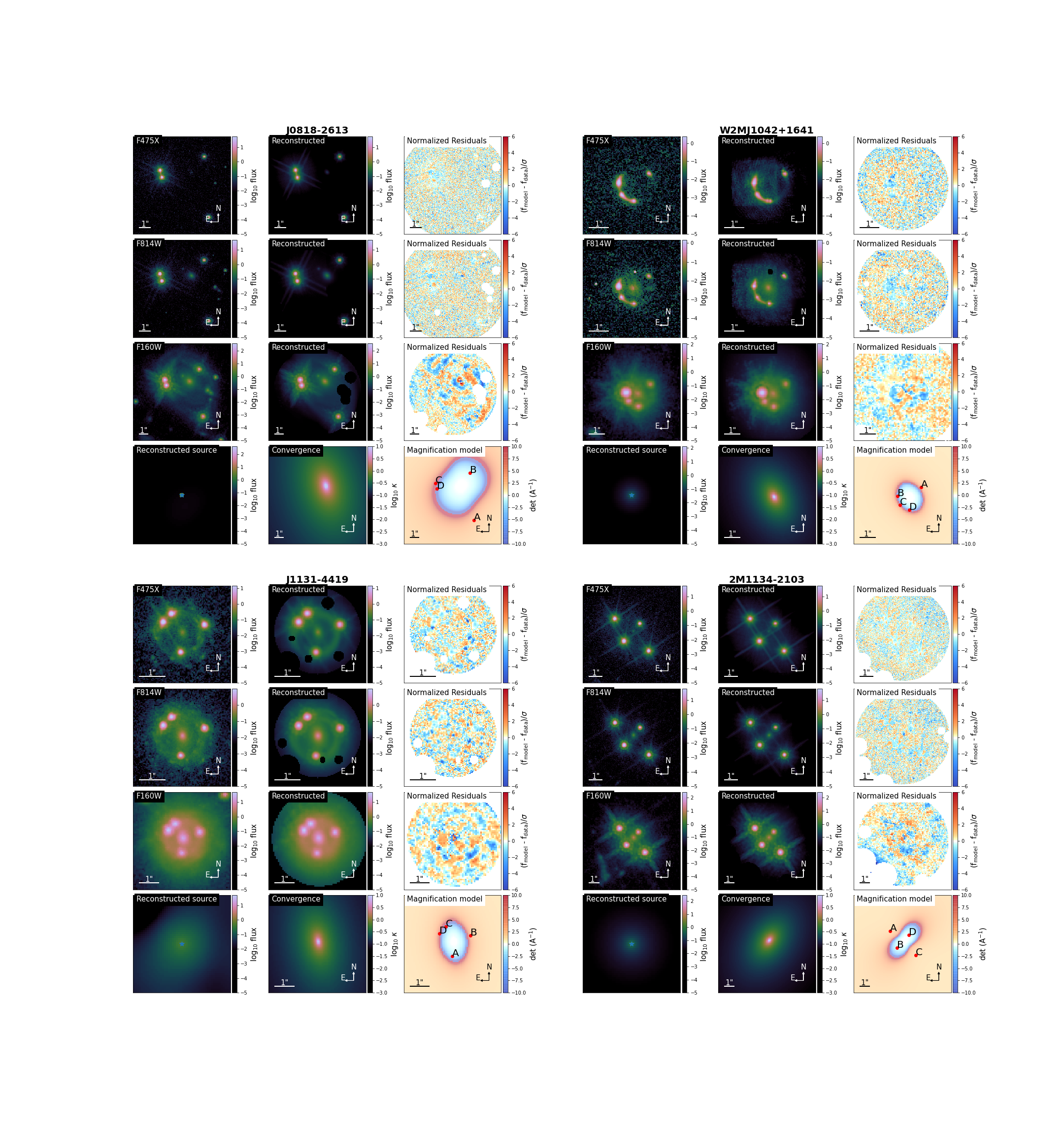}
 \caption{Comparison of observations with the reconstructed model for J0818-2613 (top left), W2M J1042+1641 (top right), J1131-4419 (bottom left), and 2M1134-2103 (bottom right), in {\it{HST}} bands F475X (first row), F814W (second row), and F160W (third row). Also shown are the respective normalized residual for each band, after the subtraction of the data from the model. The last row shows the reconstructed source using information from the F160W band (column 1), a plot of the unitless convergence, $\kappa(\theta)$ (column 2), and a model plotting the magnification as well as the position of the lensed quasar images (column 3).}
 \label{fig:model_plots_4}
\end{figure*}

\begin{figure*}
 \includegraphics[width=\textwidth]{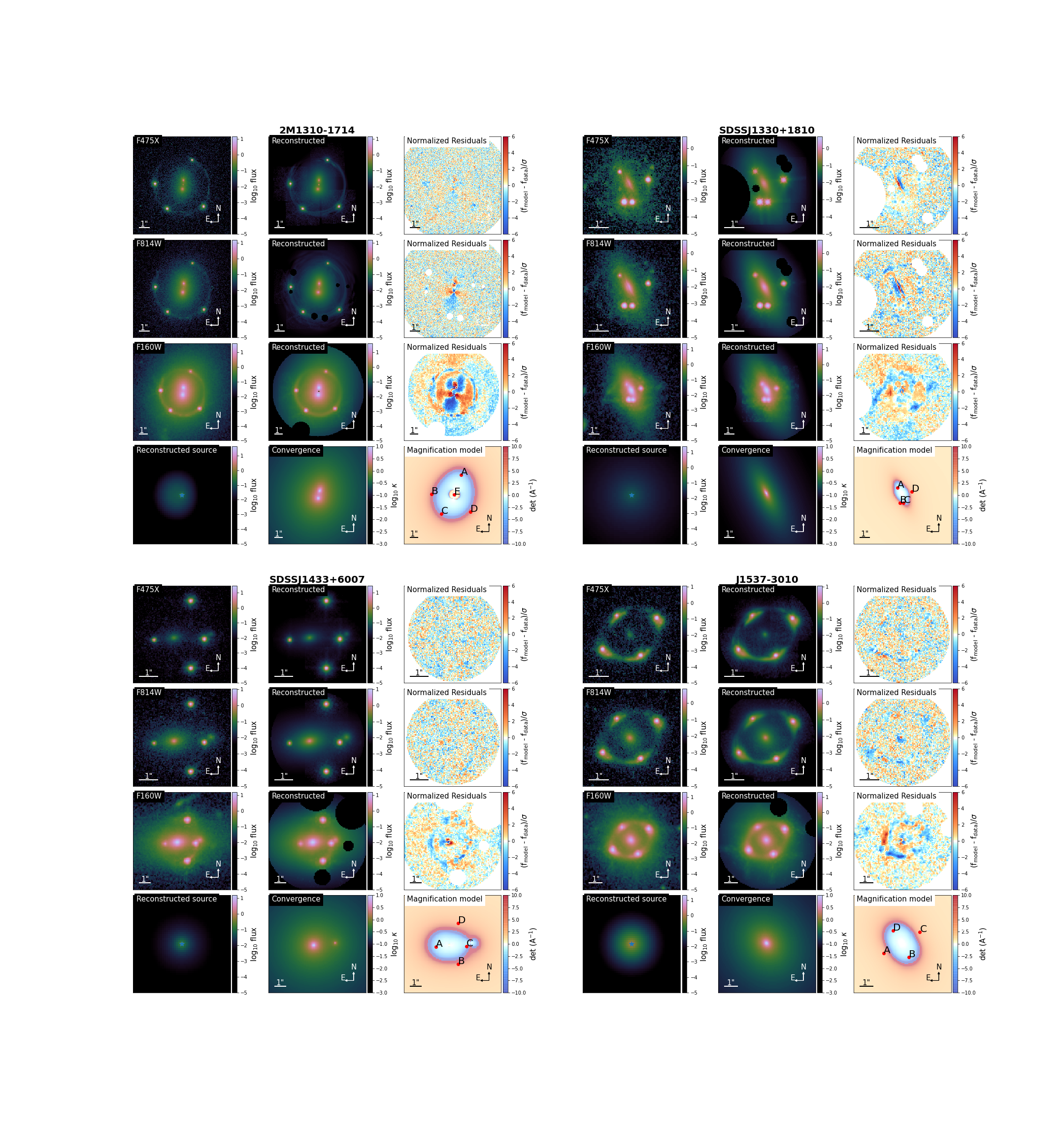}
 \caption{Comparison of observations with the reconstructed model for 2M1310-1714 (top left), SDSS J1330+1810 (top right), SDSS J1433+6007 (bottom left), and J1537-3010 (bottom right), in {\it{HST}} bands F475X (first row), F814W (second row), and F160W (third row). Also shown are the respective normalized residual for each band, after the subtraction of the data from the model. The last row shows the reconstructed source using information from the F160W band (column 1), a plot of the unitless convergence, $\kappa(\theta)$ (column 2), and a model plotting the magnification as well as the position of the lensed quasar images (column 3).}
 \label{fig:model_plots_5}
\end{figure*}

In addition to the models shown in Figure~\ref{fig:J0248} and in Figure~\ref{fig:J1251}, this section provides the remaining model plots for the lenses in our sample in Figures~\ref{fig:model_plots_1},~\ref{fig:model_plots_2},~\ref{fig:model_plots_3},~\ref{fig:model_plots_4},~\ref{fig:model_plots_5},~\ref{fig:model_plots_6}, and~\ref{fig:model_plots_7}.
The model plot for the failure mode is given in Appendix~\ref{Failure_modes}.

\begin{figure*}
 \includegraphics[width=\textwidth]{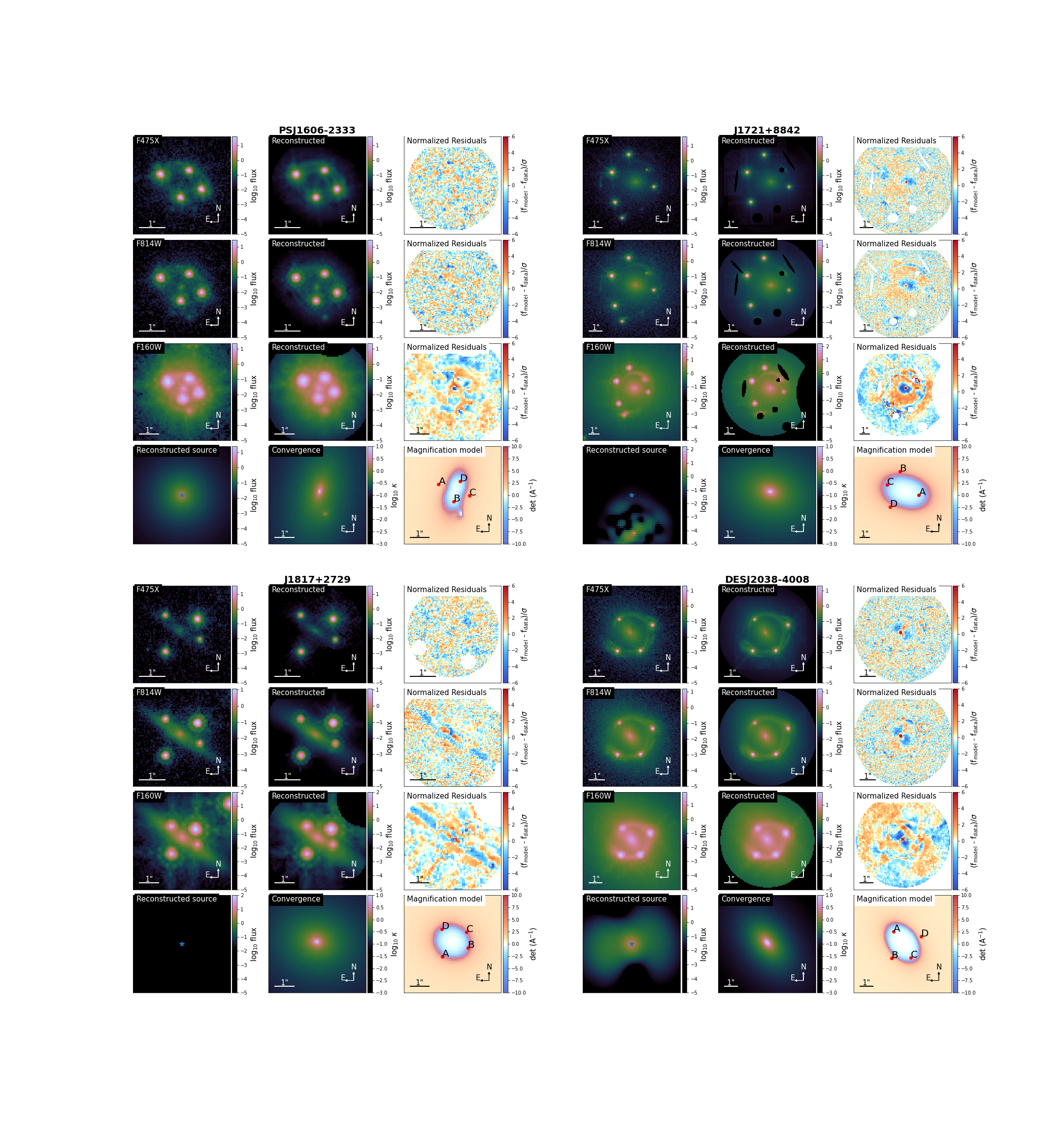}
 \caption{Comparison of observations with the reconstructed model for PS J1606-2333 (top left), J1721+8842 (top right), J1817+2729 (bottom left), and DES J2038-4008 (bottom right), in {\it{HST}} bands F475X (first row), F814W (second row), and F160W (third row). Also shown are the respective normalized residual for each band, after the subtraction of the data from the model. The last row shows the reconstructed source using information from the F160W band (column 1), a plot of the unitless convergence, $\kappa(\theta)$ (column 2), and a model plotting the magnification as well as the position of the lensed quasar images (column 3).}
 \label{fig:model_plots_6}
\end{figure*}

\begin{figure*}
 \includegraphics[width=\textwidth]{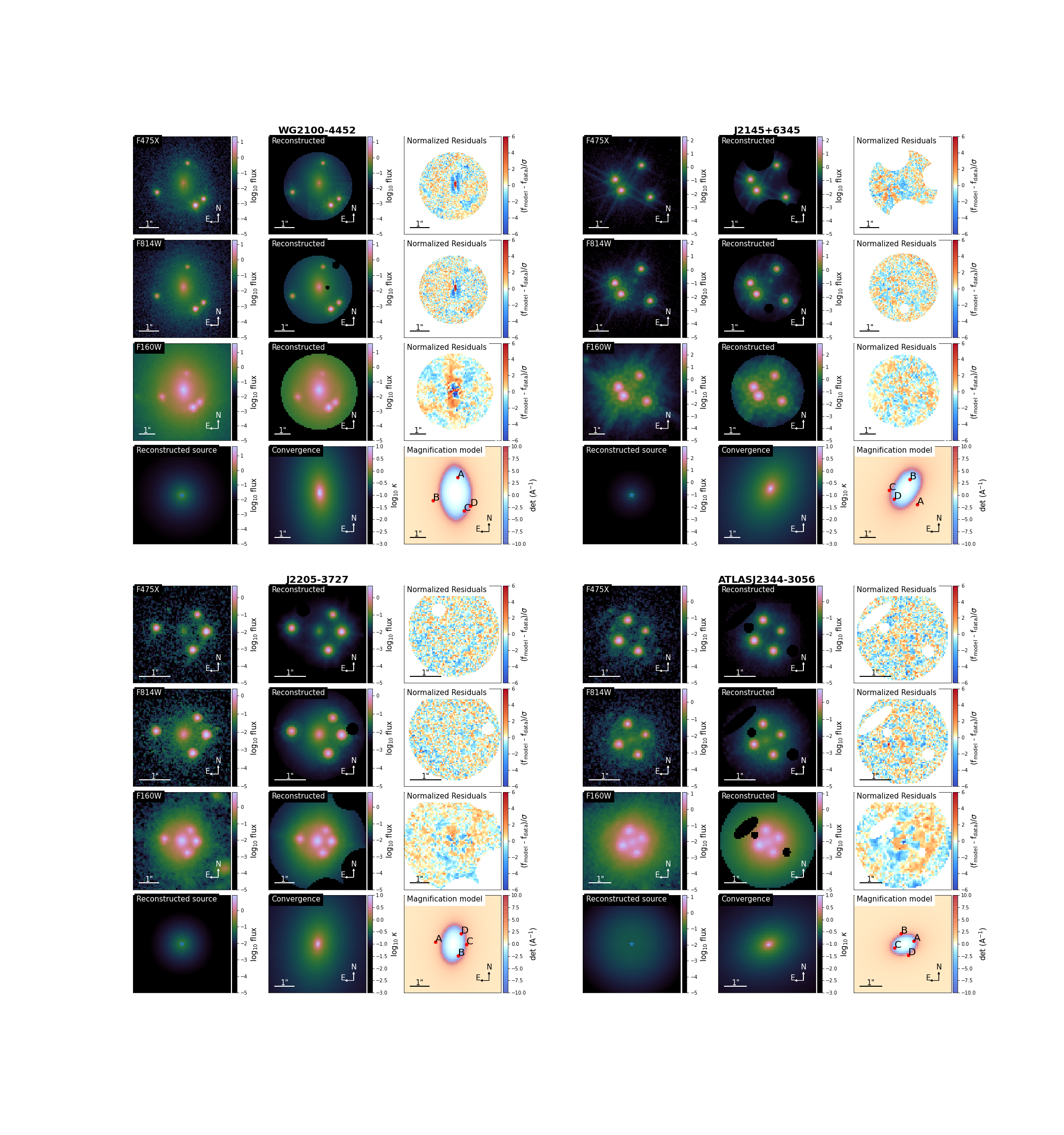}
 \caption{Comparison of observations with the reconstructed model for WG2100-4452 (top left), J2145+6345 (top right), J2205-3727 (bottom left), and ATLAS J2344-3056 (bottom right), in {\it{HST}} bands F475X (first row), F814W (second row), and F160W (third row). Also shown are the respective normalized residual for each band, after the subtraction of the data from the model. The last row shows the reconstructed source using information from the F160W band (column 1), a plot of the unitless convergence, $\kappa(\theta)$ (column 2), and a model plotting the magnification as well as the position of the lensed quasar images (column 3).}
 \label{fig:model_plots_7}
\end{figure*}

\section{Failure modes} \label{Failure_modes}

\begin{figure}
 \includegraphics[width=0.5\textwidth]{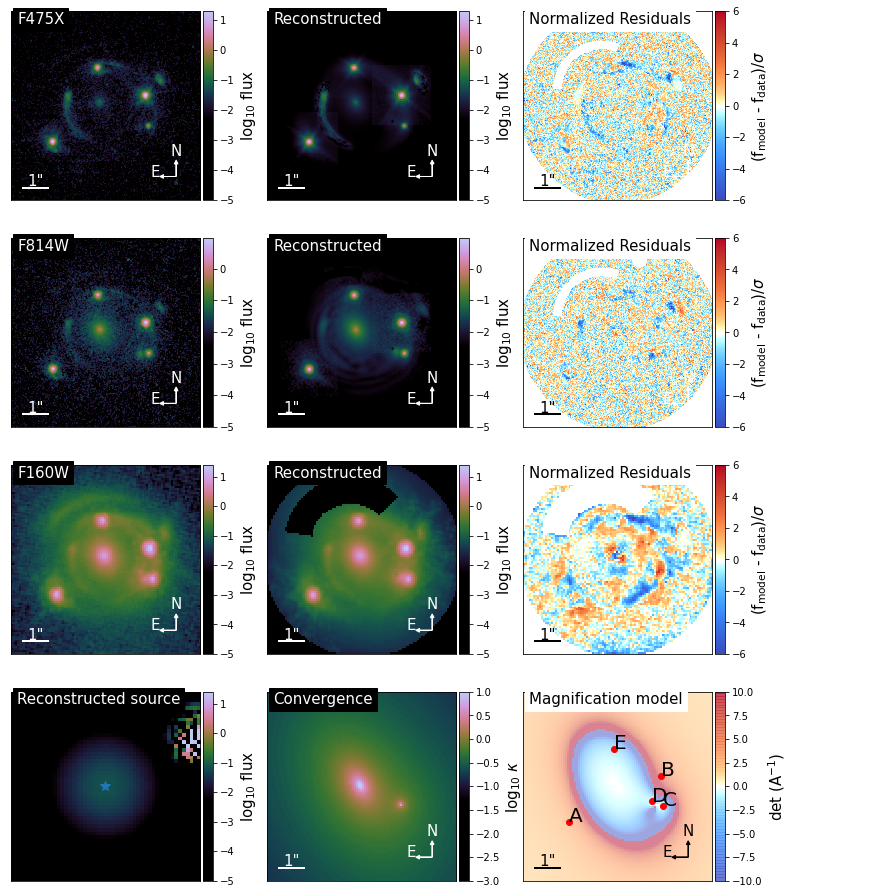}
 \caption{Comparison of observations with the reconstructed model for DES J0408-5354 in {\it{HST}} bands F475X (first row), F814W (second row), and F160W (third row). Also shown are the respective normalized residual for each band, after the subtraction of the data from the model. The last row shows the reconstructed source using information from the F160W band (column 1), a plot of the unitless convergence, $\kappa(\theta)$ (column 2), and a model plotting the magnification as well as the position of the lensed quasar images (column 3).}
 \label{fig:J0408}
\end{figure}

Our pipeline failed to produce a model for DES J0408-5354 with a sufficiently large p-vlaue or a $\chi^2$-value below the threshold of 1.10. The main reason for the failure is the secondary lensed source, which has a different redshift than the primary primary lensed source that holds the QSO. As the pipeline in its current form is limited to a single source plane, the two lensed sources are modeled to be at the same redshift (or in the same plane), causing the secondary source to appear slightly offset in the lens plane. This can be seen in the NW residuals, particularly visible in the UVIS bands. A model plot of the lens reconstruction using the final PSO iteration is included in Figure~\ref{fig:J0408}.

% \section{Convergence, shear, and stellar convergence}

% \section{Time delays}

% \section{Lens models}

%%%%%%%%%%%%%%%%%%%%%%%%%%%%%%%%%%%%%%%%%%%%%%%%%%

% Don't change these lines
\bsp	% typesetting comment
\label{lastpage}
\end{document}